%% file: main.tex
\documentclass[10pt,journal]{IEEEtran}
\newif\ifshort
\newif\ifanonymous

\ifCLASSOPTIONcompsoc
  \usepackage[nocompress]{cite}
\else
  \usepackage{cite}
\fi

\usepackage{etex}
\usepackage{graphicx}
\usepackage{amsmath}
\usepackage{amssymb}
\usepackage{color}
\usepackage{tikz}
\usepackage{url}
\usepackage{placeins}
\usepackage{float}
\usepackage{slashbox}
\usepackage{multirow}
\usepackage{array}
\usepackage{footmisc}
\usepackage[bf,sc,small]{caption} 
\usepackage[bf]{subfigure}
\usepackage{listings}
\usepackage[boxed]{algorithm}
\usepackage{algorithmic}
\usepackage{layouts}
\usepackage{tabularx}
\usepackage{mathtools}
\usepackage{adjustbox}
\usepackage{enumitem}

\usepackage{xspace}
\usepackage{changepage}
\usepackage{epsfig}
\usepackage{booktabs}
\usepackage{hyperref}
\usepackage{microtype}  	

\hypersetup{colorlinks=true, citecolor=blue, linkcolor = red}

\DeclareMathAlphabet{\mathitbf}{OML}{cmm}{b}{it}
\DeclareMathAlphabet{\mathbfit}{OML}{cmm}{b}{it}

\definecolor{mygreen}{rgb}{0,0.6,0}
\definecolor{myred}{rgb}{1,0.333,0.333}

\renewcommand{\figurename}{Figure}

\newcommand\autorefeq[1]{\hyperref[#1]{Equation~\eqref{#1}}}%
\newcommand\autorefapp[1]{\hyperref[#1]{Appendix~\ref{#1}}}%
\newcommand\autorefalg[1]{\hyperref[#1]{Algorithm~\ref{#1}}}%
\newcommand\autorefcor[1]{\hyperref[#1]{Corollary~\ref{#1}}}%
\newcommand\autorefprop[1]{\hyperref[#1]{Proposition~\ref{#1}}}%
\newcommand\autorefproperty[1]{\hyperref[#1]{Property~\ref{#1}}}%

\usepackage[utf8]{inputenc}
\usepackage{pict2e,picture}
\newsavebox\CBox
\newlength\CLength
\def\Circled#1{\sbox\CBox{#1}%
  \ifdim\wd\CBox>\ht\CBox \CLength=\wd\CBox\else\CLength=\ht\CBox\fi
    \makebox[1.2\CLength]{\makebox(0,1.2\CLength){\put(0,0){\circle{1.4\CLength}}}%
    \makebox(0,1.2\CLength){\put(-.5\wd\CBox,0){#1}}}}

\newif\ifanonymous

\begin{document}

\title{Side-Channel Hardware Trojan for Provably-Secure SCA-Protected Implementations}

\ifanonymous
\author{}
\institute{}
\else
\author{Samaneh~Ghandali,
        Thorben~Moos, 
        Amir~Moradi, 
        Christof~Paar,~\IEEEmembership{Fellow,~IEEE}%
\thanks{S. Ghandali is with the Department
	of Electrical and Computer Engineering, University of Massachusetts, Amherst,
	MA,01003 USA\protect\\E-mail: samaneh@umass.edu.}
\thanks{T. Moos, A. Moradi and C. Paar are with the Horst G\"ortz Institute for IT-Security, Ruhr University Bochum, Bochum, Germany\protect\\E-mail: \{firstname.lastname\}@rub.de.}}

\fi
\renewcommand\footnotemark{}

\maketitle


\input{abstract}


\IEEEpeerreviewmaketitle

\input{intro}

\input{background}

\input{technique}

\input{application}
\input{ASIC_Implementation}

\input{results}

\input{conclude}


\ifanonymous
{ }
\else

\section*{Acknowledgments}

The work described in this paper has been supported in part by the Deutsche Forschungsgemeinschaft (DFG, German Research Foundation) under Germany's Excellence Strategy - EXC 2092 CASA - 390781972 and through the project 271752544 ``NaSCA: Nano-Scale Side-Channel Analysis''.

\fi




\bibliographystyle{IEEEtran}
\bibliography{bib/abbrev3,bib/abbrev2,bib/abbrev1,bib/crypto_custom,bib/eprint,bib/SCA,bib/trojan,bib/HT,bib/other,bib/cryptanalysis}
\vspace{-22 mm}
\begin{IEEEbiography}[{\includegraphics[width=1in,height=1.25in,clip,keepaspectratio]{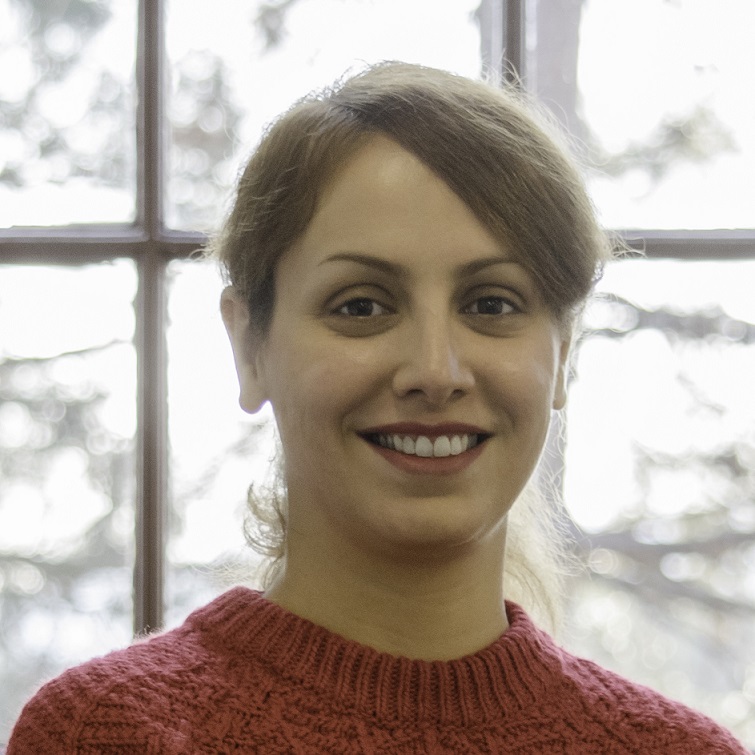}}
]{Samaneh Ghandali}
received the MSc degree in computer engineering from Shahid Beheshti University, Tehran, Iran, in 2009. Afterwards, till 2015 she worked as a graduate research assistant at  the University of Tehran, Tehran, Iran. She is currently working toward the PhD degree in computer engineering under the supervision of Prof. Christof Paar at the University of Massachusetts, Amherst, USA. Her current research interest is hardware security with a special focus on  physical security of embedded systems, hardware Trojans, side-channel analysis attacks and  the corresponding countermeasures.
\end{IEEEbiography}
\vspace{-22 mm}
\begin{IEEEbiography}[{\includegraphics[width=1in,height=1.25in,clip,keepaspectratio]{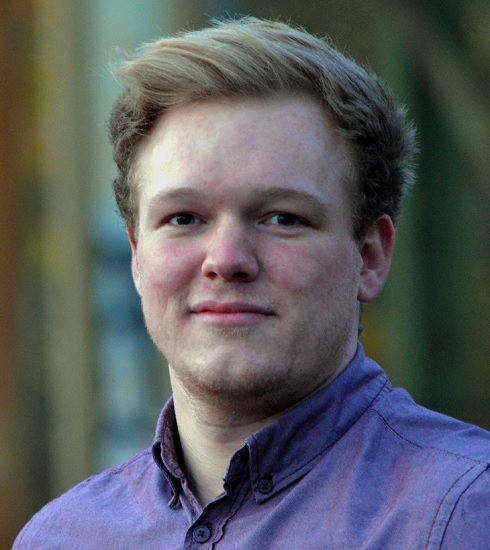}}]
    {M.Sc. Thorben Moos}
    received the B.Sc. and M.Sc. degrees in IT-Security from
Ruhr-Universit\"at Bochum, Germany in 2014 and 2016, respectively.
Currently, he is a Ph.D. student and scientific research assistant at the chair
for Embedded Security, Horst-Görtz Institute for IT-Security, Ruhr-Universit\"{a}t Bochum, Germany.
    His research interests include physical security of embedded devices
with specialization in nano-scale side-channel analysis and secure ASIC implementation.
\end{IEEEbiography}
\vspace{-22 mm}
\begin{IEEEbiography}[{\includegraphics[width=1in,height=1.25in,clip,keepaspectratio]{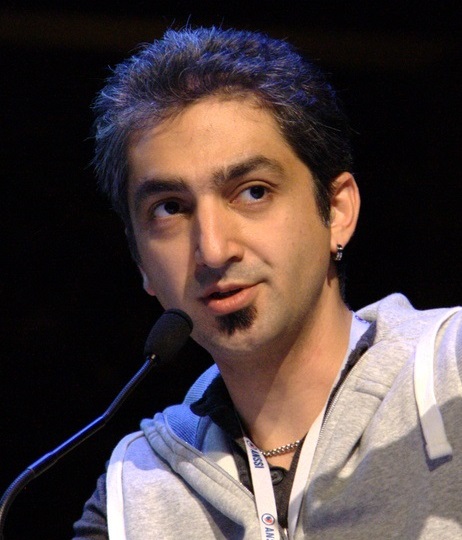}}]
{Priv.-Doz. Dr. Amir Moradi}
received the M.Sc. and Ph.D. degrees in computer engineering from Sharif University of Technology, Tehran, Iran, in 2004 and 2008 respectively. 
Afterwards, till 2015 he worked as a Post-Doctoral researcher at the chair for Embedded Security, Ruhr Universit\"{a}t Bochum, Germany. Since 2016, after obtaining the Habilitation degree, he has become a senior researcher and faculty member at the faculty of electrical engineering and information technology at Ruhr University Bochum. 
His current research interests include physical security of embedded systems, passive side-channel analysis attacks, and the corresponding countermeasures. 
He has published over 85 peer-reviewed journal articles and conference papers, in both destructive and constructive aspects of side-channel analysis. 
He also served as Program Committee Member (and the Chair) of several security- and cryptography-related conferences and workshops.	
\end{IEEEbiography}
\vspace{-22 mm}

\begin{IEEEbiography}[{\includegraphics[width=1in,height=1.25in,clip,keepaspectratio]{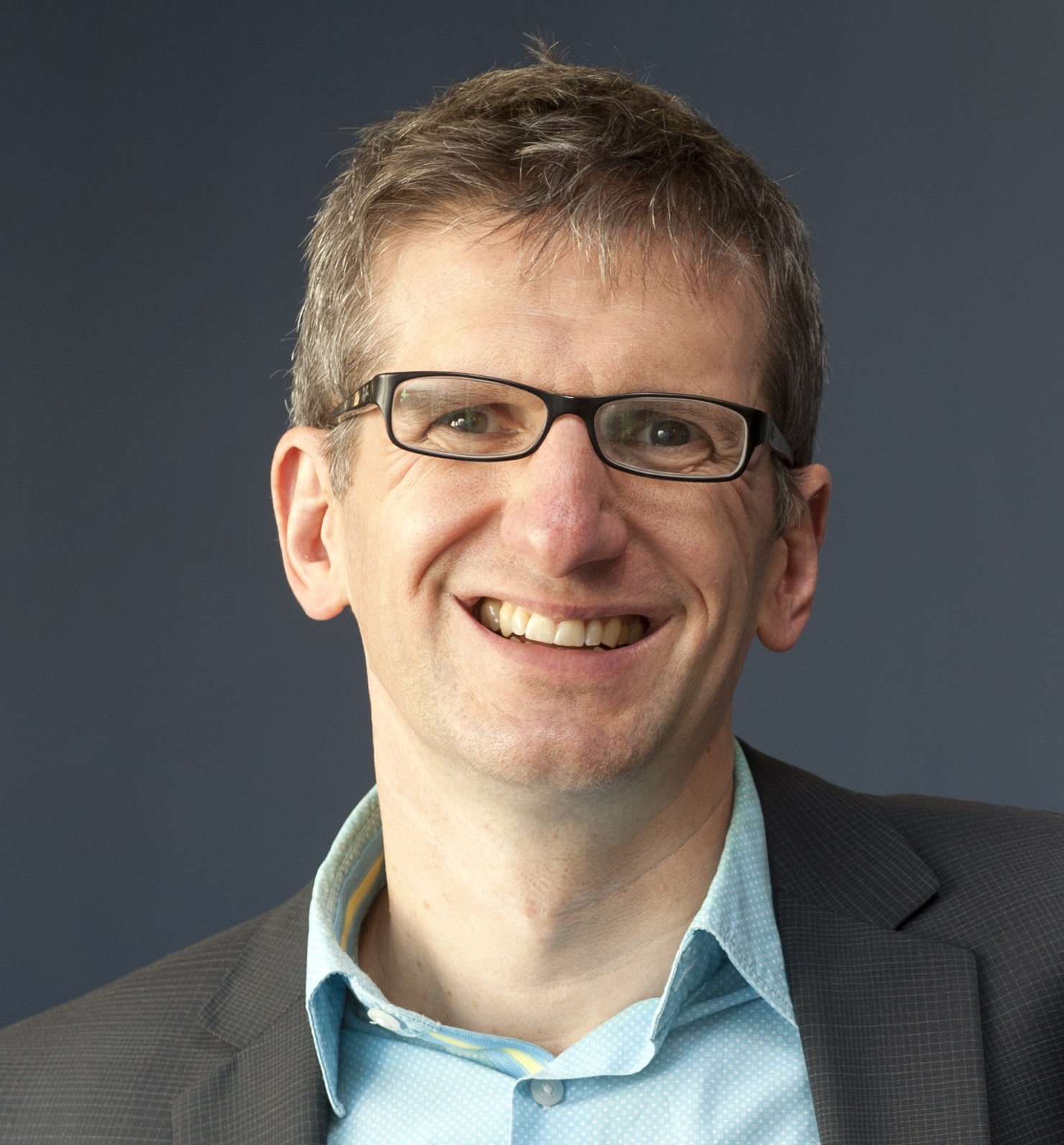}}
]{Prof. Dr.-Ing. Christof Paar (Fellow, IEEE)}
received his M.Sc. degree from the University of Siegen and the Ph.D. degree from the Institute for Experimental Mathematics at the University of Essen, both in Germany.
He holds the Chair for Embedded Security at Ruhr University Bochum, Germany, and is Affiliated Professor at the University of Massachusetts Amherst, USA. He co-founded, with C. Koc, the Conference on Cryptographic Hardware and Embedded Systems (CHES). He has over 200 peer-reviewed publications and is coauthor of the textbook Understanding Cryptography (New York, NY, USA: Springer-Verlag, 2010).  He is a co-founder of ESCRYPT -- Embedded Security, a leading consultancy firm in applied security that is now part of Bosch. His research interests include the efficient realizations of cryptography, hardware Trojans, physical security and security evaluation of real-world systems.
\end{IEEEbiography}

\end{document}

%% file: abstract.tex
\begin{abstract}

Hardware Trojans have drawn the attention of academia, industry and government agencies. 
Effective detection mechanisms and countermeasures against such malicious designs can only be developed when there is a deep understanding of how hardware Trojans can be built in practice, in particular Trojans specifically designed to avoid detection. In this work, we present a mechanism to introduce an extremely stealthy hardware Trojan into cryptographic primitives equipped with provably-secure first-order side-channel countermeasures.
Once the Trojan is triggered, the malicious design exhibits exploitable side-channel leakage, leading to successful key recovery attacks.
Generally, such a Trojan requires neither addition nor removal of any logic which makes it extremely hard to detect.
On ASICs, it can be inserted by subtle manipulations at the sub-transistor level and on FPGAs by changing the routing of particular signals, leading to \textbf{zero} logic overhead.
The underlying concept is based on modifying a securely-masked hardware implementation in such a way that running the device at a particular clock frequency violates one of its essential properties, leading to exploitable leakage. We apply our technique to a Threshold Implementation of the PRESENT block cipher realized in two different CMOS technologies, and show that triggering the Trojan makes the ASIC prototypes vulnerable. 

\end{abstract}

%% file: intro.tex
\section{Introduction}\label{sec:intro}

Cryptographic primitives are often the most trusted components in modern security solutions, ranking from network routers to IoT devices. Unfortunately, this makes cryptographic algorithms an attractive target for subversion by malicious actors. Manipulating hardware implementations as opposed to software implementations can lead to cryptographic Trojans that are particularly difficult to detect. It is widely believed that such Trojans are of special interest to nation-state adversaries. 

Hardware Trojans have moved in the focus of academia, industry and government agencies over the last decade. 
In particular, Trojan detection 
has become an active research area. At the same time, the development of effective detection mechanisms and countermeasures require a thorough understanding of how hardware Trojans can be built. This contribution is concerned with cryptographic Trojans which possess zero overhead in terms of logic resources and are thus, extremely stealthy. 


There are several paths to introduce a Trojan into an IC during the design cycle. Three general approaches are: insertion ($i$) by an untrusted semiconductor foundry during manufacturing, ($ii$) by the original hardware designer, possibly pressured by a government body or through subverted design tools, and ($iii$) through third-party IP cores.
Most hardware Trojans require the modification or insertion of additional logic resources (which can be done at different abstraction levels).
In most cases, adversaries attempt to design and implement Trojans in such a way that the chance of detection becomes very low. 
Our focus in this contribution are Trojans which disclose crucial secrets through side channels. 
The first such Trojan has been introduced in~\cite{DBLP:conf/iccad/LinBP09, DBLP:conf/ches/LinKGPB09}, which stealthily leaks out the cryptographic key through a power side channel. The underlying Trojan construction is independent of the cryptographic algorithm and is focused on key-leakage. The Trojan, based on a moderately large circuit including an LFSR and leaking circuit, is inserted at the netlist or HDL level. Such Trojans, however, can be detected through standard VLSI analysis techniques, e.g., through imaging-based reverse engineering. Another attack vector for hardware Trojans is the subversion of side-channel countermeasures. Cryptographic implementations are often threatened by side-channel analysis (SCA).
Two decades after the introduction of SCA attacks~\cite{DBLP:conf/crypto/Kocher96,dpa_kocher}, integration of dedicated countermeasures is a must in many applications.
In a follow-up work to reference \cite{DBLP:conf/ches/LinKGPB09}, a related concept that subverts an SCA-protected implementation is introduced \cite{DBLP:journals/jce/KasperMBMGPB12}. 
The technique is based on inserting a logical circuit forming an LFSR-based Trojan, which leaks the internal state of the PRNG used for masking. 
The adversary can now detect the internal state of the PRNG by means of SCA leakages, and can conduct DPA attacks based on her knowledge of the mask. 
It should be noted that products which need to be protected against physical attacks are often evaluated by a third-party certification body, e.g., through a Common Criteria evaluation lab.
Therefore, due to its relatively large circuit, such a Trojan will likely be detected by an inspector. Another relevant prior work is reference~\cite{DBLP:conf/ches/BeckerRPB13}, where a Trojan is inserted by changing the dopant polarity of a few transistors in a circuit that realizes the DPA-resistant logic style iMDPL~\cite{DBLP:conf/ches/PoppKZM07}.
However, iMPDL (and related logic styles) do not provide perfect SCA protection, and the leakage of an iMDPL circuit can still be exploited by ordinary SCA adversaries, even without a Trojan~\cite{DBLP:journals/tvlsi/MoradiKEP12}. In reference~\cite{yang2016a2} analog and RF Trojans are introduced, which can detect an extremely rare sequential event as a trigger with just a handful of transistors added to the circuit. This poses a real threat to the IC supply chain. However, since such analog Trojans are triggered by high frequency wire-flops in the processor, abnormal toggling detection methods may detect them~\cite{hou2018r2d2}.

\clearpage
\noindent
\textbf{Our contribution:}
Integrating a side-channel Trojan into an SCA-protected design is extremely challenging if the device will be evaluated by a third-party certification body. In this practical setting, the device should provide the desired SCA protection in a white-box scenario, i.e., all design details including the netlist are known to the evaluation lab. 
In this work, we present a mechanism to design a provably- and practically-secure SCA-protected implementation which can be turned into an unprotected implementation by a Trojan adversary.
In many cases, our general Trojan concept does not require the addition of any logic (even a single gate) to the design, making it extremely hard to detect.
In case of ASIC platforms, the Trojan may be introduced by slightly changing the characteristics of a few transistors, and for FPGA platforms by changing the routing of particular signals.
Unlike previous work, it does not affect the provable-security feature of the underlying design unless the Trojan is triggered. Also, our technique is not based on the leakage of the PRNG. 
More precisely, our technique injects a \textit{parametric} Trojan that can be triggered, i.e., under normal condition the device does not exhibit any SCA leakage to be detected by an evaluation lab.
By increasing the clock frequency of the subverted device (or by decreasing its supply voltage) the Trojan is triggered and exhibits exploitable leakage.
In order to avoid accidental triggering during regular operation and to make detection by evaluation labs unlikely, we choose a trigger frequency that is beyond the specified maximum operational frequency of the device (which is smaller than the actual maximum frequency of the design due to the inserted Trojan).
As shown in the following sections, there is a carefully chosen gap between the specified maximum clock frequency of the device and the clock frequency where the Trojan is triggered.
In other words, by increasing the clock frequency such that the critical path delay is violated, the device starts to operate faulty; by further increasing the clock frequency, the device operates again correctly while exhibiting SCA leakage, i.e., our inserted Trojan becomes active.\\
As a case study, we integrate this Trojan into a threshold implementation of the PRESENT cipher, which is a popular block cipher for embedded applications \cite{DBLP:conf/ches/BogdanovKLPPRSV07}. In contrast to the results presented in~\cite{DBLP:conf/asiacrypt/EnderG0P17}, we do not target FPGA platforms, but ASICs here. We succeeded in implementing the malicious design in two different low-power CMOS process technologies, 90\,nm and 65\,nm, as part of side-channel test chips. We present SCA evaluations of both ASICs based on real-silicon measurements, both when the Trojan is triggered and when it is not, which confirm the soundness of our approach. However, during the design process we identified a number of obstacles to overcome when integrating such a Trojan into lightweight ciphers implemented in advanced CMOS technology. As a result, these first Trojan implementations on ASIC platforms do not come at exactly zero overhead, but require the addition of a few cells (less than half a percent of the cipher circuit's area). This result, however, does not contradict the potential stealthiness, in terms of zero overhead and negligible frequency reduction, of the general approach. In this regard, we detail in which cases such a Trojan can indeed be implemented at zero overhead and in which cases a few additional gates are~needed. 

\vspace{2mm}
\noindent
\textbf{Outline:}
Section~\ref{sec:background} deals with necessary background and definitions in the areas of hardware Trojans and threshold implementations as an SCA countermeasure.
Afterwards, in Section~\ref{sec:technique} we express our core idea how to insert our Trojan into a secure threshold implementation. In Section~\ref{sec:application} we give details on how to apply such a technique on a threshold implementation of the PRESENT cipher, and in Section~\ref{sec:ASIC_Implementation} we explain in detail how to realize such an implementation in ASIC technology. Finally, in Section~\ref{sec:result} the corresponding results of an ASIC-based case study including SCA evaluations of the manufactured devices are presented.

%% file: background.tex
\section{Background}
\label{sec:background}

\subsection{Hardware Trojans}

In general, a hardware Trojan is a back-door that can be inserted into an integrated circuit as an undesired and malicious modification, which makes  the behavior of the IC incorrect. There are many ways to categorize Trojans such as categorizing based on physical characteristics, design phase, abstraction level, location, triggering mechanism, and functionality. One common Trojan categorization is based on the activation mechanism (Trojan trigger) and the effect on the circuit functionality (Trojan payload). A set of conditions that cause a Trojan to be activated is called trigger. Trojans can combinationally or sequentially be triggered. An attacker chooses a rare trigger condition so that the Trojan would not be triggered during conventional design-time verification and manufacturing test. Sequentially-triggered Trojans (time bombs) are activated by the occurrence of a sequence of rare events, or after a period of continuous operation~\cite{DBLP:conf/hldvt/ChakrabortyNB09}.

The goal of the Trojan can be achieved by a payload which can change the circuit functionally or leak its secret information. In~\cite{DBLP:conf/host/JinM08} a categorization method according to how the payload of a Trojan works has been defined; some Trojans after triggering, propagate internal signals to output ports which can reveal secret information to the attackers (explicit payload). Other Trojans may make the circuit malfunction or destroy the whole chip (implicit payload). Another categorization for actions of hardware Trojans has been presented in~\cite{DBLP:conf/dft/WangSTP08}, in which the actions can be categorized into classes of \emph{modify functionality}, \emph{modify specification}, \emph{leak information}, and \emph{denial of service}.

The work presented in~\cite{DBLP:conf/ches/BeckerRPB13} is concerned with building stealthy Trojans at the layout level. A hardware Trojan was inserted into a cryptographically-secure PRNG and into a side-channel resistant Sbox by manipulating the dopant polarity of a few registers. 
Another class of hardware Trojans -- called Malicious Off-chip Leakage Enabled by Side-channels (MOLES) -- has been presented in~\cite{DBLP:conf/iccad/LinBP09}, which can retrieve secret information through side channels. They formulated the mechanism and detection methods of MOLES in theory and provided a verification process for multi-bit key extractions. 
In~\cite{DBLP:conf/ches/GhandaliBHP16} a design methodology for building stealthy parametric hardware Trojans and its application to Bug Attacks~\cite{DBLP:conf/crypto/BihamCS08,biham2016bug} has been proposed. The Trojans are based on increasing the delay of gates of a very rare-sensitized path in a combinatorial circuit, such as an arithmetic multiplier circuit. The Trojans are stealthy and have rare trigger conditions, so that the faulty behavior of the circuit under attack only occurs for very few combinations of the input vectors. Also an attack on the ECDH key agreement protocol by this Trojan has been presented in this work.
\subsection{Threshold Implementations}\label{sec:background:TI}
It can definitely be said that \textit{masking} is the most-studied countermeasure against SCA attacks.
It is based on the concept of \textit{secret sharing}, where a secret $\mathbfit{x}$ (e.g., intermediate values of a cipher execution) is represented by a couple of shares $(\mathbfit{x}^1,\ldots,\mathbfit{x}^n)$.
In case of an ($n$, $n$)-threshold secret sharing scheme, having access to $t<n$ shares does not reveal any information about $\mathbfit{x}$. 
One of such schemes is Boolean secret sharing, also known as Boolean masking in the context of SCA, where $\mathbfit{x}=\bigoplus\limits_{i=1}^n \mathbfit{x}^i$.
Hence, if the entire computation of a cipher is conducted on such a shared representation, its SCA leakage will be (in average) independent of the secrets as long as no function (e.g., combinatorial circuit) operates on all $n$ shares.

Due to the underlying Boolean construction, application of a linear function $\mathcal{L}(.)$ over the shares is straightforward since $\mathcal{L}(\mathbfit{x})=\bigoplus\limits_{i=1}^n \mathcal{L}(\mathbfit{x}^i)$.
All the difficulties belong to implementing non-linear functions over such a shared representation.
This concept has been applied in hardware implementations of AES (mainly with $n=2$) with no success~\cite{DBLP:conf/fse/OswaldMPR05,DBLP:conf/ches/MangardPO05,DBLP:conf/acns/CanrightB08,DBLP:conf/ches/MoradiME10} until the Threshold Implementation (TI) -- based on sound mathematical foundations -- has been introduced in~\cite{DBLP:journals/joc/NikovaRS11}, which defines the minimum number of shares $n\geq t+1$ with $t$ the algebraic degree of the underlying non-linear function.
For simplicity (and as our case study is based on) we focus on quadratic Boolean functions, i.e., $t=2$, and minimum number of shares $n=3$.
Suppose that the TI of the non-linear function $y=\mathcal{F}(\mathbfit{x})$ is desired, i.e., $(\mathbfit{y}^1, \mathbfit{y}^2, \mathbfit{y}^3)=\mathcal{F}^*(\mathbfit{x}^1, \mathbfit{x}^2, \mathbfit{x}^3)$, where
\[
 \mathbfit{y}^1 \oplus \mathbfit{y}^2 \oplus \mathbfit{y}^3 = \mathcal{F}(\mathbfit{x}^1 \oplus \mathbfit{x}^2 \oplus \mathbfit{x}^3).
\]
Indeed, each output share $y^{i\in\{1,2,3\}}$ is provided by a component function $\mathcal{F}^i(.,.)$ which receives only two input shares. 
In other words, one input share is always missing in every component function.
This, which is a requirement defined by TI as \textit{non-completeness}, supports the aforementioned concept that ``no function (e.g., combinatorial circuit) operates on all $n$ shares'', and implies the given formula $n\geq t+1$.
Therefore, three component functions $\left(\mathcal{F}^1\left(\mathbfit{x}^2,\mathbfit{x}^3\right), \mathcal{F}^2\left(\mathbfit{x}^3,\mathbfit{x}^1\right), \mathcal{F}^3\left(\mathbfit{x}^1,\mathbfit{x}^2\right)\right)$ form the shared output $(\mathbfit{y}^1, \mathbfit{y}^2, \mathbfit{y}^3)$.

In order to fulfill the above-given statement that ``having access to $t<n$ shares does not reveal any information about $\mathbfit{x}$'', the shares need to follow a uniform distribution. 
For simplicity suppose that $n=2$, and the shares $(\mathbfit{x}^1, \mathbfit{x}^2)$ represent secret $\mathbfit{x}$.
If the distribution of $x^1$ has a bias (i.e., not uniform) which is known to the adversary, he can observe the distribution of $\mathbfit{x}^2=\mathbfit{x} \oplus \mathbfit{x}^1$ and guess $\mathbfit{x}$.
Hence, the security of masking schemes\footnote{Except those which are based on low-entropy masking~\cite{CarletDGM12,MaghrebiGD11}.} relies on the uniformity of the masks.
More precisely, when $\mathbfit{x}^1=\mathbfit{m}$, $\mathbfit{x}^2=\mathbfit{x} \oplus \mathbfit{m}$, and $\mathbfit{m}$ is taken from a randomness source (e.g., a PRNG), the distribution of $\mathbfit{m}$ should be uniform (or let say with full entropy).

The same holds for higher-order masking, i.e., $n>2$.
However, not only the distribution of every share but also the joint distribution of every $t<n$ shares is important.
In case of $\mathcal{F}^*(.,.,.)$ as a TI of a bijective function $\mathcal{F}(.)$, the \textit{uniformity} property of TI is fulfilled if $\mathcal{F}^*(.,.,.)$ forms a bijection.
Otherwise, the security of such an implementation cannot be guaranteed. 
Note that fulfilling the uniformity property of TI constructions is amongst its most difficult challenges, and it has been the core topic of several articles like~\cite{DBLP:journals/ccds/BilginNNRTV15,DBLP:journals/joc/PoschmannMKLWL11,DBLP:conf/asiacrypt/BilginGNNR14,DBLP:journals/joc/NikovaRS11,DBLP:BeyneB16}.
Alternatively, the shares can be remasked at the end of every non-uniformly shared non-linear function (see~\cite{DBLP:conf/africacrypt/BilginGNNR14,DBLP:conf/eurocrypt/MoradiPLPW11}), which requires a source to provide fresh randomness at every clock cycle. 
Along the same line, another type of masking in hardware (which reduces the number of shares) has been developed in~\cite{DBLP:conf/crypto/ReparazBNGV15,DBLP:conf/ctrsa/GrossMK17}, which (almost always) needs fresh randomness to fulfill the uniformity.

We should emphasize that the above given expressions illustrate only the first-order TI of bijective quadratic functions. 
For other cases including higher-order TI we refer the interested reader to the original articles~\cite{DBLP:journals/joc/NikovaRS11,DBLP:conf/asiacrypt/BilginGNNR14,DBLP:journals/ccds/BilginNNRTV15}.

%% file: technique.tex
\section{Technique}
\label{sec:technique}

As explained in the former section -- by means of TI -- it is possible to realize hardware cryptographic devices secure against certain SCA attacks.
Our goal is to create a situation where an SCA-secure device becomes insecure while it still operates correctly.
Such a dynamic transition from secure to insecure should be available and known only to the Trojan attacker. 
To this end, we target the uniformity property of a secure TI construction. 
More precisely, we plan to construct a secure and uniform TI design which becomes non-uniform (and hence insecure) at particular environmental conditions. 
In order to trigger the Trojan (or let say to provide such a particular environmental conditions) for example we select a \underline{higher clock frequency} than the specified maximum operational frequency of the device, or a \underline{lower power supply} than the device nominal supply voltage.
It should not be forgotten that under such conditions the underlying device should still maintain its correct functionality.

To realize such a scenario -- inspired from the stealthy parametric Trojan introduced in~\cite{DBLP:conf/ches/GhandaliBHP16} -- we intentionally lengthen certain paths of a combinatorial circuit.
This is done in such a way that -- by increasing the device's clock frequency or lowering its supply voltage -- such paths become faulty earlier than the other paths.
We would achieve our goal if $i$) the faults cancel each others' effect, i.e., the functionality of the design is not altered, and $ii$) the design does not fulfill the uniformity property anymore.

In order to explain our technique -- for simplicity without loss of generality~-- we focus on a 3-share TI construction.
As explained in Section~\ref{sec:background:TI} -- ignoring the uniformity -- achieving a non-complete shared function $\mathcal{F}^*(.,.,.)$ of a given quadratic function $\mathcal{F}(.)$ is straightforward.
Focusing on one output bit of $\mathcal{F}(\mathbfit{x})$, and representing $\mathbfit{x}$ by $s$ input bits $\langle x_s,\ldots,x_1\rangle$, we can write
\begin{align*}
\mathcal{F}_i(\langle x_s,\ldots,x_1\rangle)= & k_0 \oplus k_1x_1 \oplus k_2x_2 \oplus \ldots \oplus k_sx_s \oplus\\
& k_{1,2}x_1x_2 \oplus k_{1,3}x_1x_3 \oplus \ldots \oplus k_{s-1,s}x_{s-1}x_s.
\end{align*}
The coefficients $k_0,\ldots,k_{s-1,s}\in\{0,1\}$ form the Algebraic Normal Form (ANF) of the quadratic function $\mathcal{F}_i:\{0,1\}^s\rightarrow\{0,1\}$.
By replacing every input bit $x_i$ by the sum of three corresponding shares $x^1_i \oplus x^2_i \oplus x^3_i$, the remaining task is just to split the terms in the ANF to three categories in such a way that each category is independent of one share.
This can be done by a method denoted by \textit{direct sharing}~\cite{DBLP:journals/ccds/BilginNNRTV15} as
\begin{itemize}
	\item $\mathcal{F}^1_i(.,.)$ contains the linear terms $x^2_i$ and the quadratic terms $x^2_ix^2_j$ and $x^2_ix^3_j$.
	\item $\mathcal{F}^2_i(.,.)$ contains the linear terms $x^3_i$ and the quadratic terms $x^3_ix^3_j$ and $x^3_ix^1_j$.
	\item $\mathcal{F}^3_i(.,.)$ contains the linear terms $x^1_i$ and the quadratic terms $x^1_ix^1_j$ and $x^1_ix^2_j$.
\end{itemize}
The same is independently applied on each output bit of $\mathcal{F}(.)$ and all three component functions $\mathcal{F}^1\left(\mathbfit{x}^2,\mathbfit{x}^3\right)$, $\mathcal{F}^2\left(\mathbfit{x}^3,\mathbfit{x}^1\right)$, $\mathcal{F}^3\left(\mathbfit{x}^1,\mathbfit{x}^2\right)$ are constructed that fulfill the non-completeness, but nothing about its uniformity can be said. 

There are indeed two different ways to obtain a uniform TI construction:
\begin{itemize}[leftmargin=*]
	\item If $s$ (the underlying function size) is small, i.e., $s \leq 5$, it can be found that $\mathcal{F}(.)$ is affine equivalent to which $s$-bit class.
	More precisely, there is a quadratic class $\mathcal{Q}$ which can represent $\mathcal{F}$ as $\mathcal{A}' \circ \mathcal{Q} \circ \mathcal{A}$ (see~\cite{DBLP:conf/eurocrypt/BiryukovCBP03} for an algorithm to find $\mathcal{A}$ and $\mathcal{A}'$ given $\mathcal{F}$ and $\mathcal{Q}$).
	A classification of such classes for $s=3$ and $s=4$ are shown in~\cite{DBLP:journals/ccds/BilginNNRTV15} and for $s=5$ in~\cite{DBLP:journals/tosc/BozilovBS17}.
	Since the number of existing quadratic classes are restricted, a uniform TI can be found by exhaustive search.
	Note that while for many quadratic classes the direct sharing (explained above) can reach to a uniform TI, for some quadratic classes no uniform TI exists unless the class is represented by a composition of two other quadratic classes~\cite{DBLP:journals/ccds/BilginNNRTV15}.
	Supposing that $\mathcal{Q}^*(.,.,.)$ is a uniform TI of $\mathcal{Q}(.)$, applying the affine functions $\mathcal{A}'$ and $\mathcal{A}$ accordingly on each input and output of the component function $\mathcal{Q}^*$ would give a uniform TI of $\mathcal{F}(.)$:
\begin{align*}	
\mathcal{F}^1(\mathbfit{x}^2,\mathbfit{x}^3)=&\mathcal{A}' \circ \mathcal{Q}^1\left(\mathcal{A}\left(\mathbfit{x}^2\right),\mathcal{A}\left(\mathbfit{x}^3\right)\right),\\
\mathcal{F}^2(\mathbfit{x}^3,\mathbfit{x}^1)=&\mathcal{A}' \circ \mathcal{Q}^2\left(\mathcal{A}\left(\mathbfit{x}^3\right),\mathcal{A}\left(\mathbfit{x}^1\right)\right),\\
\mathcal{F}^3(\mathbfit{x}^1,\mathbfit{x}^2)=&\mathcal{A}' \circ \mathcal{Q}^3\left(\mathcal{A}\left(\mathbfit{x}^1\right),\mathcal{A}\left(\mathbfit{x}^2\right)\right).
\end{align*}
	This scenario has been followed in several works, e.g.,~\cite{DBLP:conf/asiacrypt/0001S16,DBLP:conf/ches/MoradiW15,DBLP:conf/sacrypt/SasdrichMG15,DBLP:conf/ches/BilginBKMW13,DBLP:conf/ches/BilginNNRS12}.
		
	\item Having a non-uniform TI construction, e.g., obtained by direct sharing, we can add \textit{correction terms} to the component functions in such a way that the correctness and non-completeness properties are not altered, but the uniformity may be achieved. 
	For example, the linear terms $x^2_i$ and/or the quadratic terms $x^2_ix^2_j$ as correction terms can be added to the same output bit of \textbf{both} component functions $\mathcal{F}^1\left(\mathbfit{x}^2,\mathbfit{x}^3\right)$ and $\mathcal{F}^3\left(\mathbfit{x}^1,\mathbfit{x}^2\right)$. 
	Addition of any correction term changes the uniformity of the design.
	Hence, by repeating this process -- up to examining all possible correction terms and their combination, which is not feasible for large functions -- a uniform construction might be obtained.
	Such a process has been conducted in~\cite{DBLP:journals/joc/PoschmannMKLWL11,DBLP:conf/cardis/BilginDNNRA13
	} to construct uniform TI of PRESENT and Keccak non-linear functions.\\
	We should here refer to a similar approach called remasking~\cite{DBLP:conf/eurocrypt/MoradiPLPW11,DBLP:journals/ccds/BilginNNRTV15} where -- instead of correction terms -- fresh randomness is added to the output of the component functions to make the outputs uniform. 
	In this case, obviously a certain number of fresh mask bits are required at every clock cycle (see~\cite{DBLP:conf/eurocrypt/MoradiPLPW11,DBLP:journals/tcad/BilginGNNR15}).
\end{itemize}

\begin{figure}[t]%
\centering
\includegraphics[width=0.5\columnwidth]{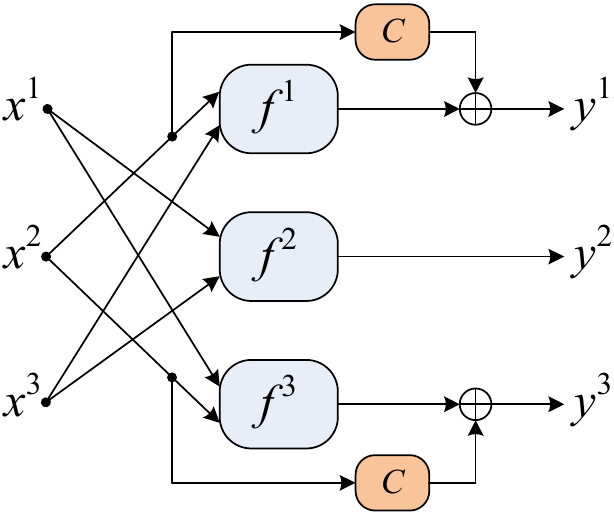}%
\caption{Exemplary TI construction with a correction term $C$.}%
\label{fig:blockDiagram}%
\end{figure}

Our technique is based on the second scheme explained above. 
If we make the paths related to the correction terms the longest, by increasing the clock frequency such paths are the first whose delays are violated.
As illustrated, each correction term must be added to two component functions (see Figure~\ref{fig:blockDiagram}).
The paths must be very carefully altered in such a way that the path delay of both instances of the targeted correction term are the longest in the entire design and relatively the same.
Hence, at a particular clock frequency both instances of the correction terms are not correctly calculated while all other parts of the design are fault free.
This enables the design to still work properly, i.e., it generates correct ciphertexts assuming that the underlying design realizes an encryption function. 
It means that the design operates like an alternative design where no correction terms exist.
Hence, the uniformity of the TI construction is not fulfilled and SCA leakage can be exploited.
To this end, we should keep a margin between $i$) the path delay of the correction terms and $ii$) the critical path delay of the rest of the circuit, i.e., that of the circuit without correction terms.
This margin guarantees that at a certain high clock frequency the correction terms are canceled out but the critical path delay of the remaining circuit is not violated.

We would like to emphasize that in an implementation of a cipher once one of the TI functions generates non-uniform output (by violating the delay of correction terms), the uniformity is not maintained in the next TI functions and it leads to first-order leakage in all further rounds. 
If the uniformity is achieved by remasking (e.g., in~\cite{DBLP:conf/dsd/GrossWDE15}), the above-expressed technique can have the same effect by making the XOR with fresh mask the longest path.
Hence, violating its delay in one TI function would make its output non-uniform, but the fresh randomness may make the further rounds of the cipher again uniform.


\begin{figure}[h]%
\centering
\includegraphics[width=0.9\columnwidth]{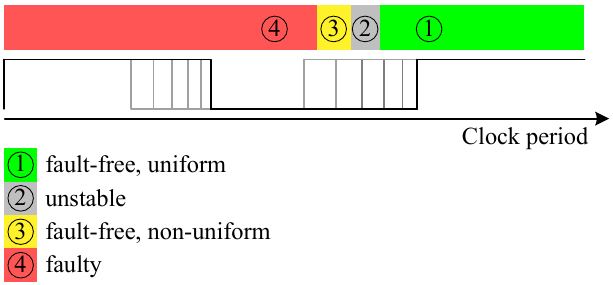}%
\caption{Status of the design with Trojan at different clock frequencies.}%
\label{fig:timingDiagram}%
\end{figure}

Based on Figure~\ref{fig:timingDiagram}, which shows a corresponding timing diagram, the device's status can be categorized into four states:
\begin{itemize}[leftmargin=*]
	\item at a low clock frequency (denoted by \colorbox{green}{\Circled{1}}) the device operates fault free and maintains the uniformity,
	\item by increasing the clock frequency (in the \colorbox{lightgray}{\Circled{2}} period), the circuit first starts to become unstable, when indeed the correction terms do not fully cancel each others' effect, and the hold time and/or setup time of the registers are violated, 
	\item by further increasing the clock frequency (in the \colorbox{yellow}{\Circled{3}} period), the delay of both instances of the correction term are violated and the circuit operates fault free, but does not maintain the uniformity, and
	\item by even further increasing the clock frequency (marked by \colorbox{myred}{\Circled{4}}), the clock period becomes smaller than the critical path delay of the rest of the circuit, and the device does not operate correctly.
\end{itemize}
The aforementioned margin defines the length of the \colorbox{lightgray}{\Circled{2}} and \colorbox{yellow}{\Circled{3}} periods, which are of crucial importance.
If they are very wide, the maximum operation frequency of the resulting circuit is obviously reduced, and the likelihood of the inserted Trojan to be detected by an evaluator is increased. 
Correct functionality of the circuit is required in period \colorbox{yellow}{\Circled{3}}, to make sure that the correct sensitive intermediate values are processed by the circuit and revealed to the Trojan attacker through their side-channel leakage.

%% file: application.tex
\section{Application}
\label{sec:application}
In order to show an application of our technique, we focus on a first-order TI design of the PRESENT cipher~\cite{DBLP:conf/ches/BogdanovKLPPRSV07} as a case study.
The PRESENT Sbox is a 4-bit cubic bijection $\mathcal{S}:\mathtt{C56B90AD3EF84712}$.
Hence, its first-order TI needs at least $n=4$ shares. 
Alternatively, it can be decomposed to two quadratic bijections $\mathcal{S}:\mathcal{F} \circ \mathcal{G}$ enabling the minimum number of shares $n=3$ at the cost of having an extra register between $\mathcal{F}^*$ and $\mathcal{G}^*$ (i.e., TI of $\mathcal{F}$ and $\mathcal{G}$).
As shown in~\cite{DBLP:journals/ccds/BilginNNRTV15}, $\mathcal{S}$ is affine equivalent to class $\mathcal{C}_{266}:\mathtt{0123468A5BCFED97}$, which can be decomposed to quadratic bijections with uniform TI. 
The works reported in~\cite{DBLP:conf/ches/MoradiW15,DBLP:conf/sacrypt/SasdrichMG15, DBLP:conf/ctrsa/Sasdrich0G17} have followed this scenario and represented the PRESENT Sbox as $\mathcal{S}:\mathcal{A}'' \circ \mathcal{Q}' \circ \mathcal{A}' \circ \mathcal{Q} \circ \mathcal{A}$, with many possibilities for the affine functions $\mathcal{A}''$, $\mathcal{A}'$, $\mathcal{A}$ and the quadratic classes $\mathcal{Q}'$ and $\mathcal{Q}$ whose uniform TI can be obtained by direct sharing (see Section~\ref{sec:technique}).

However, the first TI of PRESENT has been introduced in~\cite{DBLP:journals/joc/PoschmannMKLWL11}, where the authors have decomposed the Sbox by $\mathcal{G}:\mathtt{7E92B04D5CA1836F}$ and $\mathcal{F}:\mathtt{08B7A31C46F9ED52}$.
They have accordingly provided a uniform TI of each of such 4-bit quadratic bijections.
We focus on this decomposition, and select $\mathcal{G}$ as the target where our Trojan is implemented.
Contrary to all other related works, we first try to find a \textbf{non-uniform} TI of $\mathcal{G}(.)$, and we later make it uniform by means of correction terms.
We start with the ANF of $\mathcal{G}(\langle d,c,b,a\rangle)=\langle g_3,g_2,g_1,g_0\rangle$:
\begin{align*}
g_0 & = 1 \oplus a \oplus dc \oplus db \oplus cb, & g_2 & = 1 \oplus c \oplus b,\\
g_1 & = 1 \oplus d \oplus b \oplus ca \oplus ba,  & g_3 & = c \oplus b \oplus a.
\end{align*}
One possible sharing of $\mathbfit{y}=\mathcal{G}(\mathbfit{x})$ can be represented by $\left(\mathbfit{y}^1,\mathbfit{y}^2,\mathbfit{y}^3\right)= \left(\mathcal{G}^1\left(\mathbfit{x}^2,\mathbfit{x}^3\right),\mathcal{G}^2\left(\mathbfit{x}^3,\mathbfit{x}^1\right),\mathcal{G}^3\left(\mathbfit{x}^1,\mathbfit{x}^2\right)\right)$ as
\begin{align*}
 y^1_0 = & 1 \oplus a^2 \oplus d^2  c^3 \oplus d^3  c^2	\oplus d^2  b^3 \oplus d^3  b^2 \oplus c^2  b^3 \oplus c^3  b^2 \oplus \\
         & d^2  c^2 \oplus d^2  b^2 \oplus c^2  b^2\\
y^1_1  = & 1 \oplus b^2 \oplus d^3 \oplus c^2  a^3 \oplus c^3  a^2 \oplus b^2  a^3 \oplus b^3  a^2 \oplus c^2  a^2 \oplus b^2  a^2\\ 
y^1_2  = & 1 \oplus c^2 \oplus b^2\\
y^1_3  = & c^2 \oplus b^2 \oplus a^2	
\end{align*}
\begin{align*}
 y^2_0 = & a^3  \oplus  d^3  c^3  \oplus  d^1  c^3  \oplus  d^3  c^1  \oplus  d^3  b^3  \oplus  d^1  b^3  \oplus  d^3  b^1  \oplus\\
         &  c^3  b^3 \oplus c^1  b^3  \oplus  c^3  b^1\\
y^2_1  = & b^3  \oplus d^1  \oplus c^1  a^3  \oplus  c^3  a^1  \oplus  b^1  a^3  \oplus  b^3  a^1  \oplus c^3  a^3  \oplus  b^3  a^3\\ 
y^2_2  = & c^3 \oplus b^3\\ 
y^2_3  = & c^3 \oplus b^3 \oplus a^3
\end{align*}
\begin{align*}
y^3_0  = & a^1  \oplus  d^1  c^1  \oplus  d^1  c^2  \oplus  d^2  c^1  \oplus  d^1  b^1  \oplus  d^1  b^2  \oplus  d^2  b^1  \oplus \\
         & c^1  b^1 \oplus c^1  b^2  \oplus  c^2  b^1\\
y^3_1  = & b^1  \oplus  d^2  \oplus  c^1  a^2  \oplus  c^2  a^1  \oplus  b^1  a^2  \oplus  b^2  a^1 \oplus c^1  a^1  \oplus  b^1  a^1\\ 
y^3_2  = & c^1 \oplus b^1\\
y^3_3  = & c^1 \oplus b^1 \oplus a^1,
\end{align*}
with $\mathbfit{x}^{i\in\{1,2,3\}}=\langle d^i,c^i,b^i,a^i\rangle$.
This is not a uniform sharing of $\mathcal{G}(.)$, and by searching through possible correction terms we found three correction terms $c^1b^1$, $c^2b^2$, and $c^3b^3$ to be added to the second bit of the above-expressed component functions, that lead us to a uniform TI construction.
More precisely, by defining
\begin{align*}
	\mathcal{C}^1(\mathbfit{x}^2,\mathbfit{x}^3) = \textcolor{red}{c^2b^2} \oplus c^3b^3,\\
	\mathcal{C}^2(\mathbfit{x}^3,\mathbfit{x}^1) = c^1b^1 \oplus c^3b^3,\\
	\mathcal{C}^3(\mathbfit{x}^1,\mathbfit{x}^2) = c^1b^1 \oplus \textcolor{red}{c^2b^2},
\end{align*}
and adding them respectively to $y^1_1$, $y^2_1$, and $y^3_1$, the resulting TI construction becomes uniform.
If any of such correction terms is omitted, the uniformity is not maintained. 
In the following we focus on a single correction term $\textcolor{red}{c^2b^2}$ which should be added to $\mathcal{G}^1(.,.)$ and $\mathcal{G}^3(.,.)$.
\subsection{Inserting the Trojan}\label{sec:application:trojan}
We explain how to realize the Trojan functionality by path delay fault model~\cite{DBLP:conf/itc/Smith85}, without modifying the logic circuit. The Trojan can be triggered by violating the delay of the combinatorial logic paths that pass through the targeted correction terms $\textcolor{red}{c^2b^2}$. It is indeed a parametric Trojan, which does not require any additional logic. The Trojan is inserted by modifying a few gates during manufacturing, so that their delays increase and add up to the path delay faults.
	
Given in~\cite{DBLP:conf/ches/GhandaliBHP16}, the underlying method to create a triggerable and stealthy delay-based Trojan consists of two phases: path selection and delay distribution. In the first phase, a set of uniquely-sensitized paths are found that passes through a combinatorial circuit from primary inputs to the primary outputs. Controllability and observability metrics are used to guide the selection of which gates to include in the path to make sure that the path(s) are uniquely sensitized\footnote{It means that the selected paths are the only ones in the circuit whose critical delay can be violated.}. Furthermore, a SAT-based check is performed to make sure that the path remains sensitizable each time a gate is selected to be added to the path. After a set of uniquely-sensitized paths is selected, the overall delay of the path(s) must be increased so that a delay fault occurs when the path is sensitized. However, any delay added to the gates of the selected path may also cause delay faults on intersecting paths, which would cause undesirable errors and affect the functionality of the circuit. The delay distribution phase addresses this problem by smartly choosing delays for each gate of the selected path to minimize the number of faults caused by intersecting paths. At the same time, the approach ensures that the overall path delay is sufficient for the selected paths to make it faulty.

\subsubsection{ASIC Platforms}
In an ASIC platform, the necessary delays for such Trojans can be introduced in a multitude of ways. Apart from the addition of extra gates to the chosen paths there is also the attractive option to achieve the same goal by slight modifications on the sub-transistor level so that only the parameters of a few transistors of the design are changed. To increase the delays of transistors in stealthy ways, there are many possible ways in practice. Such a Trojan is very difficult to be detected by e.g., functional testing, visual inspection, or side-channel profiling, because not a single transistor is removed or added to the design and the changes to the individual gates are minor. Also, full reverse-engineering of the IC would unlikely reveal the presence of the malicious manipulation in the design. Furthermore, this Trojan would not be present at higher abstraction levels and hence cannot be detected at those levels, because the actual Trojan is inserted at the sub-transistor level. There are several stealthy ways to slightly change the parameters of transistors of a gate and make it slower. Exemplarily, we list three methods below.

\vspace{2mm}
\noindent
\textbf{Decreasing the Width:} Usually a standard cell library has different drive strengths for each logic gate type, which correspond to various transistor widths. Current of a transistor is linearly proportional to the transistor width, therefore a transistor with smaller width is slower to charge its load capacitance. One way to increase the delay of a gate is to substitute it with its weaker version in the library which has smaller width, or to create a custom version of the gate with an extremely narrow width, if the lower level information of the gate is available in the library (e.g., SPICE model). 

\vspace{2mm}
\noindent
\textbf{Raising the Threshold:} A common way of increasing the delay of a gate is to increase the threshold voltage of its transistors by body biasing or doping manipulation. Using high and low threshold voltages at the same time in a design (i.e., Dual-Vt design) is very common approach and provides the designer with more options to satisfy the speed goals of the design. Devices with low threshold voltage are fast and used where delay is critical; devices with high threshold voltage are slow and used where power consumption is important. 

\vspace{2mm}
\noindent
\textbf{Increasing the Gate Length:} Gate length biasing can increase delay of a gate by reducing the current of its transistors~\cite{DBLP:journals/tcad/GuptaKSS06}. 

\subsubsection{FPGA Platforms}
In case of the FPGAs, the combinatorial circuits are realized by Look-Up Tables (LUT), in currently-available Xilinx FPGAs, by 6-to-1 or 5-to-2 LUTs and in former generations by 4-to-1 LUTs. 
The delay of the LUTs cannot be changed by the end users; alternatively we offer using \textit{Through Switch Boxes} to make certain paths longer. The routings in FPGA devices are made by configuring the switch boxes.
Since the switch boxes are made by active components realizing logical switches, a signal which passes through many switch boxes has a longer delay compared to a short signal. 
Therefore, given a fully placed-and-routed design we can modify the routings by lengthening the selected signals.
This is for example feasible by means of \texttt{Vivado Design Suite} as a standard tool provided by Xilinx for recent FPGA families and \texttt{FPGA Editor} for the older generations. 
It in fact needs a high level of expertise, and cannot be done at HDL level.
Interestingly, the resulting circuit would not have any additional resource consumption, i.e., the number of utilized LUTs, FFs and Slices, and hence is hard to detect particularly if the utilization reports are compared.\\
Focusing on our target, the only paths which should be lengthened are both instances of $\textcolor{red}{c^2b^2}$ in $\mathcal{G}^1(.,.)$ and $\mathcal{G}^3(.,.)$. Considering Figure~\ref{fig:blockDiagram}, the XOR gate which receives the $\mathcal{F}^1$ and $\mathcal{C}$ output should be the last gate in the combinatorial circuit generating $y^1_1$, i.e., the second bit of $\mathcal{G}^1(.,.)$. The same holds for $y^3_1$, i.e., the second bit of $\mathcal{G}^3(.,.)$. In the following section we explain in detail which algorithms, metrics and heuristics need to be used to select and lengthen the correct paths in arbitrary ASIC designs of such kind.

%% file: ASIC_Implementation.tex
\section{ASIC Implementation}
\label{sec:ASIC_Implementation}

For ASIC platforms, we utilize the stealthy parametric Trojan introduced in~\cite{DBLP:conf/ches/GhandaliBHP16}. It consists of two main phases: \textit{path selection phase} and \textit{delay distribution phase}. We briefly explain each of these phases in subsections~\ref{Phase I} and~\ref{Phase II}. Our goal is to make the paths related to our target correction term, which is added to two component functions, the longest so that by increasing the clock frequency such paths are the first whose delays are violated. The paths must be very carefully selected and altered in such a way that the path delay of both instances of the targeted correction term are the longest in the entire design and relatively the same. Hence, at a particular clock frequency both instances of the correction terms are not correctly calculated while all other parts of the design are fault free. This enables the design to still work properly. 
\subsection{Rare Path Selection Phase}
\label{Phase I}
The path selection phase seeks to find a path $\pi$ through the netlist of the circuit that passes through the targeted correction term. Note that the delays are not considered in this phase of the work. Path $\pi$ is initialized to contain a transition on the targeted correction term node. 
This initial single-node path $\pi$ is then extended incrementally backwards until reaching the primary inputs, and extended incrementally forward until reaching the primary outputs. The path selection algorithm is given in Alg.~\ref{alg:path_search}. 
\begin{algorithm}[t]
	\footnotesize
	\begin{algorithmic}[1]
		
		\REQUIRE{A single node $\pi$ in the netlist of the circuit}
		\ENSURE {A sensitizable path $\pi$ starting at a primary input and ending at a primary output}
		
		\WHILE{($\pi$ does not start at a primary input)}		
		\STATE new\_node\_candidates = \{All transitions that can be prepended to $\pi$\}
		\STATE Order new\_node\_candidates by difficulty of justification.
		\FOR{(each member $n'$ of new\_node\_candidates)}		
		\STATE new\_subpath $\pi'=$  prepend $n'$ to the tail of $\pi$ 
		\IF{$(\text{check-SAT}(\pi'))$}		
		\STATE $\pi = \pi'$ 
		\STATE Exit for loop.		 
		\ENDIF	
		\ENDFOR		
		\ENDWHILE
		\WHILE{($\pi$ does not end at a primary output)}
		\STATE new\_node\_candidates = \{ALL transitions that can be appended to $\pi$\}
		\STATE Order new\_node\_candidates by difficulty of propagation.
		\FOR{(each member $n'$ of new\_node\_candidates)}
		\STATE new\_subpath $\pi'=$  append $n'$ to the head of $\pi$
		\IF{$(\text{check-SAT}(\pi'))$} 
		\STATE $\pi = \pi'$ 
		\STATE Exit for loop. 
		\ENDIF
		\ENDFOR
		\ENDWHILE
		
	\end{algorithmic}
	\caption{Extracting a hard to trigger sensitizable path passing through a specific node.}
	\label{alg:path_search}
	\normalsize
\end{algorithm} 
Starting from the first transition on the current path $\pi$, we repeatedly try to extend the path back towards the PIs by prepending one new transition to the path. To select such a transition, the algorithm creates a list of candidate transitions that can be prepended to the path, which is sorted according to the difficulty of creating the necessary conditions to justify the transition. 
Whenever a node is prepended to $\pi$ to create a candidate path $\pi'$, 
the sensitizability of $ \pi'$ is checked by calling \textit{check-SAT} function. In this function SAT-based techniques \cite{eggersgluss2013improved} are used to check sensitizability of the path 
If the SAT solver returns SAT, then path $\pi'$ is known to be a subpath of a sensitizable path from a primary input to a primary output. 
If this newly added tail node is not a primary input, then the algorithm will again try to extend it backwards. 
%
%
%

The forward propagation part is similar to the aforementioned backward propagation, except that it adds nodes to the head of the path until reaching a primary output. At each step of the algorithm, a list of candidates is again formed. In this case, 
they are ordered according to difficulty of propagation instead of difficulty of justification. Each time a new candidate path is created by adding a candidate node to the existing path, a SAT check is again performed to ensure that the nodes are only added to $\pi$ if it remains sensitizable. 

\subsection{Delay Distribution Phase}
\label{Phase II}
Once paths are selected, the delay of them must be increased so that the total path delays exceed the clock period and errors occur when the paths are sensitized. Choosing where to add delay on the paths must be done carefully, because the gates along the chosen paths are also part of many other intersecting or overlapping paths. Any delay added to the chosen paths therefore may cause errors even when the chosen paths are not sensitized. Genetic algorithm is used to smartly decide the delay of each gate along with some constraints to restrict the allowed solution space, and a fitness function for evaluating solutions. 

\vspace{2mm}
\noindent
\textbf{Total Path Delay Constraint:}
Assume each of the chosen paths $\pi$ includes $n$ gates and  target path delay is $D$. This constraint specifies that the sum of assigned delays along the path is equal to the target path delay $D$. To cause an error, $D$ must exceed the period \colorbox{myred}{\Circled{4}}.
\begin{equation}
	\label{eq:constraint_path_delay}
	D = \sum_{i=0}^n d_i
\end{equation}

\noindent
\textbf{Gate Delay Constraint:}
Assume $d'_i$ represents the nominal delay of the $i^{th}$ gate on the chosen path $\pi$, and $s_i$ represents the slack metric associated with the same gate. Each slack parameter $s_i$ describes how much delay can be added to the corresponding gate without causing the path to exceed the period \colorbox{myred}{\Circled{4}}. The slack for each gate is computed as a function of the nominal delay of the gate, data dependency, and the clock period~\cite{ghandali2015low, tang2005leakage}. Because the targeted path delay $D$ does exceed the period \colorbox{myred}{\Circled{4}}, gate delays are allowed to exceed their computed slack. The following equation shows this constraint where $ c $ is a constant.
\begin{equation}
	\label{eq:constraint_gate_delay}
	d'_i + s_i - c  \leq d_i \leq d'_i + s_i + c
\end{equation}

\noindent
\textbf{Fitness Function:}
 The cost function consists of two parts; $i$) the faults cancel each others' effect, i.e., the faults on targeted correction term in two functions $\mathcal{G}^1$ and $\mathcal{G}^3$ happen at the same time and cancel the effect of each other so that the functionality of the design is not altered, and $ii$) the design does not fulfill the uniformity property anymore. To cover both cases in our final cost function we define it as the following equation in where the first term corresponds to case (i) and the second term corresponds to case (ii). Our goal is to minimize this cost function.
\begin{equation}
\label{eq:cost_function}
\textrm{Cost}_F(d_1,...,d_n) =  \textrm{ErrorRate}_{\textrm{design}} + 1/\textrm{ErrorRate}_{\textrm{$\mathcal{G}^1$ and $\mathcal{G}^3$}}
\end{equation}
We use random simulation to evaluate the cost of any delay assignment. When the genetic algorithm in Matlab~\cite{ga2016matlab} needs to evaluate the cost of a particular delay assignment, it does so by executing a timing simulator. The timing simulator, in our case ModelSim, applies test vectors to the circuit-under-evaluation and a golden copy of the circuit and compares the respective outputs to count the number of errors. 

%% file: results.tex
\section{ASIC Practical Results}
\label{sec:result}

In this section we describe how we have designed and implemented first ASIC prototypes incorporating such a malicious design. We then verify that the resulting chips are indeed resistant against side-channel attacks when the Trojan is not triggered and that this resistance can be nullified when triggering it.

\begin{figure}[htbp]
\centering
\subfigure[Layout schematic 65~nm ASIC]{
\includegraphics[width=0.44\columnwidth]{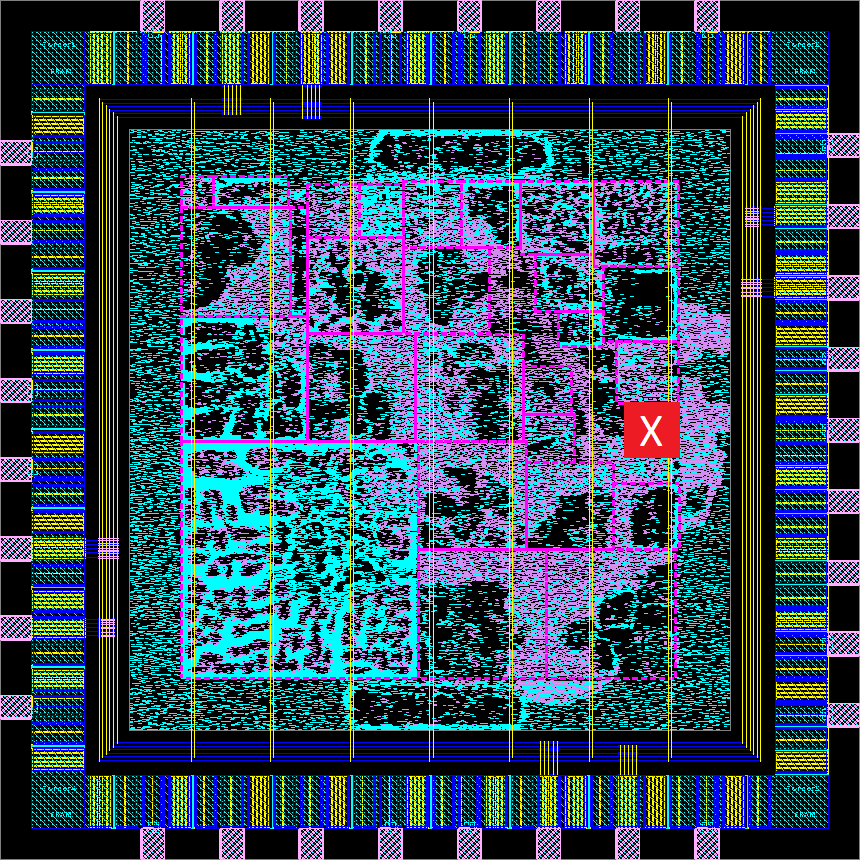}
\label{fig:65nm_Layout}}\hfill
\subfigure[Layout schematic90~nm ASIC]{
\includegraphics[width=0.44\columnwidth]{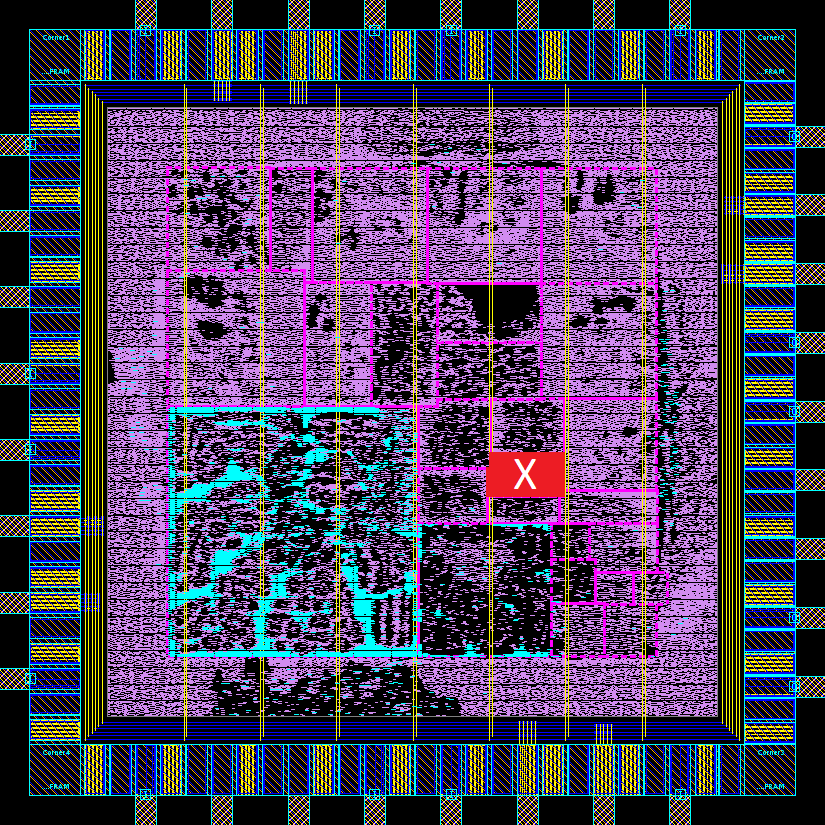}
\label{fig:90nm_Layout}}\\
\subfigure[Photo of packaged 90~nm ASIC]{
\includegraphics[height=3.8cm]{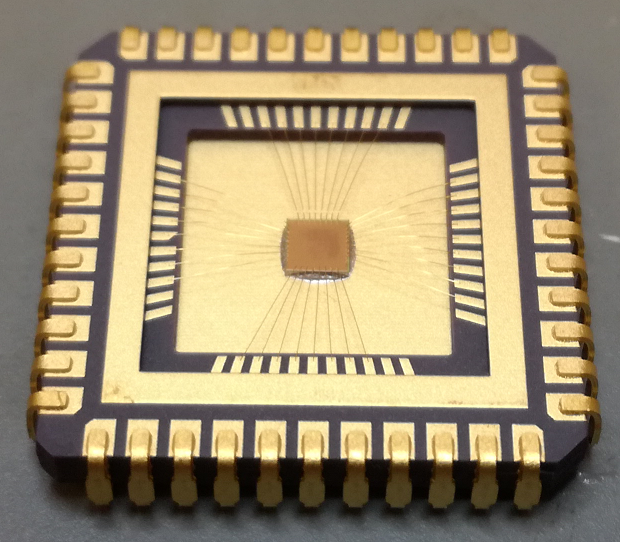}
\label{fig:90nm_JLCC}}\hfill
\subfigure[Microscope photo 90~nm ASIC]{
\includegraphics[height=3.8cm]{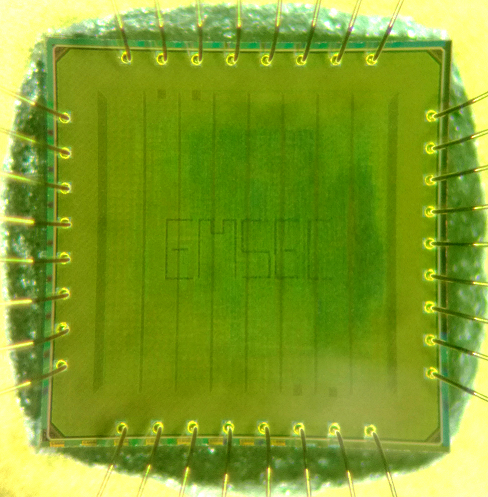}
\label{fig:90nm_Micro}}
\caption{Layout schematic and photos of the ASIC prototypes.}
\vspace{-0.1 in}
\label{fig:ASICs}
\end{figure}

Section~\ref{sec:result_fpga} demonstrated by practical experiments that the proposed hardware Trojan and the presented implementation techniques are valid on FPGA-based platforms. Here, we aim to provide a similar case study, but with respect to ASIC platforms. In this regard we carried out the described design stages and implemented the trojanized PRESENT threshold implementation circuit in two different process technologies,90~nm and 65~nm low power CMOS. Both ASICs, which can be seen in Figure~\ref{fig:ASICs}, were developed using an identical design procedure, including the usage of low, high and standard threshold voltage cells, and were manufactured by the same foundry.

The size of both chips is 2mm \texttt{x} 2mm. The side-channel resistant PRESENT TI cores containing the parametric SCA Trojans have been placed and routed in clearly delimited rectangular areas, which are marked in red color with a white cross in both layout schematics~\ref{fig:65nm_Layout} and~\ref{fig:90nm_Layout}, taken from the \emph{Synopsys IC Compiler (Version 2016.12)} software.


We made use of the malicious PRESENT TI S-Box that has been introduced in the previous sections and embedded it in a design with full encryption functionality shown in Figure~\ref{fig:arch}. 

Synthesizing the unaltered threshold implementation of the PRESENT S-Box (i.e., without the inserted Trojan) in the 90\,nm and 65\,nm target libraries revealed that the design could potentially meet clock frequencies in the GHz range, even when operated under worst case operating conditions (i.e., low supply voltage, high temperature). Unfortunately, no digital IO cells were available in our target technologies that could reliably propagate such a high-frequency clock into the circuit. Thus, when inserting the Trojan in the proposed way, i.e., by subtle manipulations at the sub-transistor level, and keeping period \colorbox{yellow}{\Circled{3}} small and stealthy, it could never be triggered, due to the restrictions of the IO cells and the extremely high performance of the circuit in the target technologies.

This observation already shows that implementing and testing such a design on an ASIC is more challenging than on an FPGA, due to the much higher performance of ASICs. In this regard we have to conclude that an ultra-lightweight block cipher implementation like the serialized PRESENT, implemented in an advanced CMOS technology with small propagation delays, may not be the optimal choice for integrating such a Trojan on an ASIC in the most stealthy way. Yet, to keep the results comparable to those in~\cite{DBLP:conf/asiacrypt/EnderG0P17}, we stick to this example and find a workaround for the IO restriction. 

Another difficulty when developing ASIC prototypes is the extensive amount of time and monetary resources that have to be invested. Thus, it is desirable to obtain a fully functioning prototype in the first attempt when designing a test chip. However, this is particularly difficult to achieve when the functionality of the design depends highly on the exact timing of certain signal paths in such a way that even small deviations from the predicted behavior can invalidate crucial assumptions. In such a case the designer has to trust its foundry that the characterized timing information included in the standard cell libraries and simulation models perfectly reflects the reality -- which is hardly ever possible due to process variations. Thus, even commercial IC design houses often require multiple generations of prototypes that need to be characterized and adapted between each iteration to finally end up with a marketable end-user product. Unlike FPGA platforms where a new HDL design can be synthesized and implemented within a few minutes and without any additional cost, which allows for trial \& error approaches, an IC implementation requires at least several months per tape out as well as a significant amount of money, even when sharing a wafer between multiple projects. Thus, for our case study, in order to not require multiple IC manufacturing iterations, but rather obtain a working prototype in the first attempt, we chose to limit the potential sources of error at the cost of sacrificing a part of the potential stealthiness of the Trojan. In particular, we chose to realize the delay which needs to be distributed among the selected paths partially by so-called delay gates\footnote{Those gates were required since selecting even the slowest cells (high threshold voltage, low drive strength) could not add enough delay in order to make the Trojan triggerable through the IO cells.} and optimize for a broad frequency range that triggers the Trojan while the PRESENT core still encrypts correctly (i.e., period \colorbox{yellow}{\Circled{3}}). A delay gate does not have any logical functionality but simply propagates its input signal with a certain propagation delay to its output. Clearly, inserting delay gates into the masked S-Box makes the Trojan less stealthy than sub-transistor level modifications. The same is true for a significant reduction of the overall operating frequency of the circuit as it can be observed in the results presented in the following (this reduction is neccessary due to the restrictions of the IO cells). However, we would like to stress that this case study is simply proving the conceptual soundness of the approach, in the sense that inserting this delay-based Trojan makes a side-channel resistant implementation vulnerable when increasing the clock frequency beyond a certain point. It is planned to demonstrate the stealthiness of the Trojan on ASIC platforms in further case studies. In many cases, for example targeting more complex non-linear functions (like the AES S-Box) or less advanced CMOS technologies (implying larger delays), such a use of additional delay gates will not be required since the critical path of the design actually restricts the maximum operating frequency of the design (and not the limitations of the IO cells). 
Again, we chose the PRESENT threshold implementation as a case study here to keep the results comparable to~\cite{DBLP:conf/asiacrypt/EnderG0P17}. And even in our case, where we particularly aimed for a broad range of period \colorbox{yellow}{\Circled{3}}, the overhead in terms of area is very small, even less than half a percent as apparent from Table~\ref{tab:area}.
\begin{table}
	\begin{tabular}{cccc}
	\hline
	\textbf{Technology node} & \textbf{Area w/o Trojan} & \textbf{Area w/ Trojan} & \textbf{Overhead}\\
	\hline
	65\,nm & 4988.5 GE & 5006.5 GE & +0.36\%\\
	90\,nm & 4807.8 GE & 4825.8 GE & +0.37\%\\
	\hline
	\end{tabular}
	\caption{Area comparison (post-layout) of PRESENT TI implementation with and without inserted Trojan (realized by delay gates).}
	\label{tab:area}
\end{table}
The range of clock frequencies that cause a certain state of the trojanized design can be seen in Table~\ref{tab:frequency}.
\begin{table}
	\begin{tabular}{ccccc}
	\hline
	\textbf{Status} & \textbf{65\,nm ASIC} & \textbf{90\,nm ASIC}\\
	\hline
	\colorbox{green}{\Circled{1}} & $f$ $\leq$ 33 MHz & $f$ $\leq$ 56 MHz\\
	\colorbox{lightgray}{\Circled{2}} & 33 MHz $<$ $f$ $\leq$ 38 MHz & 56 MHz $<$ $f$ $\leq$ 61 MHz\\
	\colorbox{yellow}{\Circled{3}} & 38 MHz $<$ $f$ $\leq$ appr. 1 GHz & 61 MHz $<$ $f$ $\leq$ appr. 1 GHz\\
	\colorbox{myred}{\Circled{4}} & appr. 1 GHz $<$ $f$ & appr. 1 GHz $<$ $f$\\
	\hline
	\end{tabular}
	\caption{Frequency ranges for the different design states.}
	\label{tab:frequency}
	\vspace{-0.2 in}
\end{table}
As described before, state \colorbox{yellow}{\Circled{3}} has the broadest frequency range and can easily be targeted by setting the clock frequency above 38 MHz for the 65\,nm ASIC and 56 MHz for the 90\,nm ASIC. The upper limit where the output of the circuit becomes faulty is an approximation, since it could not be determined experimentally due to the limitation of the IO cells.
\subsection{Measurement Setup}
In order to perform the SCA evaluations on the ASIC prototypes we built a simple custom measurement board. Since the ASICs have been packaged in JLCC-44 package (see Figure~\ref{fig:90nm_JLCC}), the custom board provides a corresponding PLCC-44 socket as well as connectors for a BASYS-3 FPGA board (containing an Artix-7 FPGA) to control the communication between PC and the ASIC. We measured the power consumption of the ASICs in the $V_{dd}$ path by means of a digital sampling oscilloscope at a fixed sampling rate of 200 samples per clock cycle. Since the operating frequency varies between the different scenarios (Trojan triggered or not triggered), fixing the number of samples per clock cycle (instead of per time period) is the most fair evaluation method.
\subsection{SCA Results}
We evaluate the SCA resistance of our designs in three different settings using a non-specific t-test (fixed versus random)~\cite{t_test2, ttestCHES15} to examine the existence of detectable leakage. First, to validate the correct functionality of the setup, we start with a non-specific t-test when the PRNG of the target design (used to share the plaintext for the TI PRESENT encryption) is turned off, i.e., generating always zero instead of random numbers. Afterwards, we activate the PRNG and operate the design at low frequency in order to not activate the Trojan. Then, when the PRNG is still running we increase the clock frequency in order to activate the Trojan. In the latter case we also conduct key-recovery attacks.
\subsubsection{Results on 90\,nm ASIC}

\begin{figure*}[ht!]
\centering
\vspace{-.3 in}
\subfigure{
\includegraphics[width=0.45\textwidth]{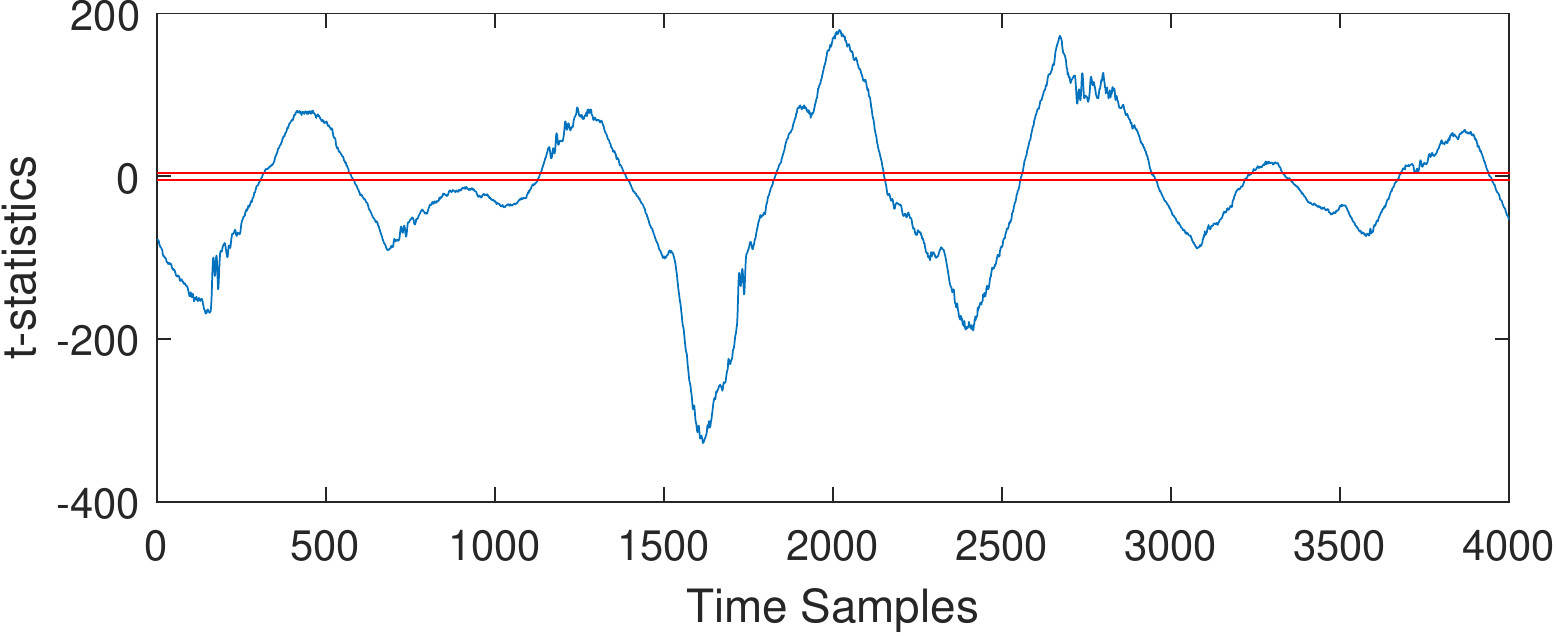}
\label{fig:90nm_PRNG_OFF_1st}}\hfill
\subfigure{
	\includegraphics[width=0.45\textwidth]{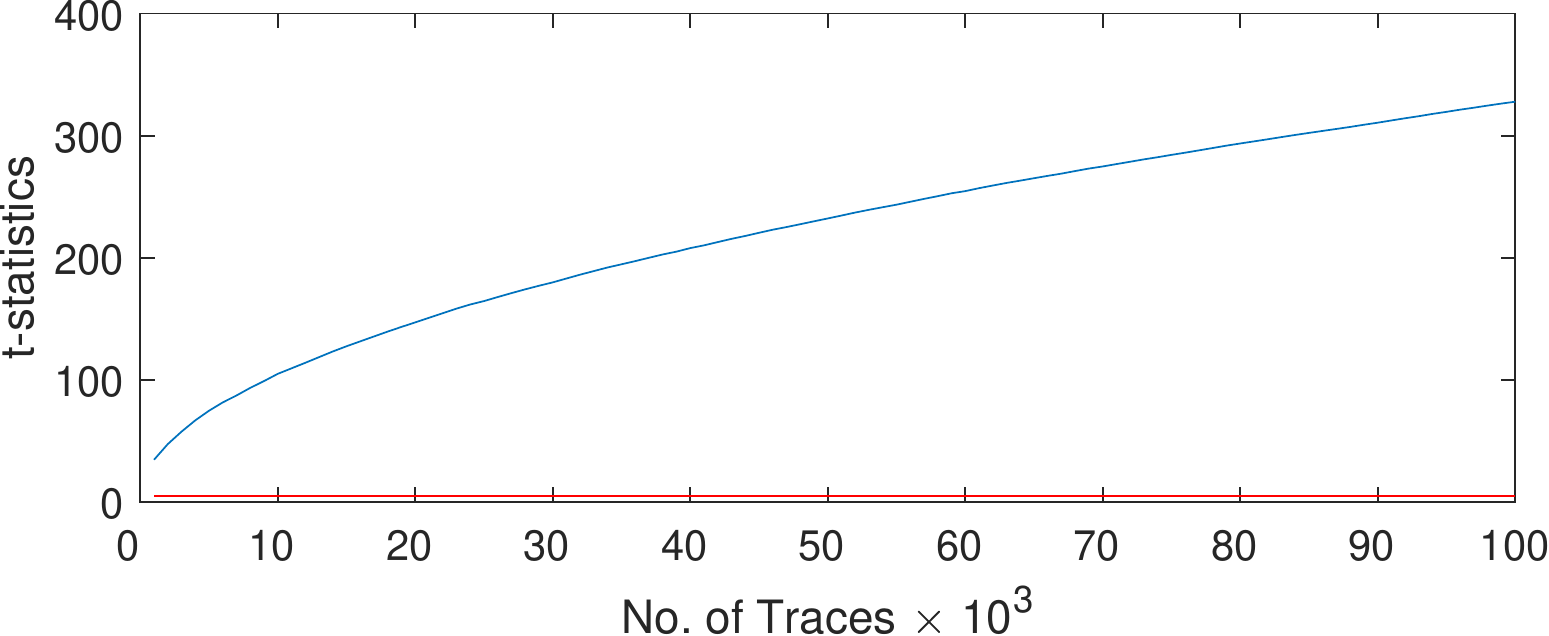}
	\label{fig:90nm_PRNG_OFF_1st_prog}}\\
\subfigure{
\includegraphics[width=0.45\textwidth]{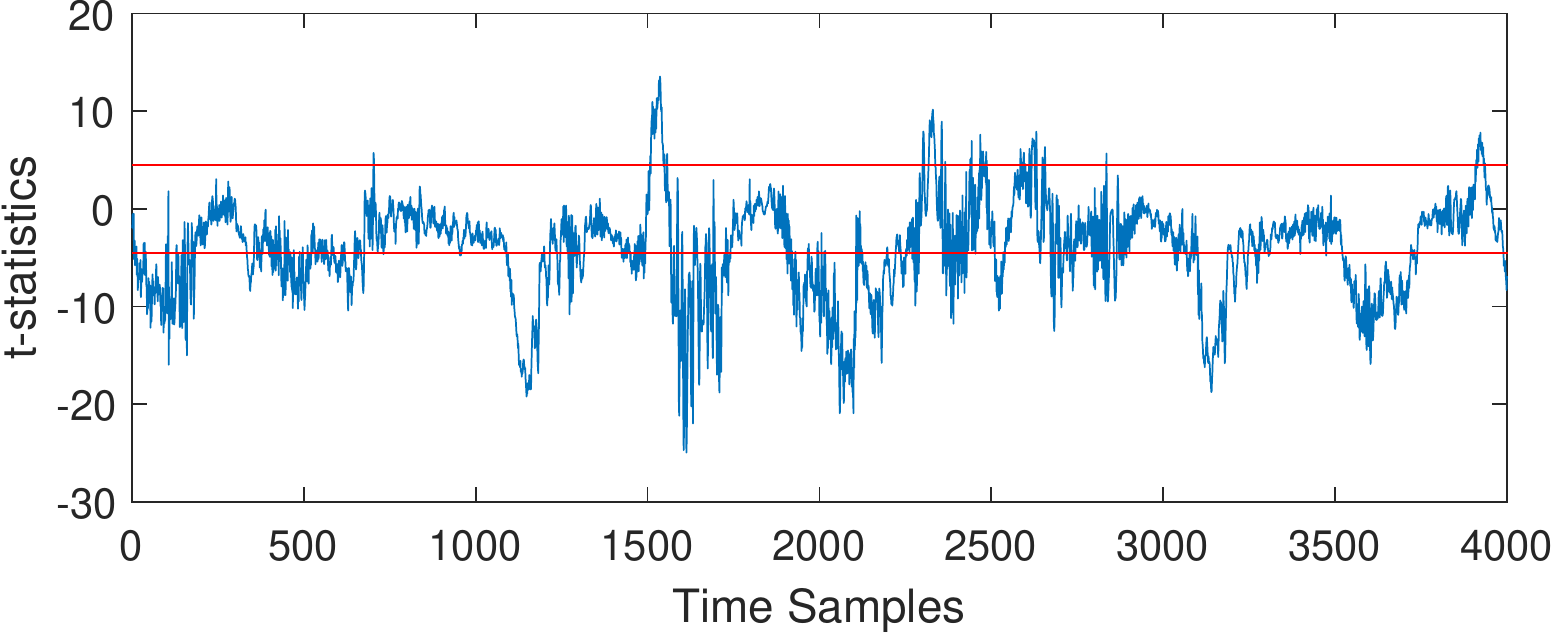}
\label{fig:90nm_PRNG_OFF_2nd}}\hfill
\subfigure{
	\includegraphics[width=0.45\textwidth]{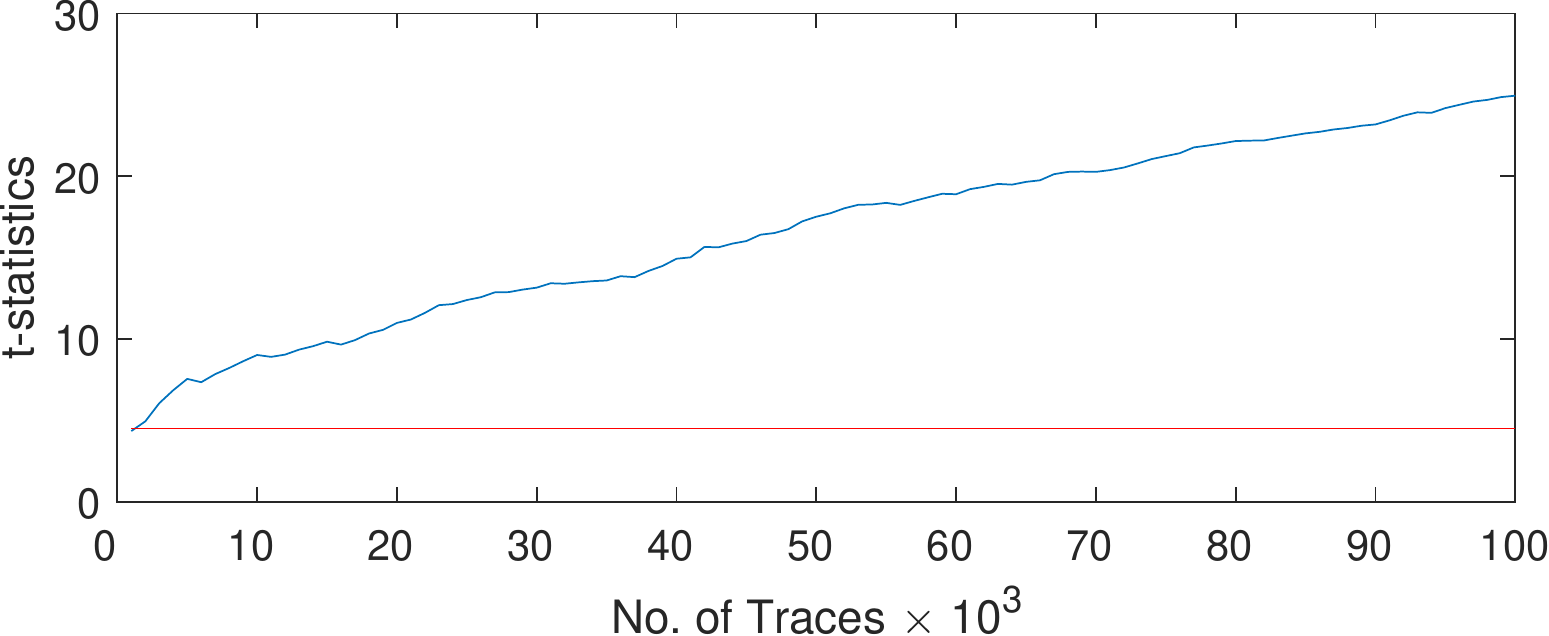}
	\label{fig:90nm_PRNG_OFF_2nd_prog}}\\
\subfigure{
\includegraphics[width=0.45\textwidth]{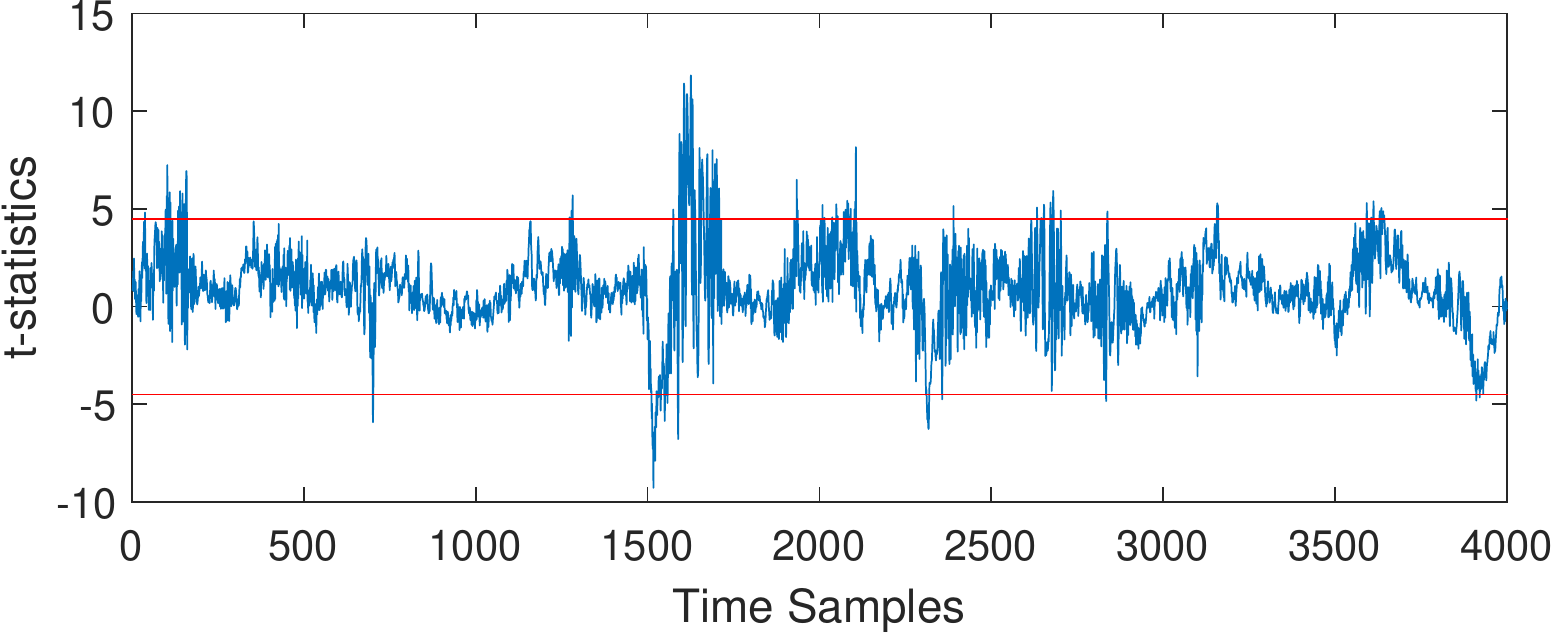}
\label{fig:90nm_PRNG_OFF_3rd}}\hfill
\subfigure{
\includegraphics[width=0.45\textwidth]{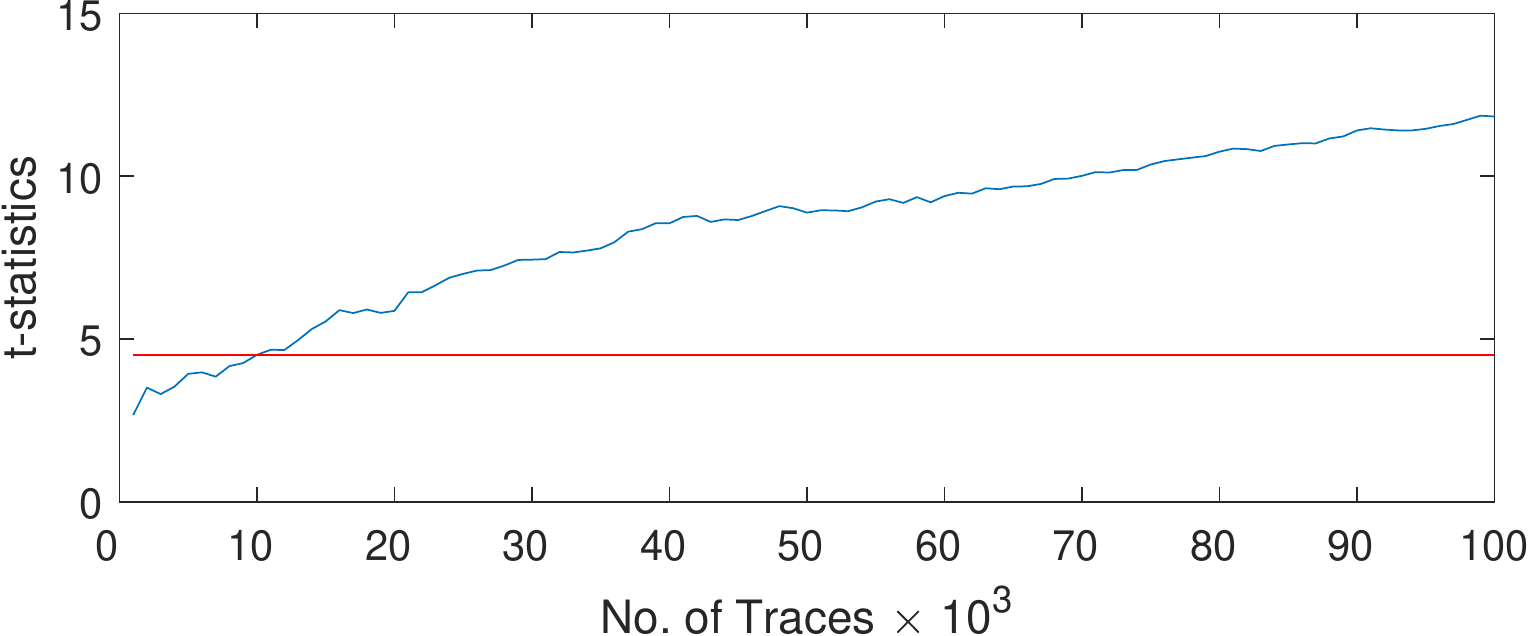}
\label{fig:90nm_PRNG_OFF_3rd_prog}}
\caption{90\,nm ASIC, PRNG off, clock frequency 25~MHz (trojan not triggered), $t$-test results with 1 million traces (left), absolute maximum $t$-value over the number of traces (right).}
\label{fig:90nm_PRNG_OFF-eps-converted-to.pdf}
\end{figure*}

We first collected 1,000,000 traces with PRNG switched off when the design is operated at 25 MHz, i.e., the Trojan is not triggered. We followed the concept given in~\cite{ttestCHES15} for the collection of traces belonging to fixed and random inputs. Figure~\ref{fig:90nm_PRNG_OFF-eps-converted-to.pdf} shows the corresponding t-test results.

As expected a significant amount of detectable leakage can be observed in all moments, confirming the validity of the setup and the developed evaluation tools.

\begin{figure*}[ht!]
\centering
\vspace{-.1 in}
\subfigure{
\includegraphics[width=0.45\textwidth]{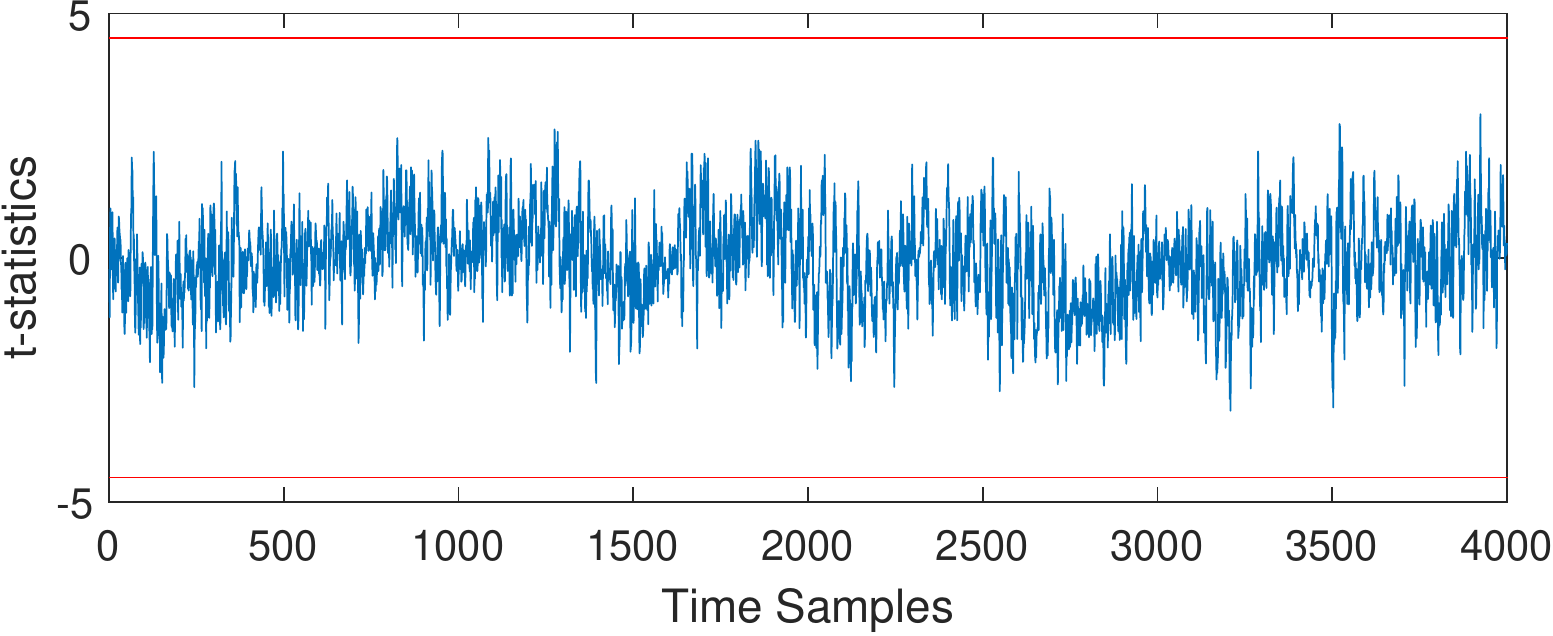}
\label{fig:90nm_PRNG_ON_1st}}\hfill
\subfigure{
	\includegraphics[width=0.45\textwidth]{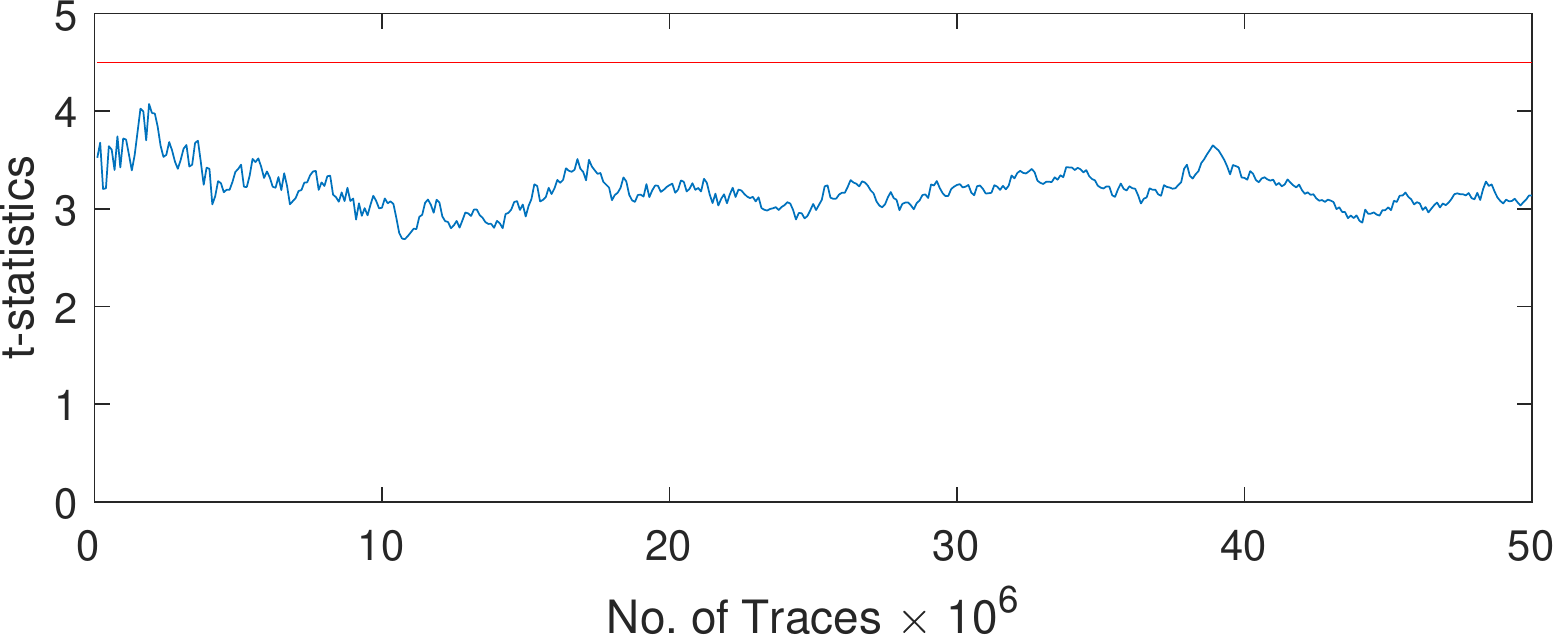}
	\label{fig:90nm_PRNG_ON_1st_prog}}\\
\subfigure{
\includegraphics[width=0.45\textwidth]{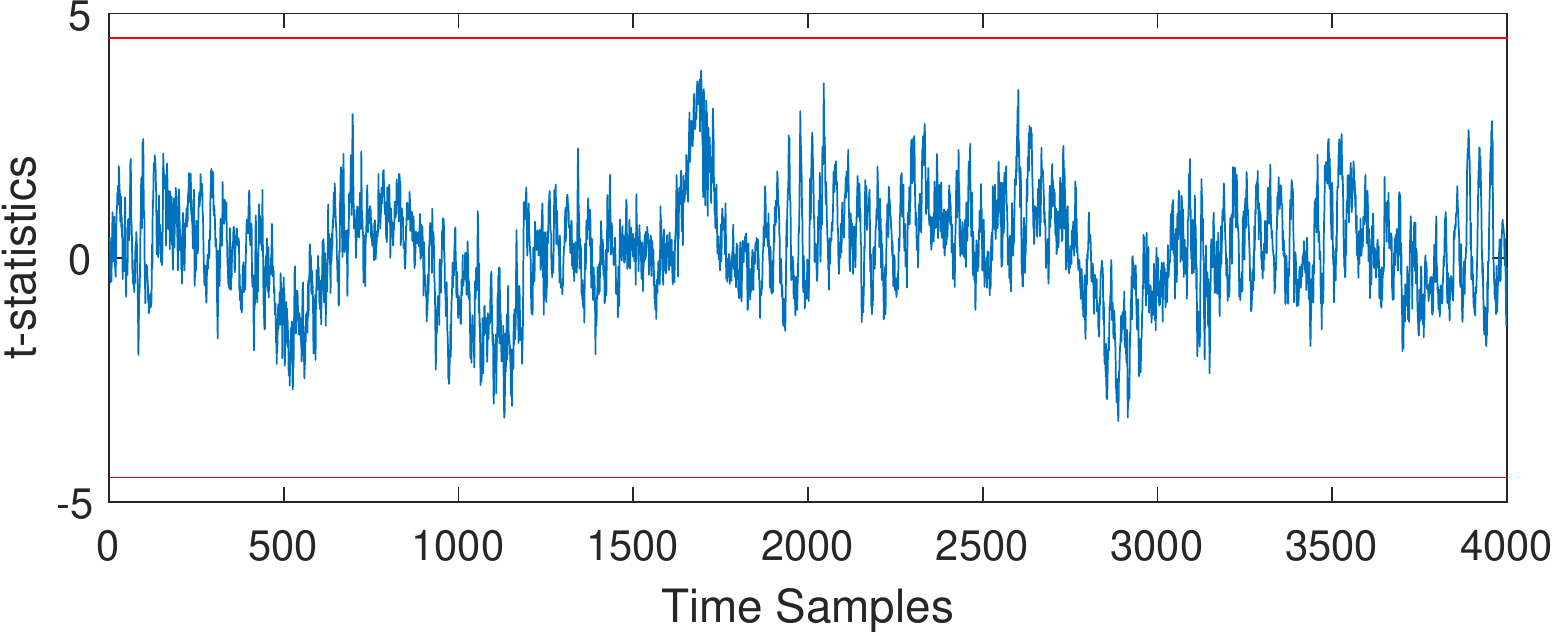}
\label{fig:90nm_PRNG_ON_2nd}}\hfill
\subfigure{
	\includegraphics[width=0.45\textwidth]{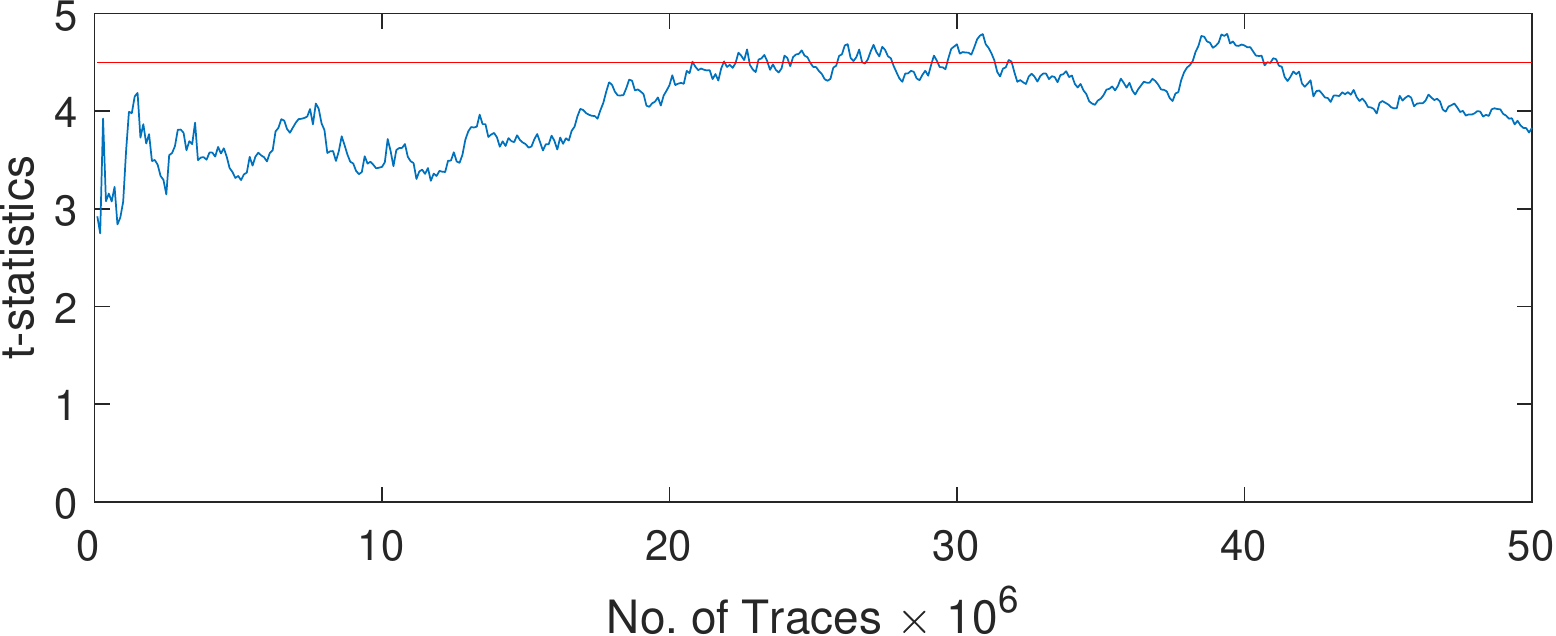}
	\label{fig:90nm_PRNG_ON_2nd_prog}}\\
\subfigure{
\includegraphics[width=0.45\textwidth]{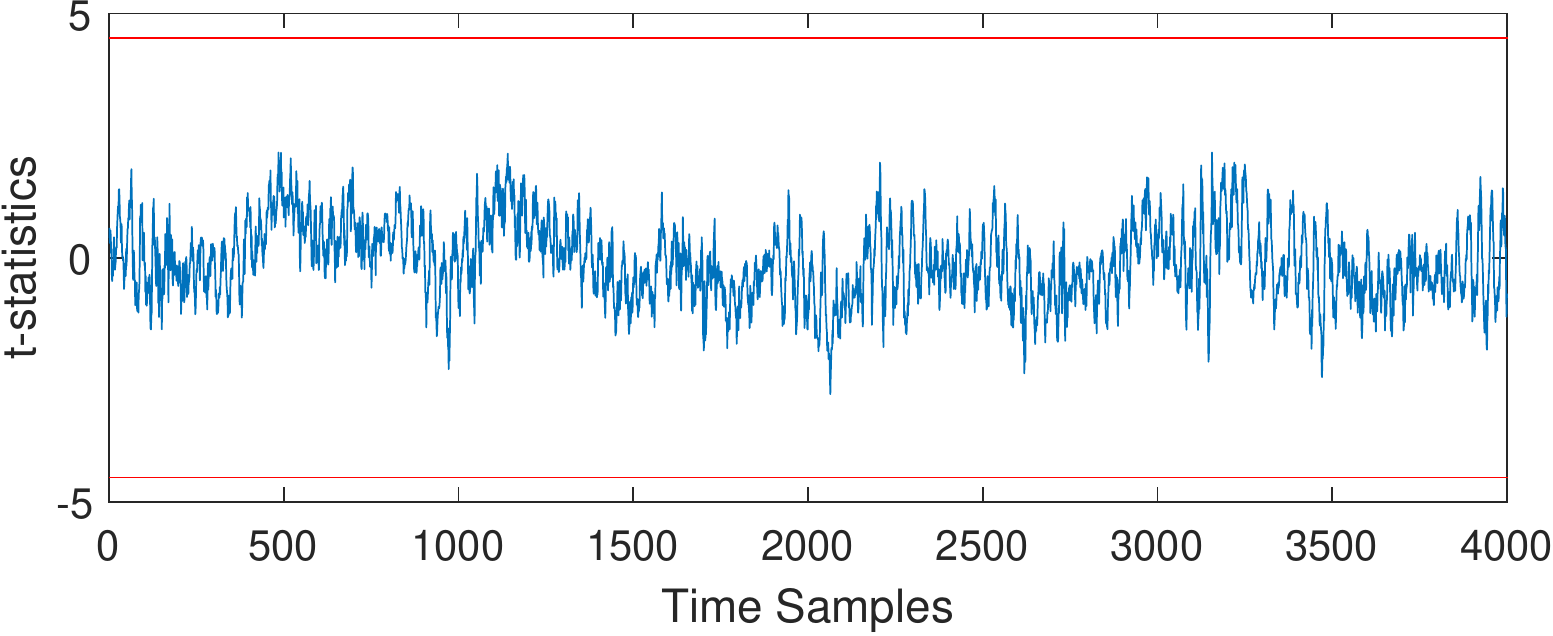}
\label{fig:90nm_PRNG_ON_3rd}}\hfill
\subfigure{
\includegraphics[width=0.45\textwidth]{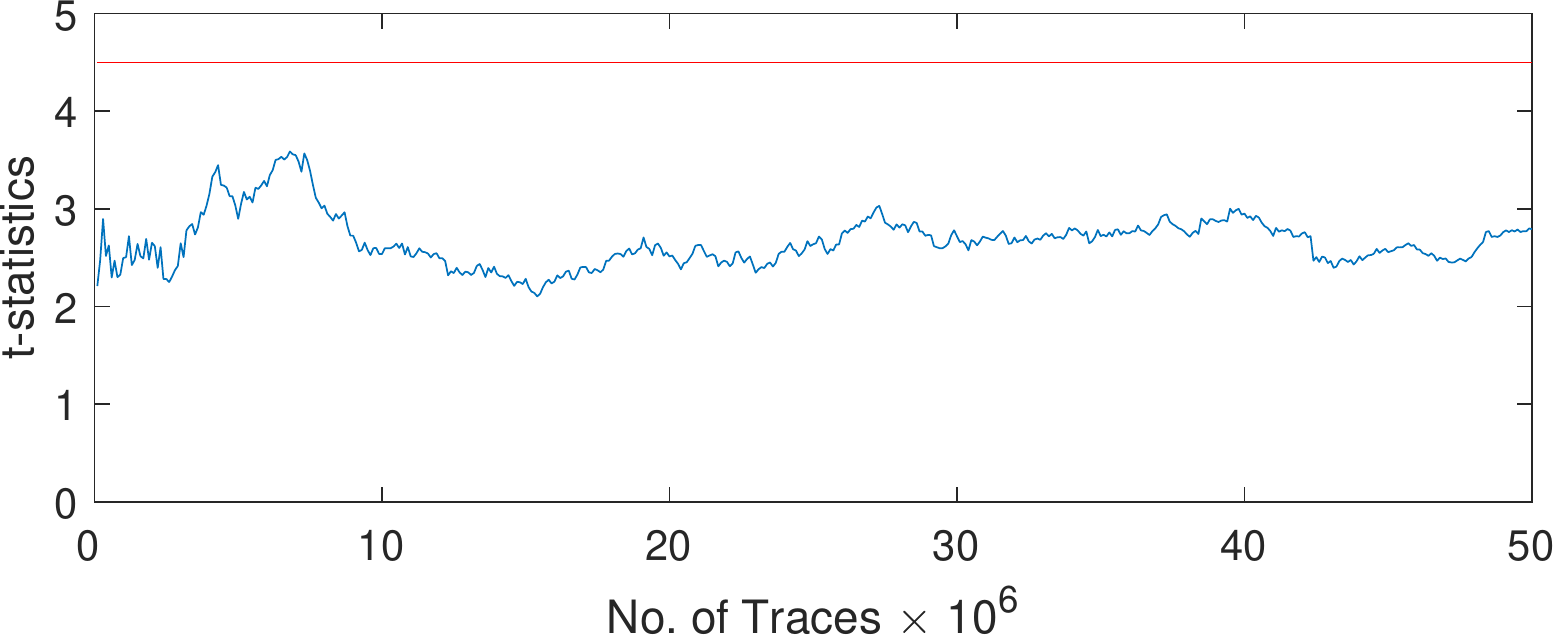}
\label{fig:90nm_PRNG_ON_3rd_prog}}
\caption{90\,nm ASIC, PRNG on, clock frequency 25~MHz (trojan not triggered), $t$-test results with 50 million traces (left), absolute maximum $t$-value over the number of traces (right).}
\label{fig:90nm_PRNG_ON}
\end{figure*}

To repeat the same process when the PRNG is turned on, i.e., the masks for initial sharing of the plaintext are randomly chosen and uniformly distributed, we collected 50,000,000 traces for non-specific t-test evaluations. In this case, the device still operates at 25 MHz, i.e., the Trojan is not triggered. The corresponding results are shown in Figure~\ref{fig:90nm_PRNG_ON}.
 
It can be seen that no leakage is detected in any of the three statistical moments after 50,000,000 traces. However, when observing the progress of the maximum absolute t-value in the second-order moment over the number of traces one may notice that the 4.5 threshold is occasionally exceeded. We should emphasize here that the underlying TI construction is a first-order masking, which can provide provable security against first-order SCA attacks. However, higher-order attacks (in this case second-order attacks already) are expected to exploit the leakage, but they are sensitive to the noise level~\cite{DBLP:journals/tc/ProuffRB09} since accurately estimating higher-order statistical moments requires huge amounts of samples compared to lower-order moments. Thus, the second-order leakage is not unexpected, but the noise level seems too large to reliably detect (or exploit) this leakage.

\begin{figure*}[ht!]
\centering
\vspace{-.1 in}
\subfigure{
\includegraphics[width=0.45\textwidth]{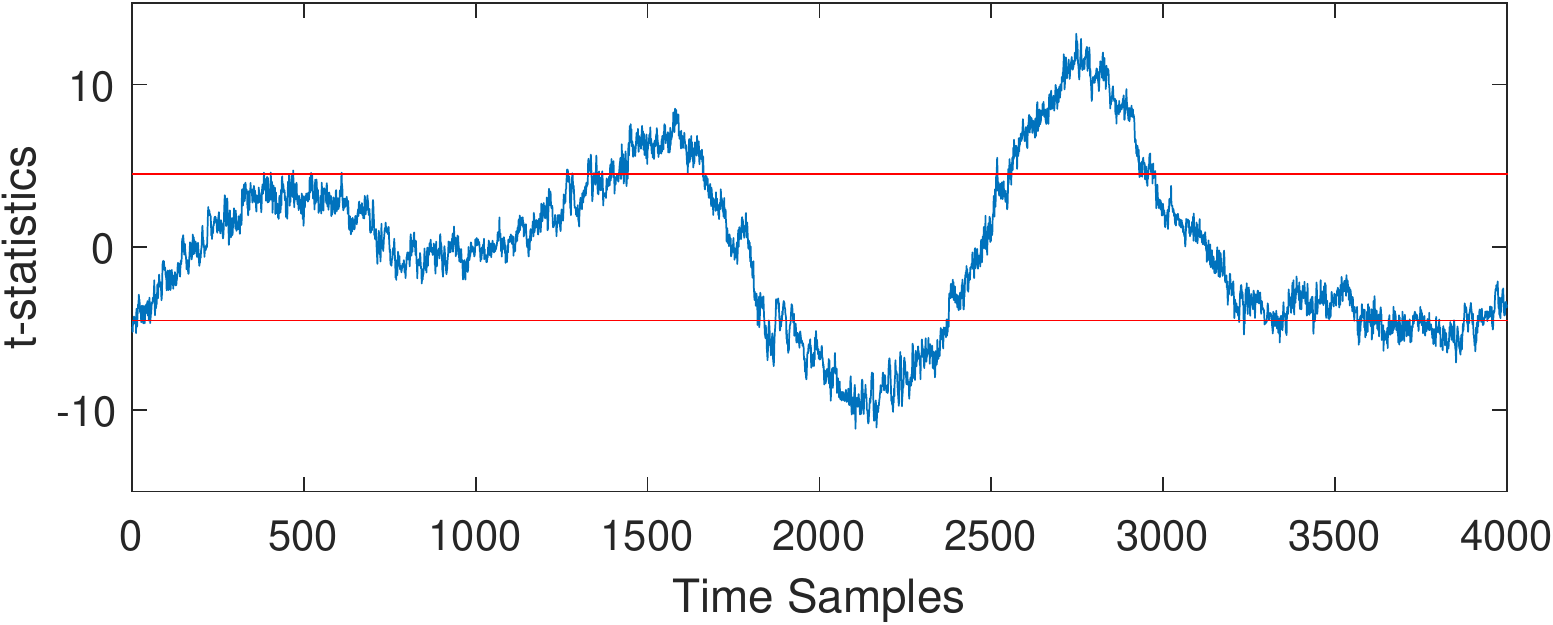}
\label{fig:90nm_PRNG_ON_Trojan_ON_1st}}\hfill
\subfigure{
	\includegraphics[width=0.45\textwidth]{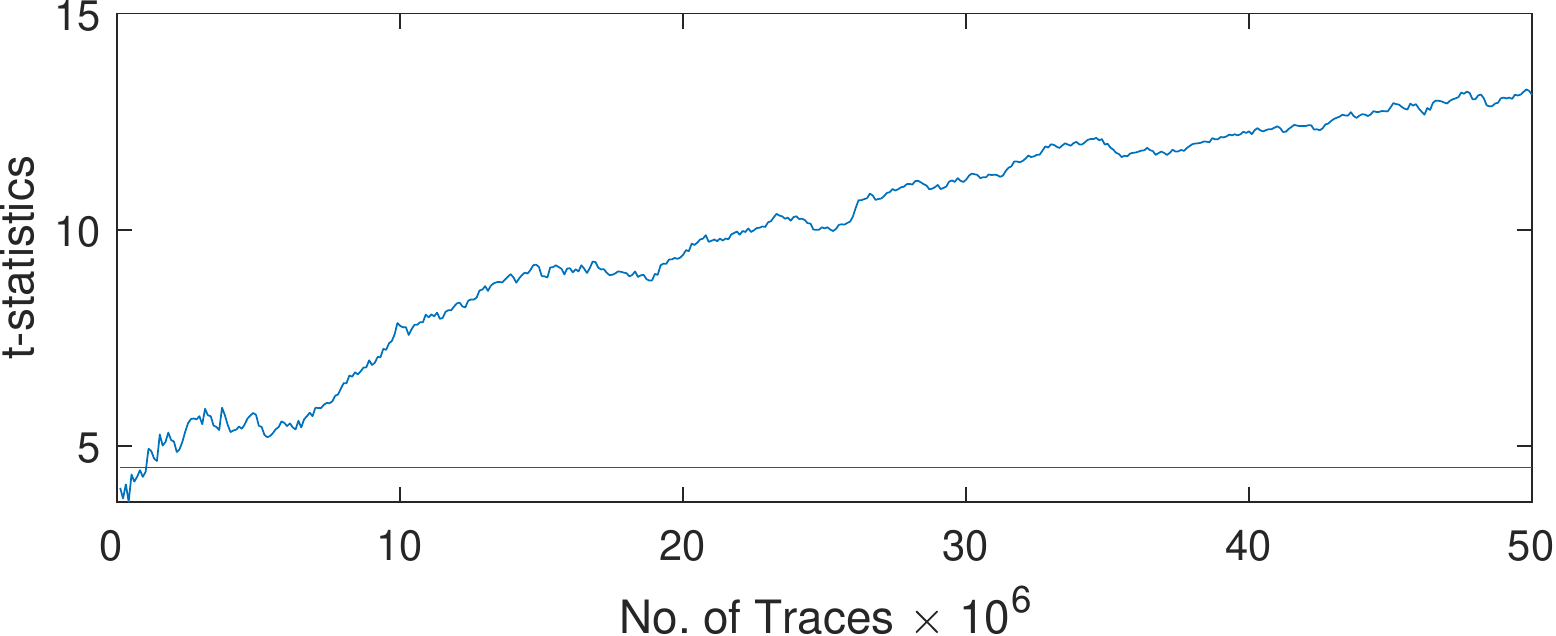}
	\label{fig:90nm_PRNG_ON_Trojan_ON_1st_prog}}\\
\subfigure{
\includegraphics[width=0.45\textwidth]{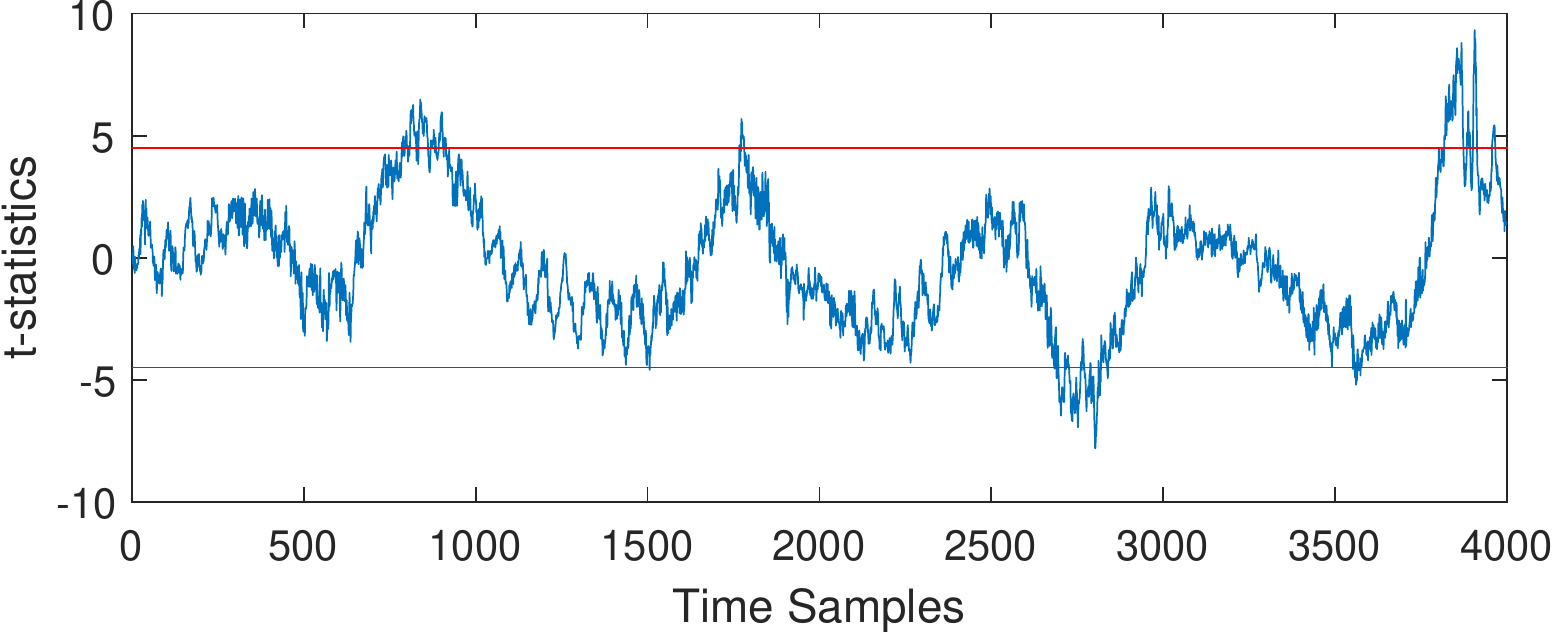}
\label{fig:90nm_PRNG_ON_Trojan_ON_2nd}}\hfill
\subfigure{
	\includegraphics[width=0.45\textwidth]{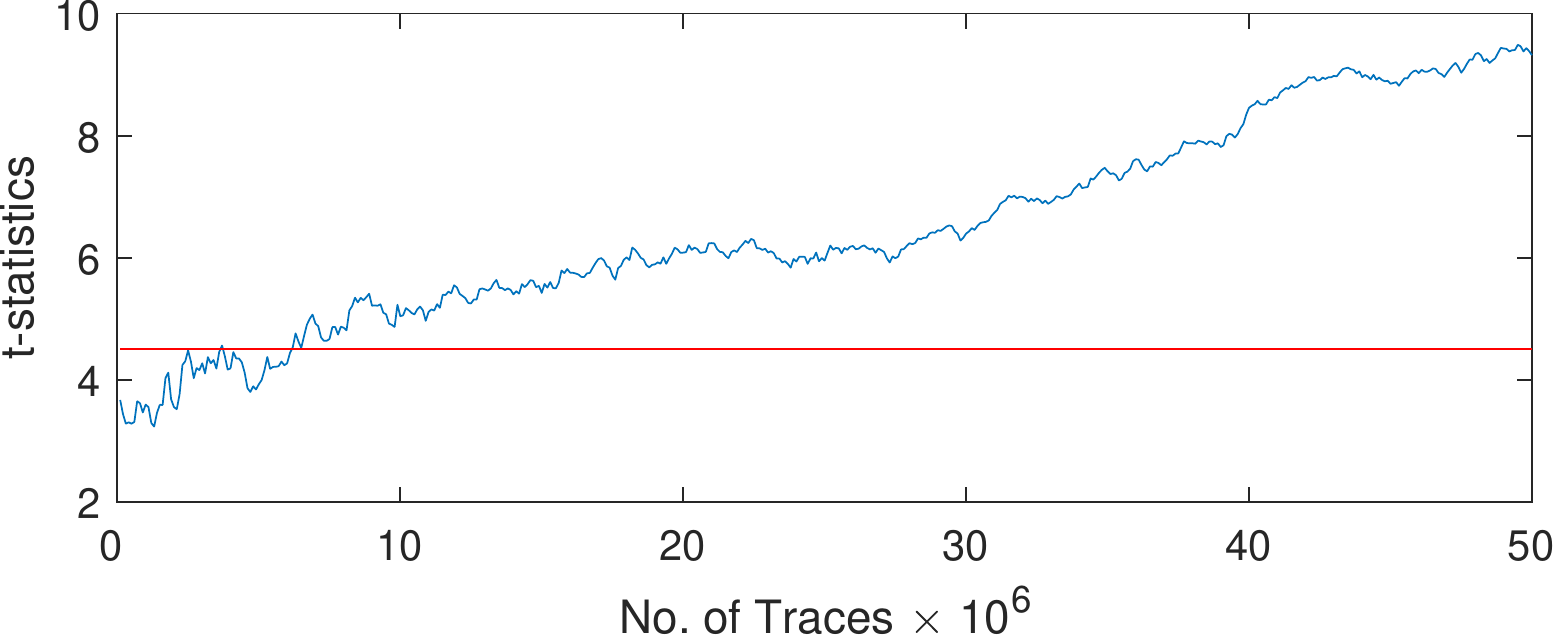}
	\label{fig:90nm_PRNG_ON_Trojan_ON_2nd_prog}}\\
\subfigure{
\includegraphics[width=0.45\textwidth]{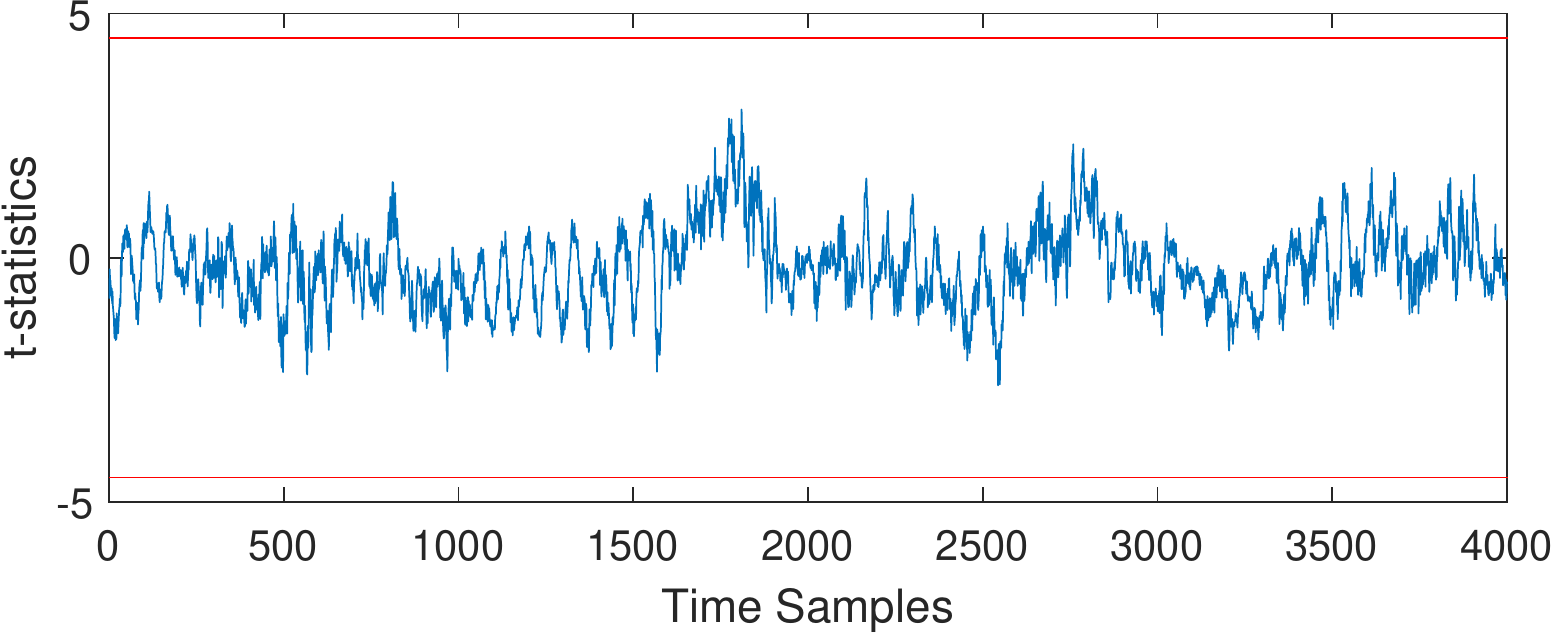}
\label{fig:90nm_PRNG_ON_Trojan_ON_3rd}}\hfill
\subfigure{
\includegraphics[width=0.45\textwidth]{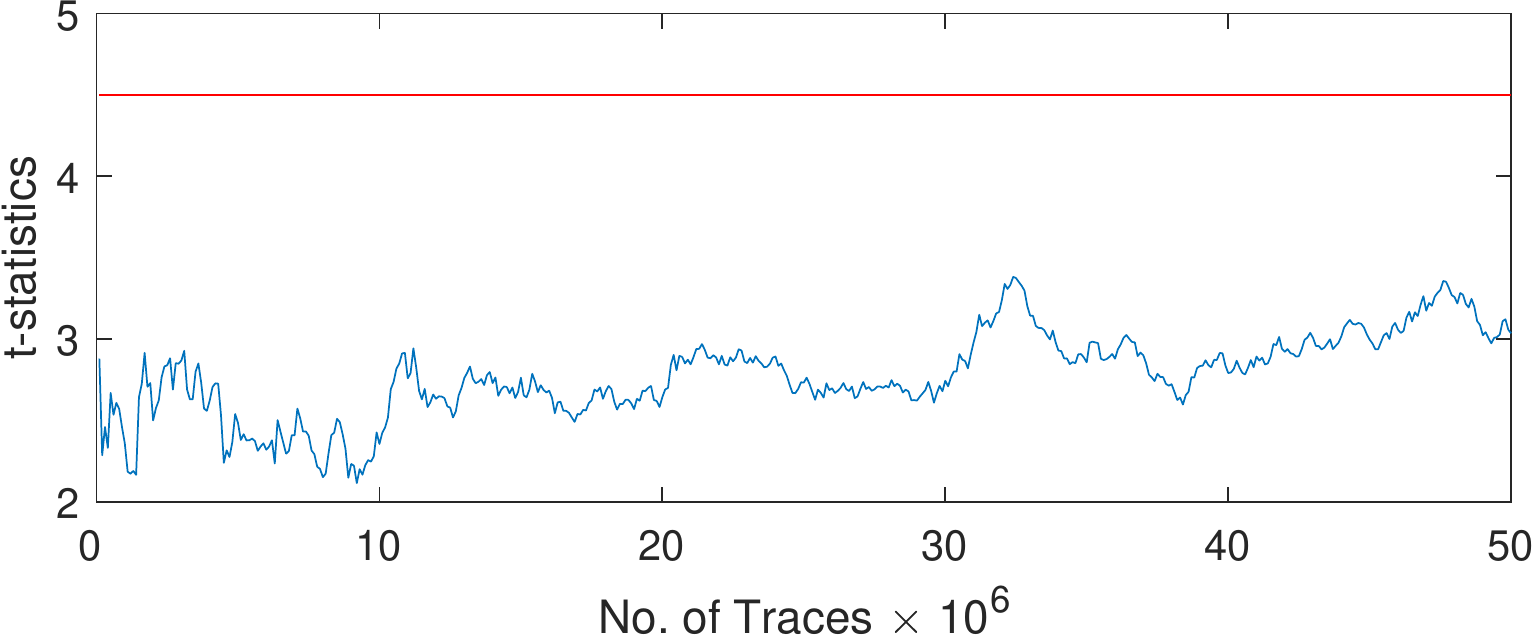}
\label{fig:90nm_PRNG_ON_Trojan_ON_3rd_prog}}
\caption{90\,nm ASIC, PRNG on, clock frequency 85~MHz (trojan triggered), $t$-test results with 50 million traces (left), absolute maximum $t$-value over the number of traces (right).}
\label{fig:90nm_PRNG_ON_Trojan_ON}
\end{figure*}

As the last step, the same scenario is repeated when the clock frequency is increased to 85 MHz, where the design is in the \colorbox{yellow}{\Circled{3}} period, i.e., with correct functionality and without uniformity. Similar to the previous experiment, we collected 50,000,000 traces for a non-specific t-test, whose results are shown in Figure~\ref{fig:90nm_PRNG_ON_Trojan_ON}.

\begin{figure*}[ht!]
	\centering
	\vspace{-.15 in}
	\subfigure{
		\includegraphics[width=0.98\columnwidth]{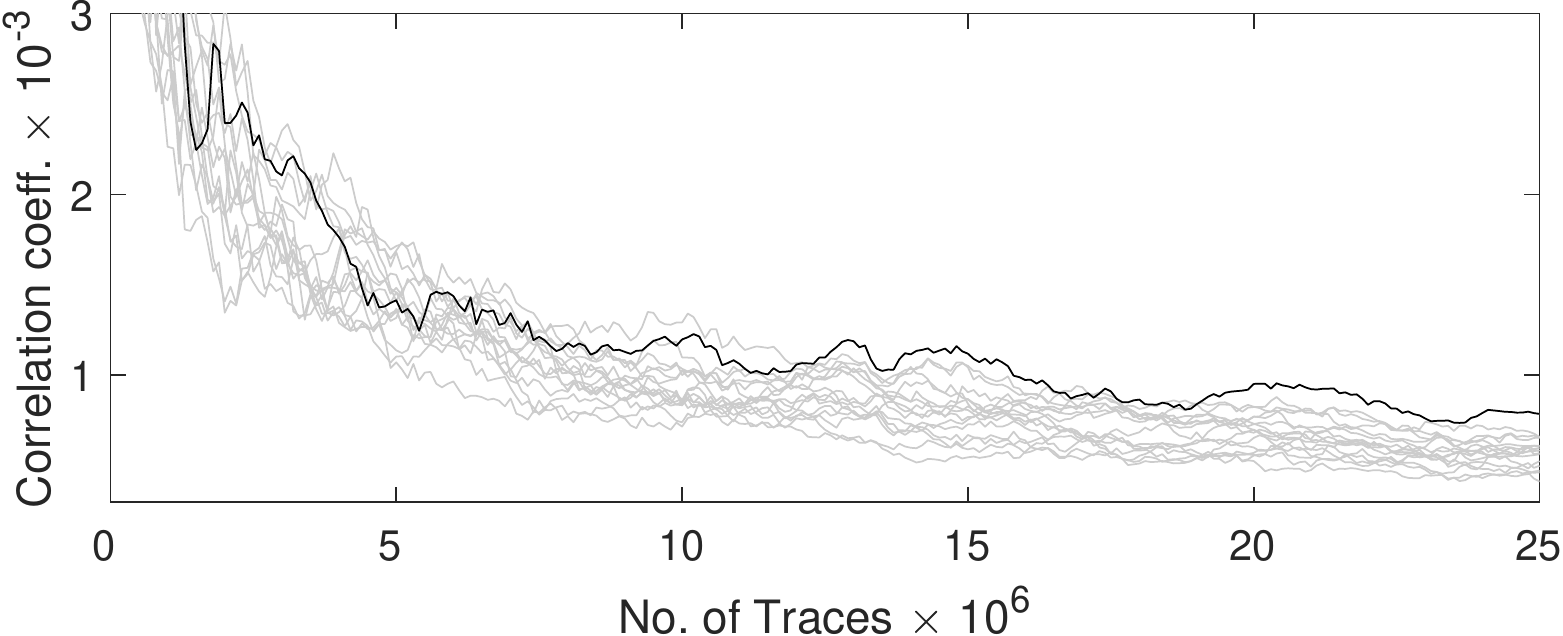}
		\label{fig:90nm_CPA_prog}}\hfill		
	\subfigure{
		\includegraphics[width=0.98\columnwidth]{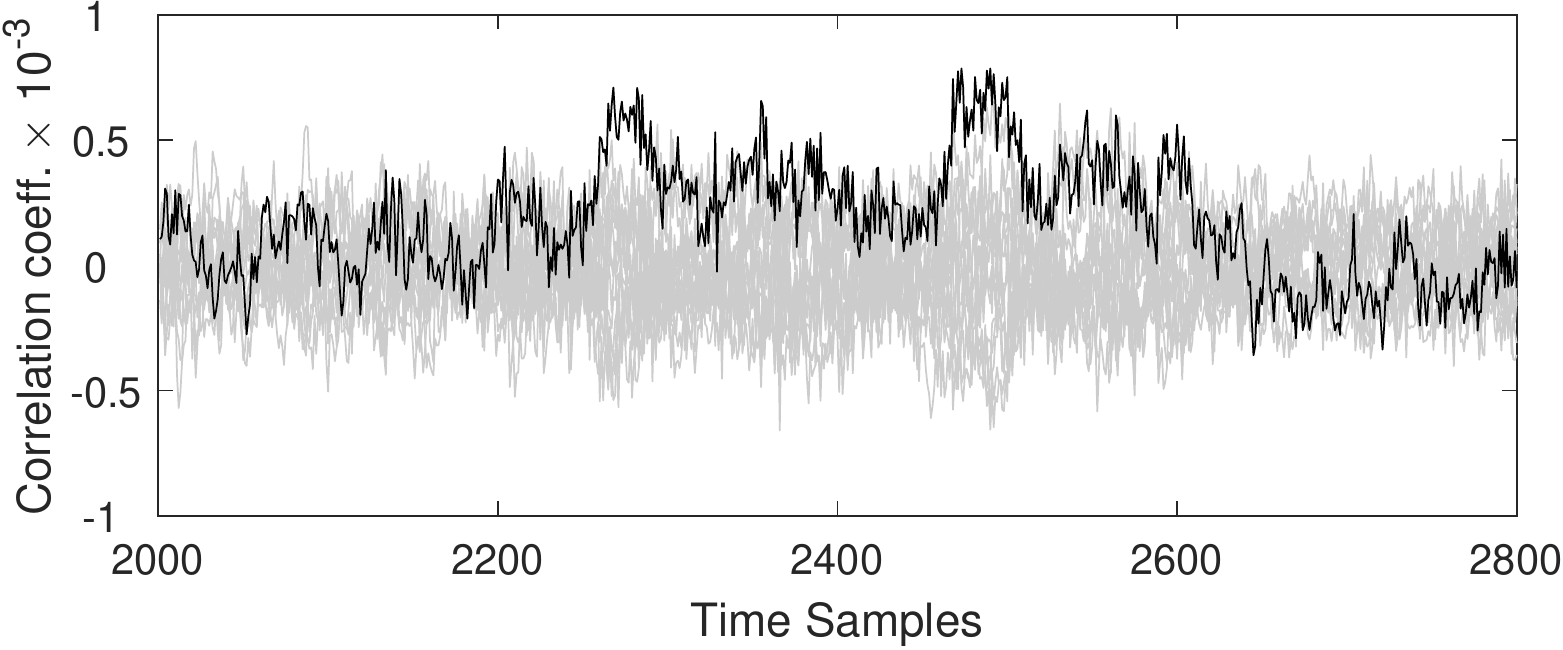}
		\label{fig:90nm_CPA_sample}}
	\caption{90\,nm ASIC, PRNG on, clock frequency 85~MHz (trojan triggered), CPA results targeting a key nibble based on an S-Box output bit with 25 million traces (right), absolute maximum correlation coefficient over the number of traces (left).}
	\label{fig:90nm_CPA}
	\vspace{-0.2 in}
\end{figure*}

As shown by the graphics, there is detectable leakage through the first and second statistical moment but with lower t-statistics compared to the case with PRNG off. Therefore, we also have to examine the feasibility of key recovery attacks. To this end, we made use of those collected traces which are associated with random inputs, i.e., around 25,000,000 traces of the last non-specific t-test. We conducted several different CPA and DPA attacks considering intermediate values of the underlying PRESENT encryption function. The most successful attack was recognized as classical DPA attack~\cite{dpa_kocher} targeting a key nibble by predicting
an S-Box output bit at the first round of the encryption. As an example, Figure~\ref{fig:90nm_CPA} presents an exemplary corresponding result.

\subsubsection{Results on 65\,nm ASIC}

\begin{figure*}[h!]
\centering
\vspace{-.3 in}
\subfigure{
\includegraphics[width=0.45\textwidth]{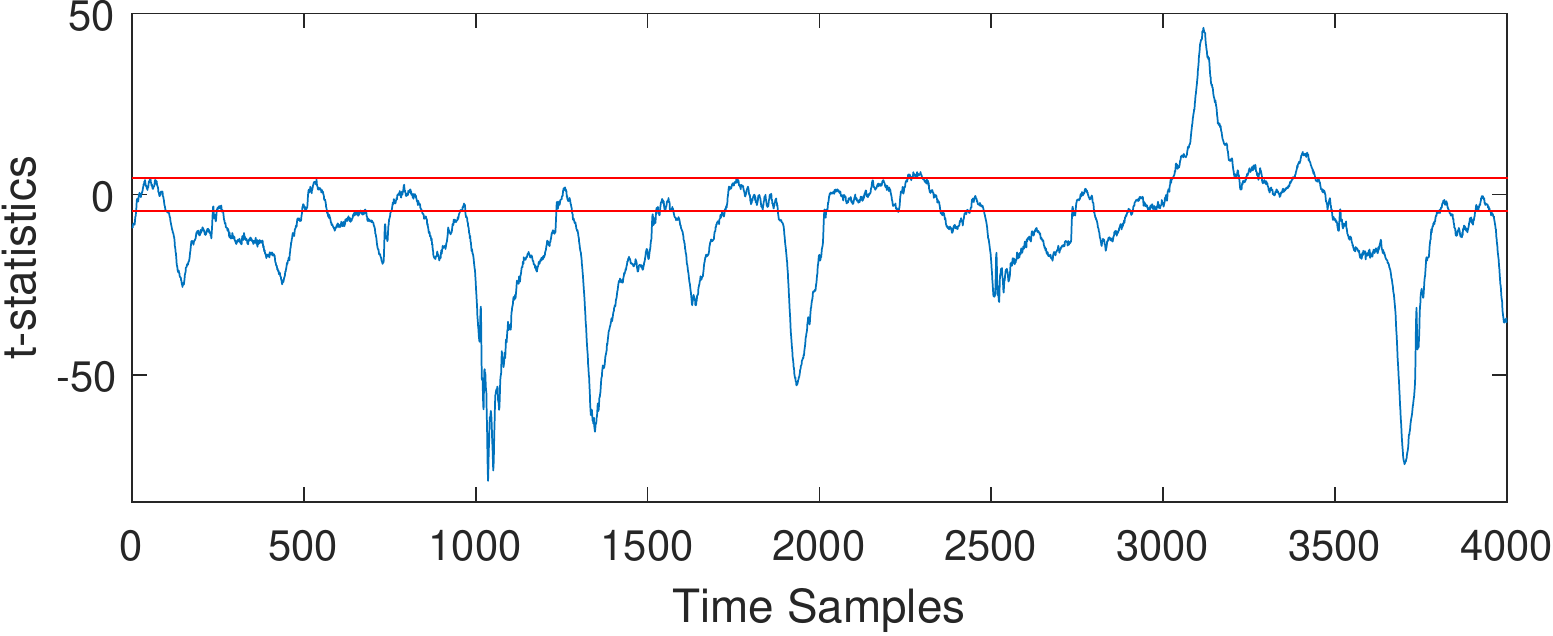}
\label{fig:65nm_PRNG_OFF_1st}}\hfill
\subfigure{
	\includegraphics[width=0.45\textwidth]{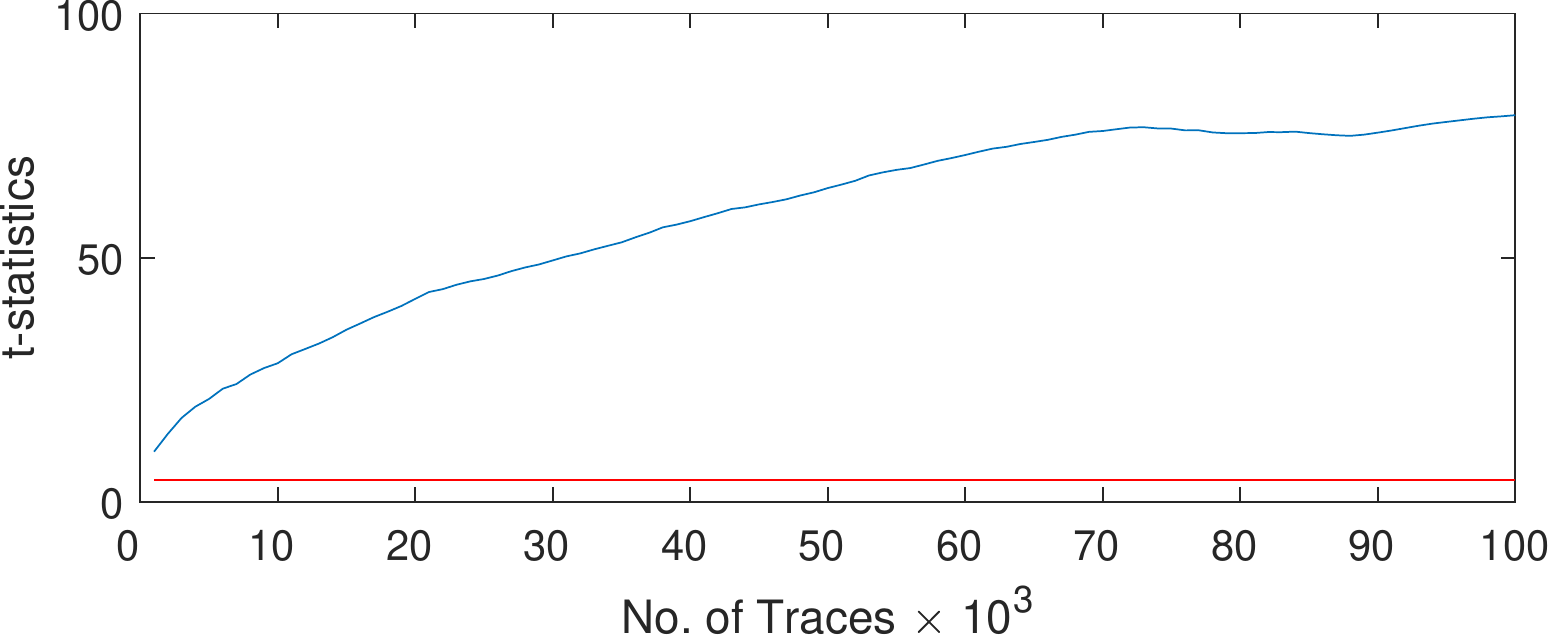}
	\label{fig:65nm_PRNG_OFF_1st_prog}}\\
\subfigure{
\includegraphics[width=0.45\textwidth]{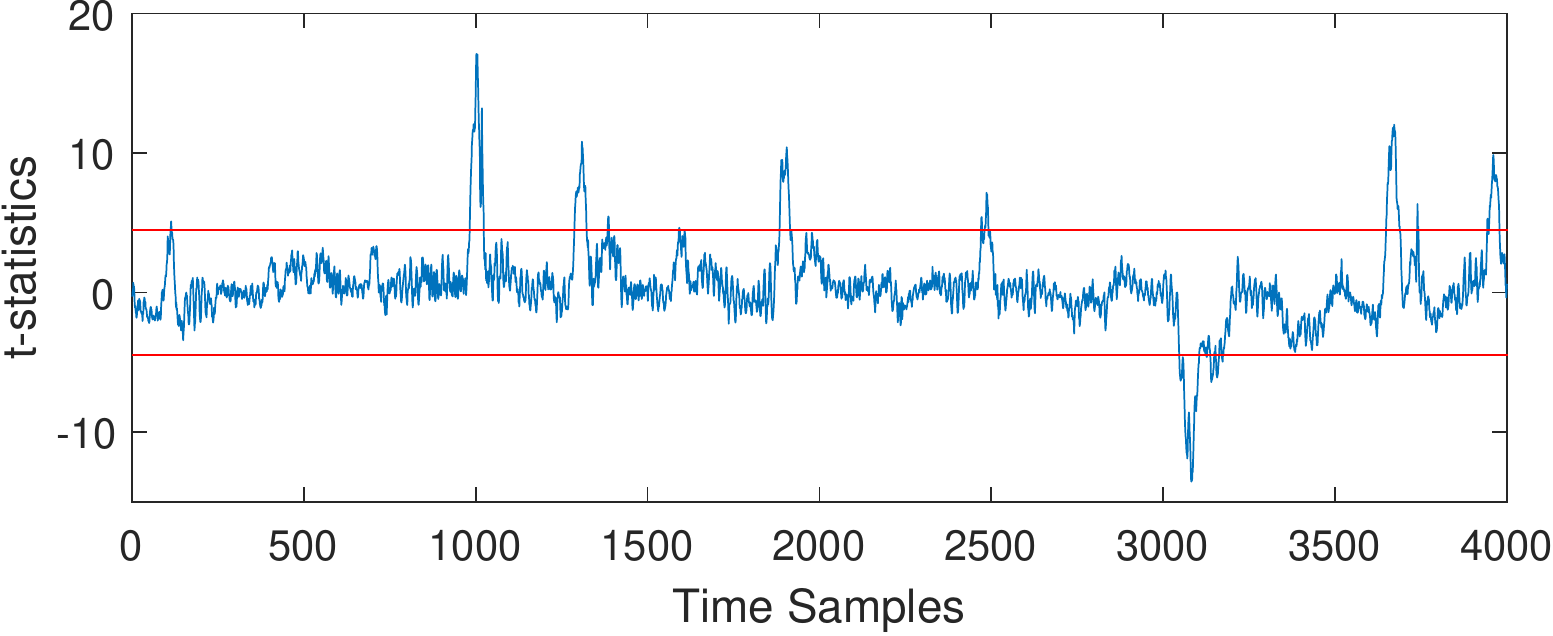}
\label{fig:65nm_PRNG_OFF_2nd}}\hfill
\subfigure{
	\includegraphics[width=0.45\textwidth]{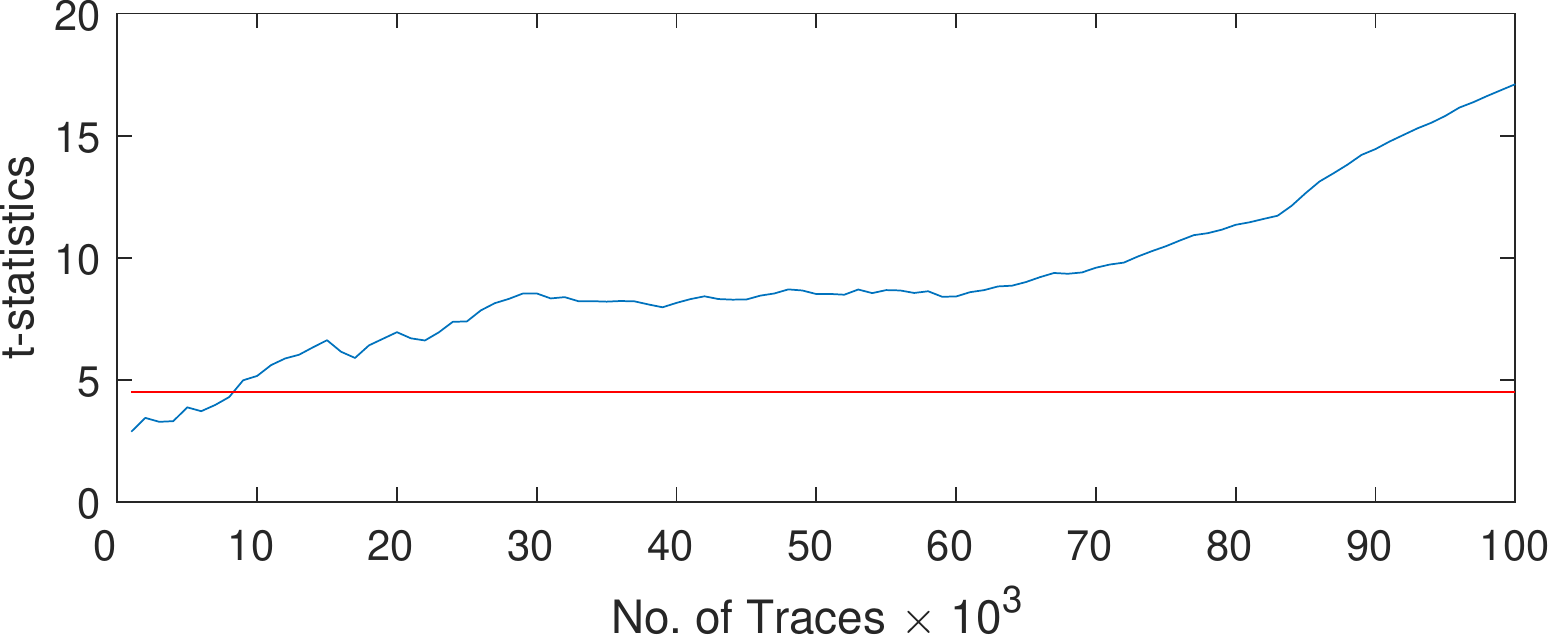}
	\label{fig:65nm_PRNG_OFF_2nd_prog}}\\
\subfigure{
\includegraphics[width=0.45\textwidth]{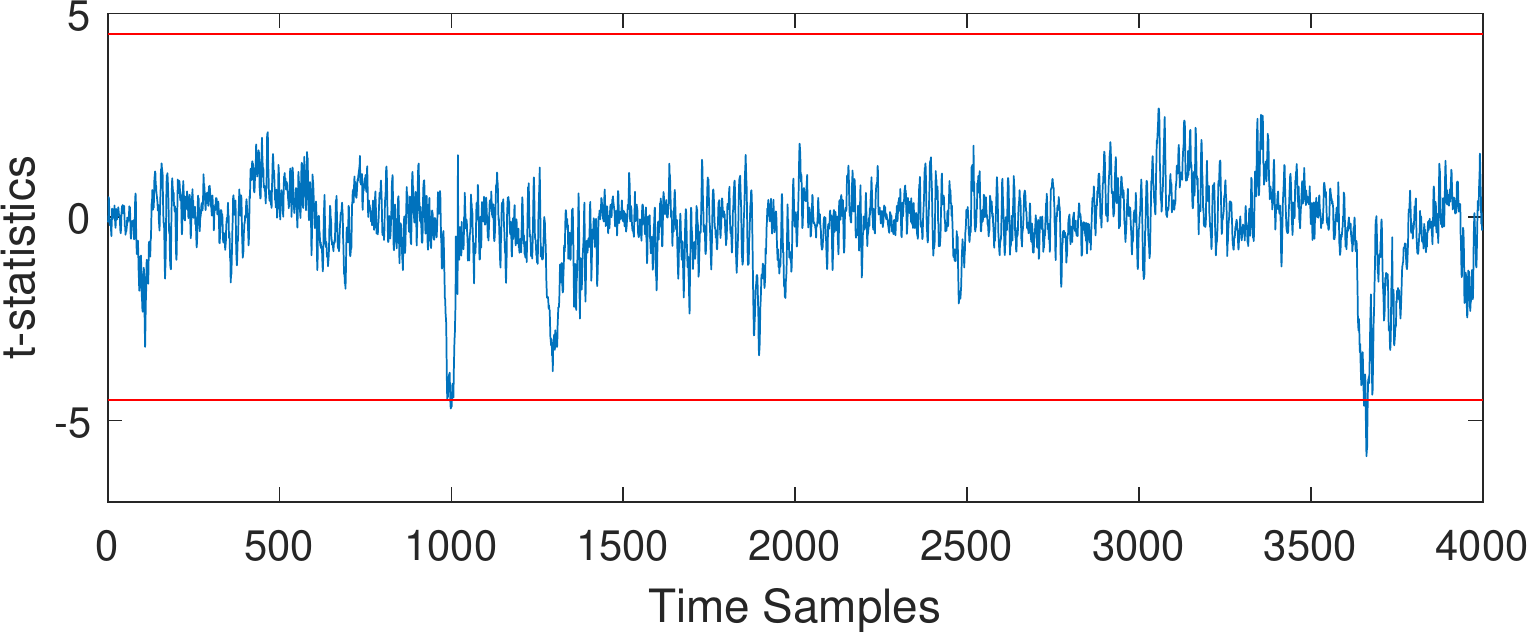}
\label{fig:65nm_PRNG_OFF_3rd}}\hfill
\subfigure{
\includegraphics[width=0.45\textwidth]{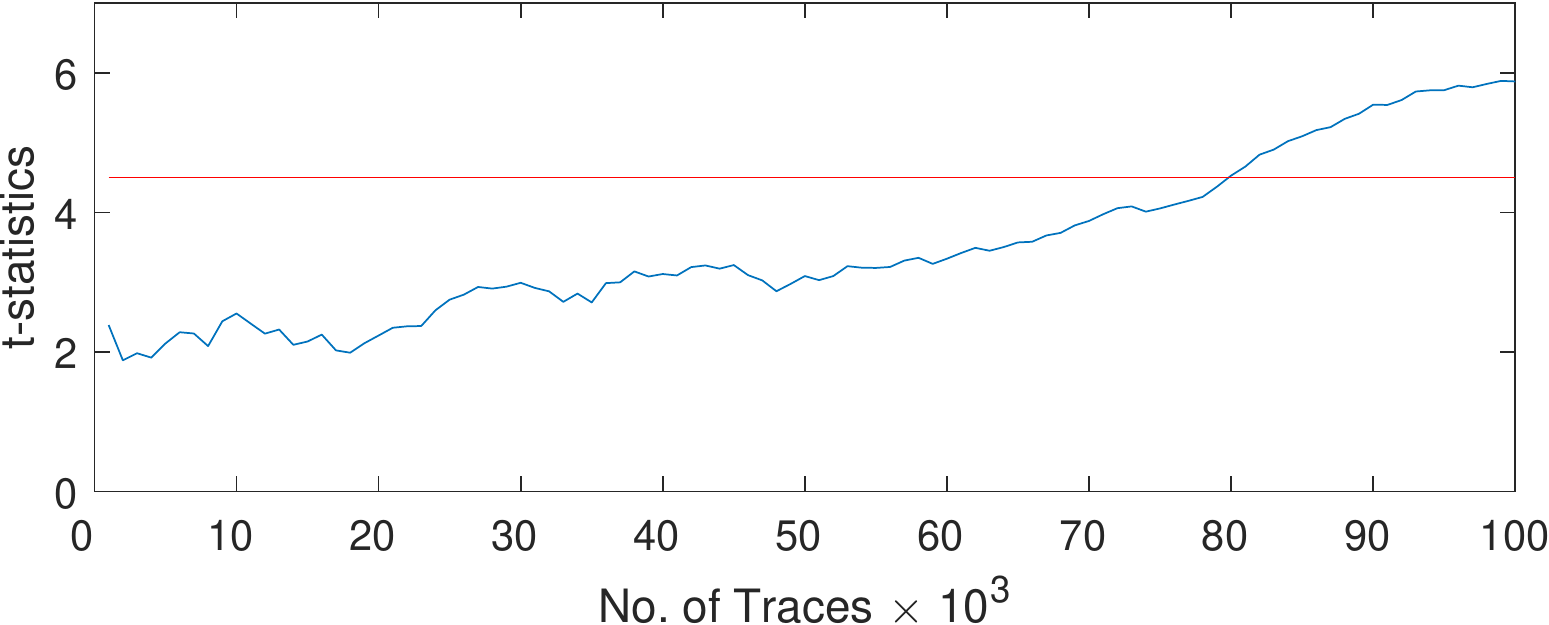}
\label{fig:65nm_PRNG_OFF_3rd_prog}}
\caption{65\,nm ASIC, PRNG off, clock frequency 25~MHz (Trojan not triggered), $t$-test results with 1 million traces (left), absolute maximum $t$-value over the number of traces (right).}
\label{fig:65nm_PRNG_OFF-eps-converted-to.pdf}
\end{figure*}

After we have seen that the Trojan indeed achieves what it has been designed for on the 90\,nm ASIC, we repeat the same kind of experiments on the 65\,nm chip. At first, the results after 1,000,000 traces with the deactivated Trojan (25 MHz) and the switched off PRNG can be seen in Figure~\ref{fig:65nm_PRNG_OFF-eps-converted-to.pdf}.

\begin{figure*}[htbp]
\centering
\vspace{-.3 in}
\subfigure{
\includegraphics[width=0.45\textwidth]{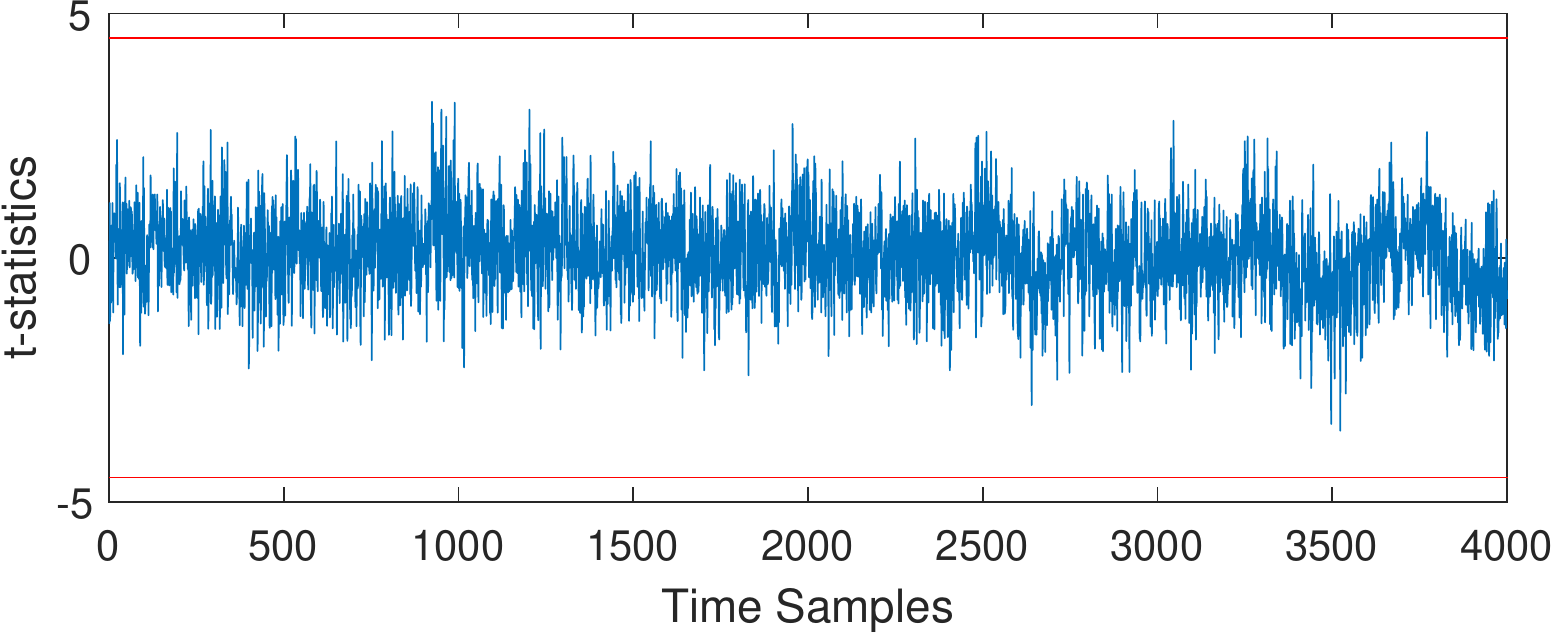}
\label{fig:65nm_PRNG_ON_1st}}\hfill
\subfigure{
	\includegraphics[width=0.45\textwidth]{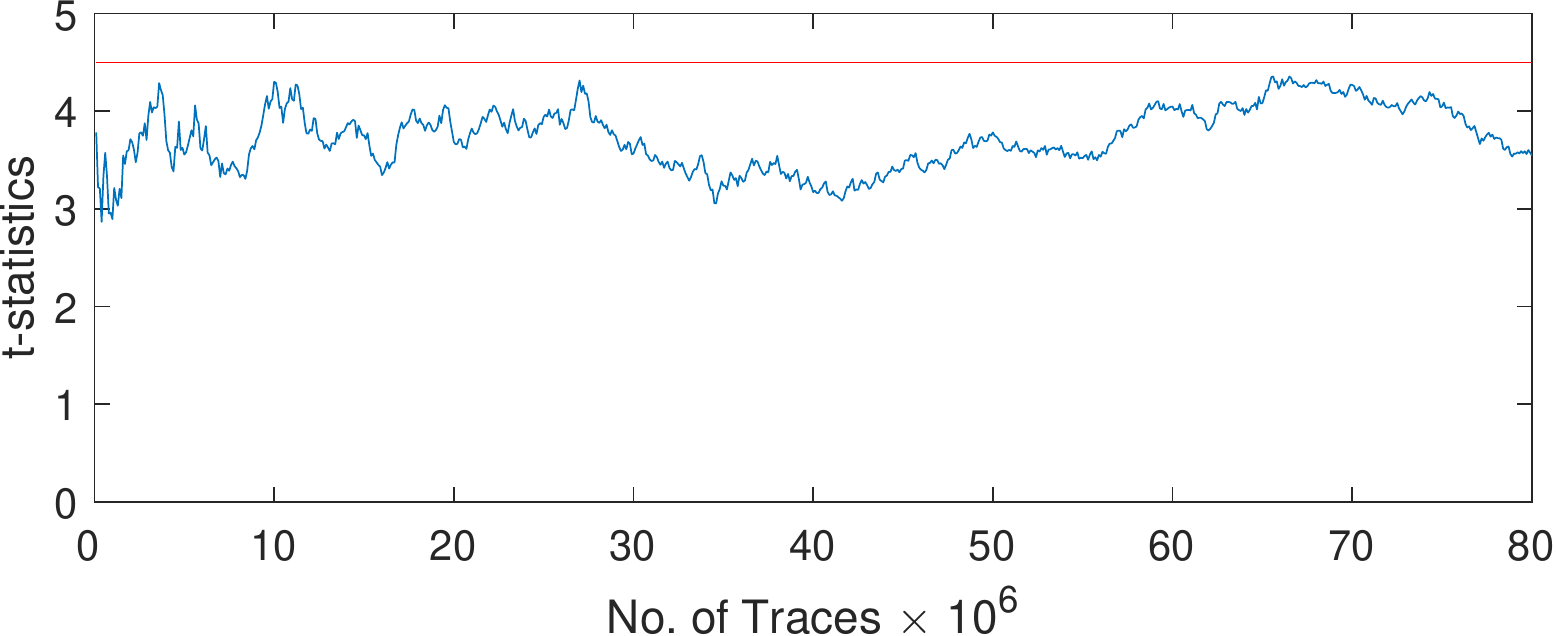}
	\label{fig:65nm_PRNG_ON_1st_prog}}\\
\subfigure{
\includegraphics[width=0.45\textwidth]{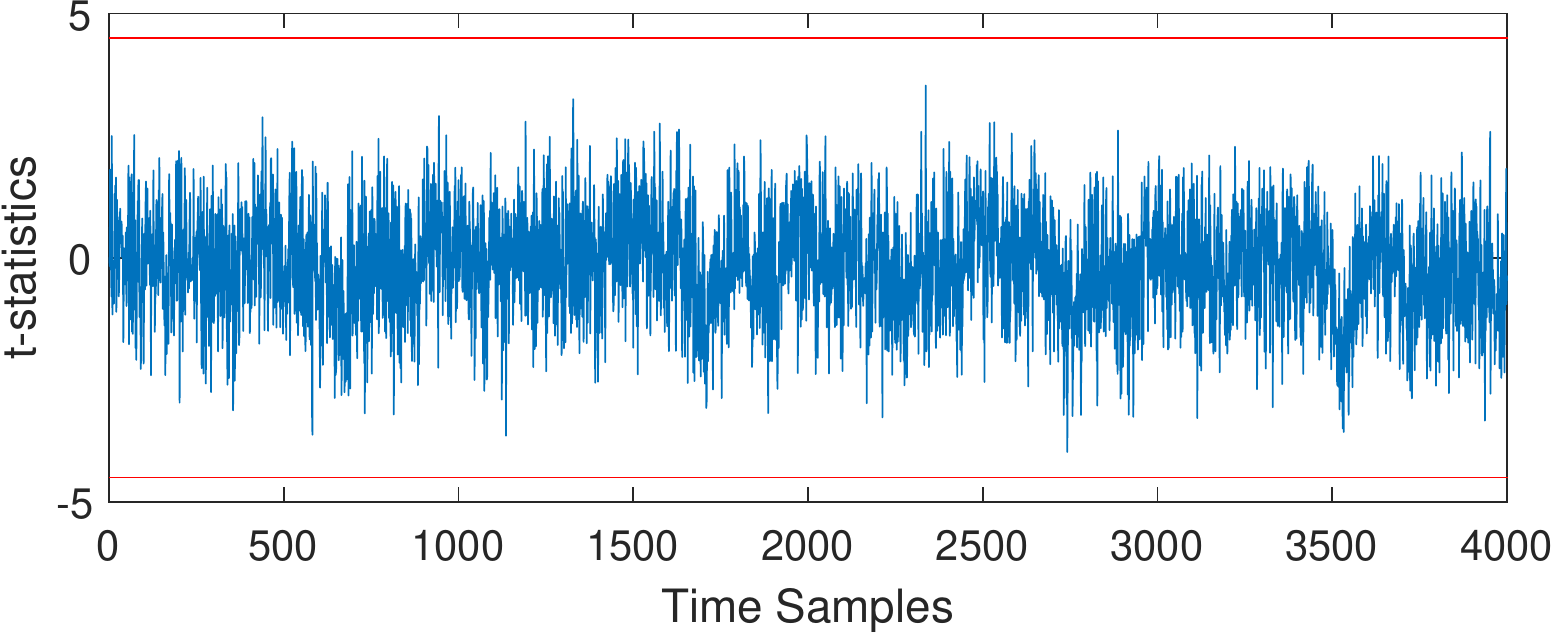}
\label{fig:65nm_PRNG_ON_2nd}}\hfill
\subfigure{
	\includegraphics[width=0.45\textwidth]{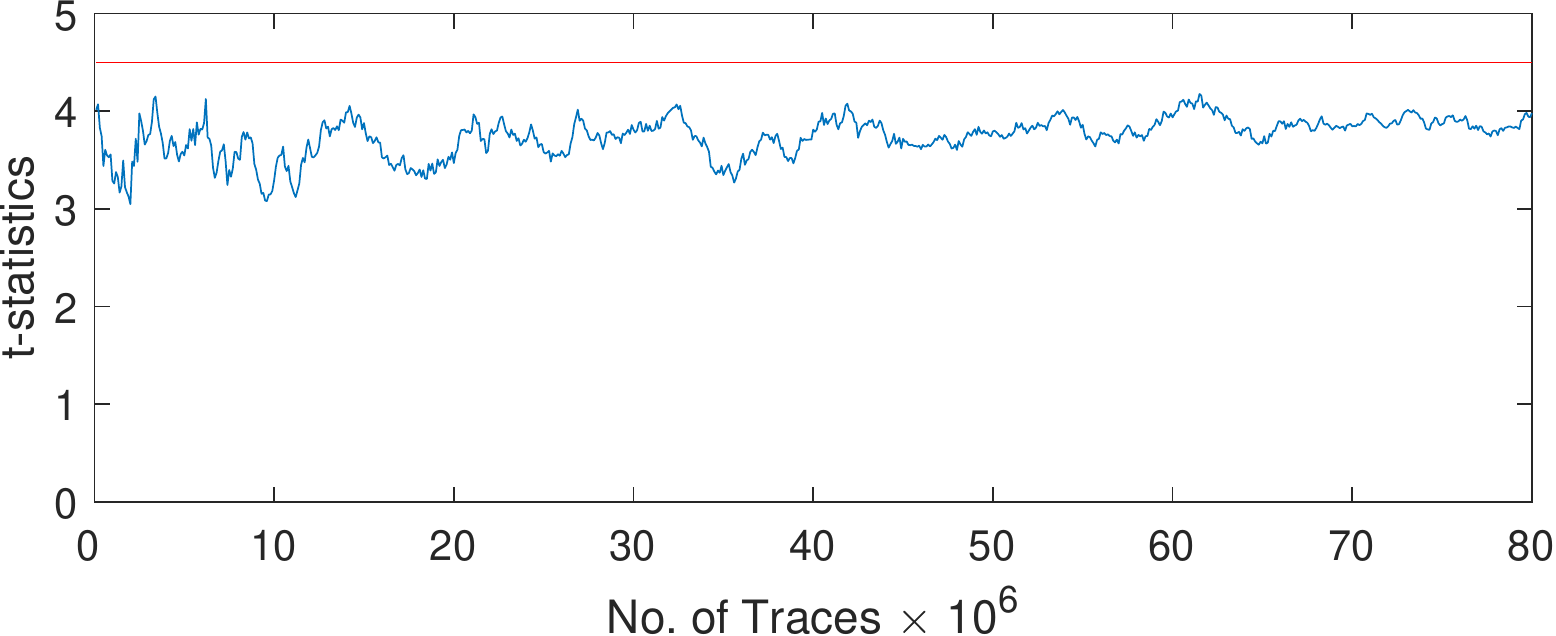}
	\label{fig:65nm_PRNG_ON_2nd_prog}}\\
\subfigure{
\includegraphics[width=0.45\textwidth]{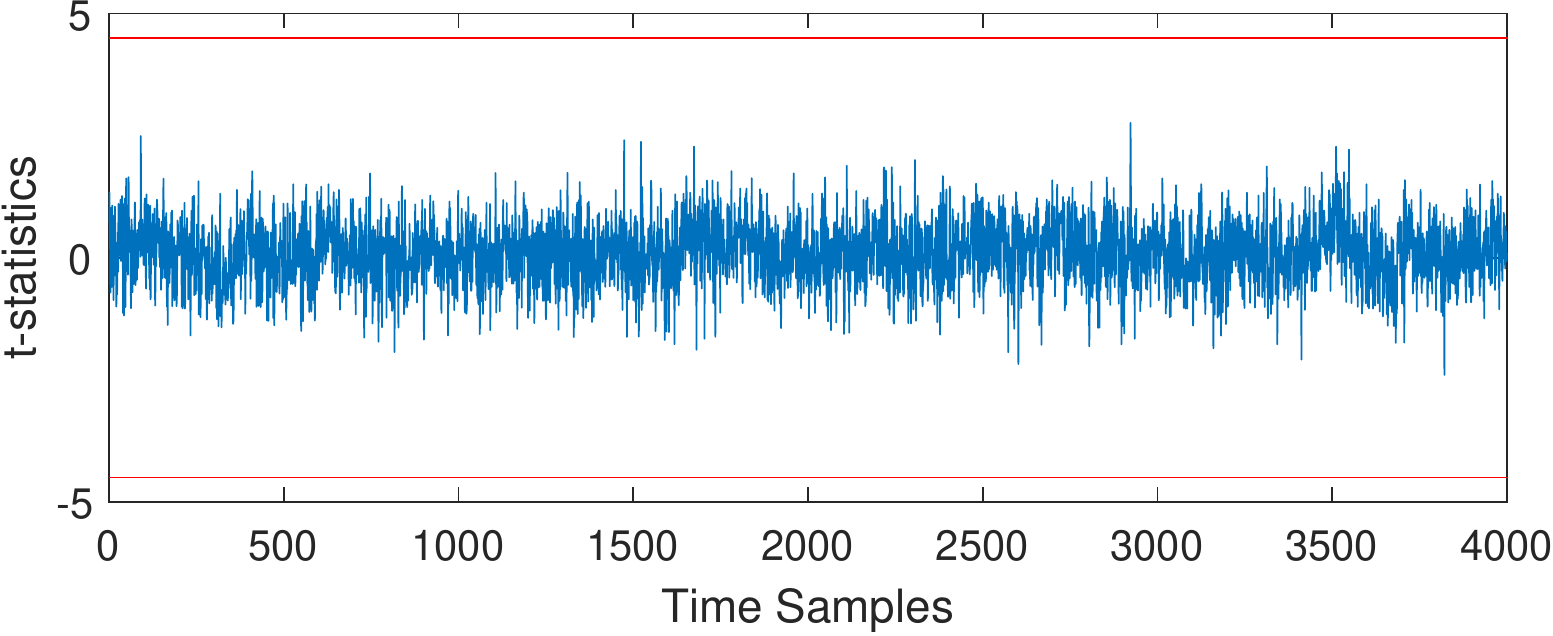}
\label{fig:65nm_PRNG_ON_3rd}}\hfill
\subfigure{
\includegraphics[width=0.45\textwidth]{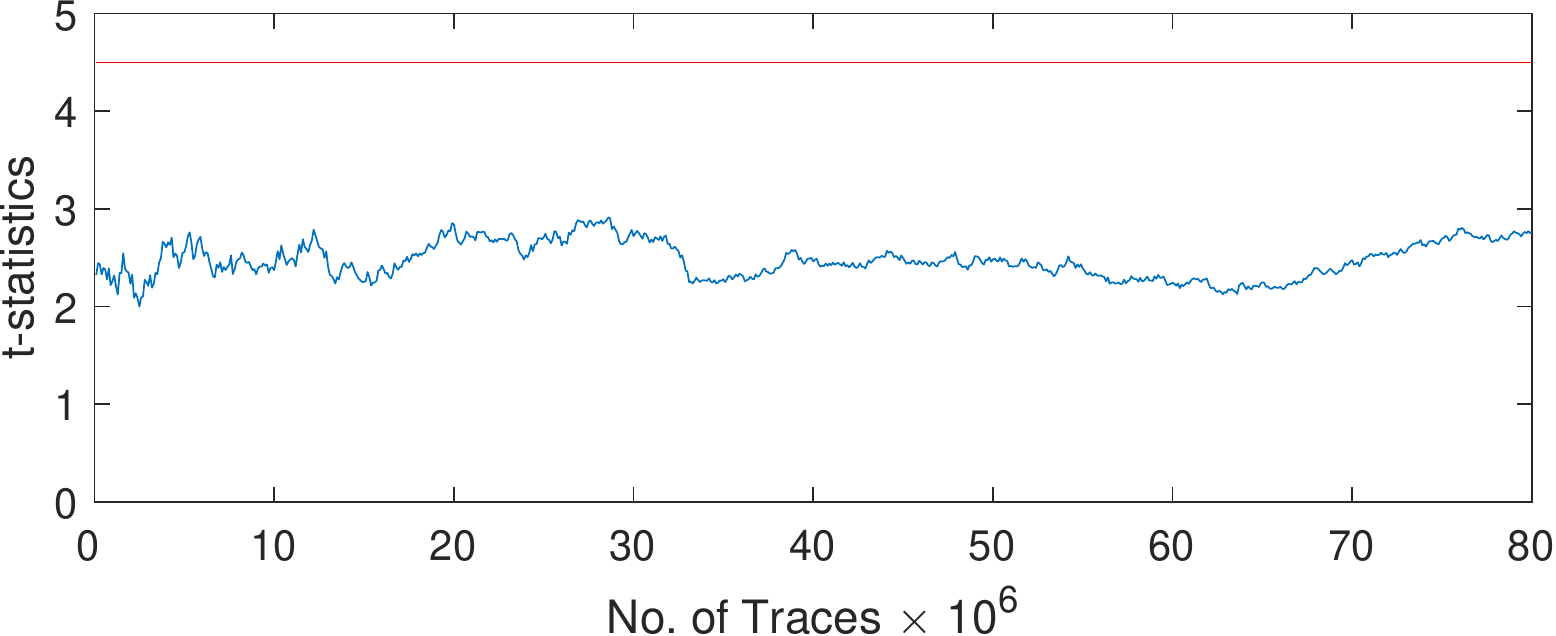}
\label{fig:65nm_PRNG_ON_3rd_prog}}
\caption{65\,nm ASIC, PRNG on, clock frequency 25~MHz (Trojan not triggered), $t$-test results with 80 million traces (left), absolute maximum $t$-value over the number of traces (right).}
\label{fig:65nm_PRNG_ON}
\end{figure*}

As before, detectable leakage is visible in all three statistical moments, but its magnitude is significantly smaller than on the 90\,nm ASIC, indicating a lower signal-to-noise ratio. Thus, for the next step with PRNG on we measured more traces than before, namely 80,000,000. The results in Figure~\ref{fig:65nm_PRNG_ON} show that with PRNG on and the Trojan not triggered at 25 MHz clock, there is no detectable leakage in any moment.

\begin{figure*}[htbp]
\centering
\vspace{-.3 in}
\subfigure{
\includegraphics[width=0.45\textwidth]{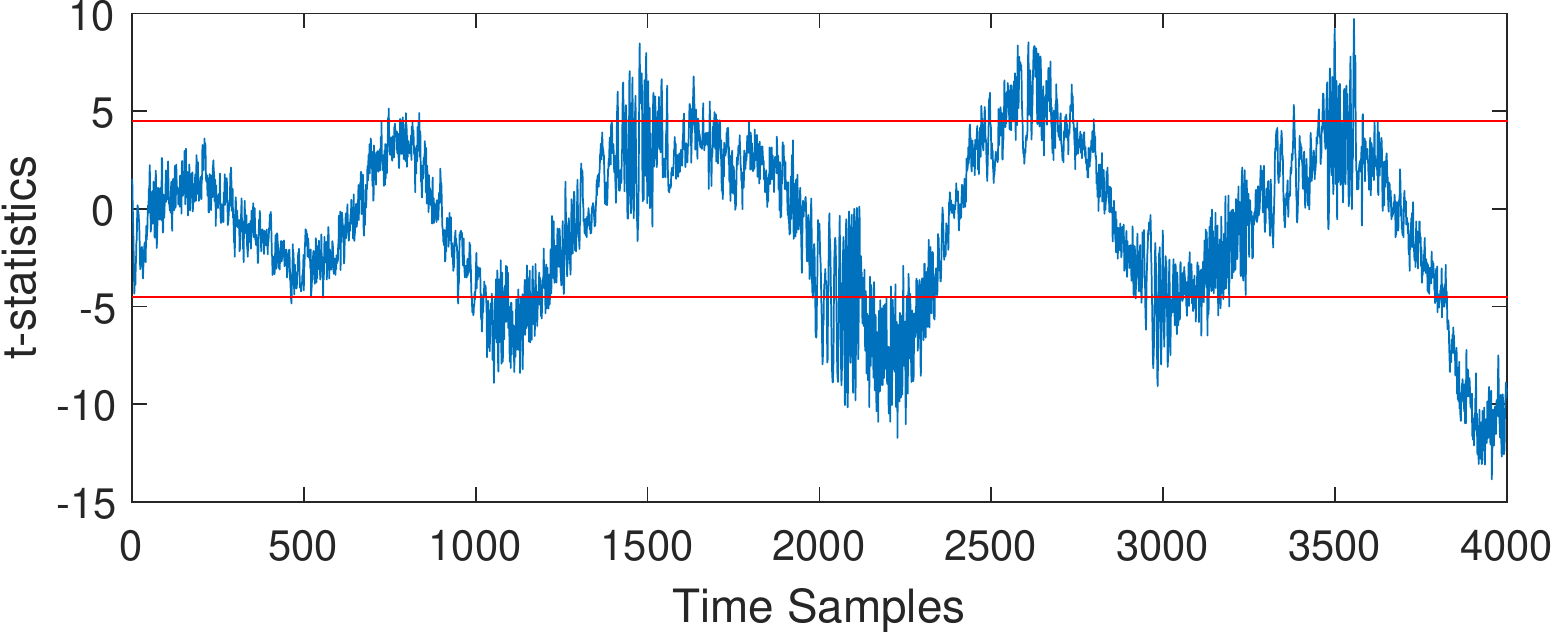}
\label{fig:65nm_PRNG_ON_Trojan_ON_1st}}\hfill
\subfigure{
	\includegraphics[width=0.45\textwidth]{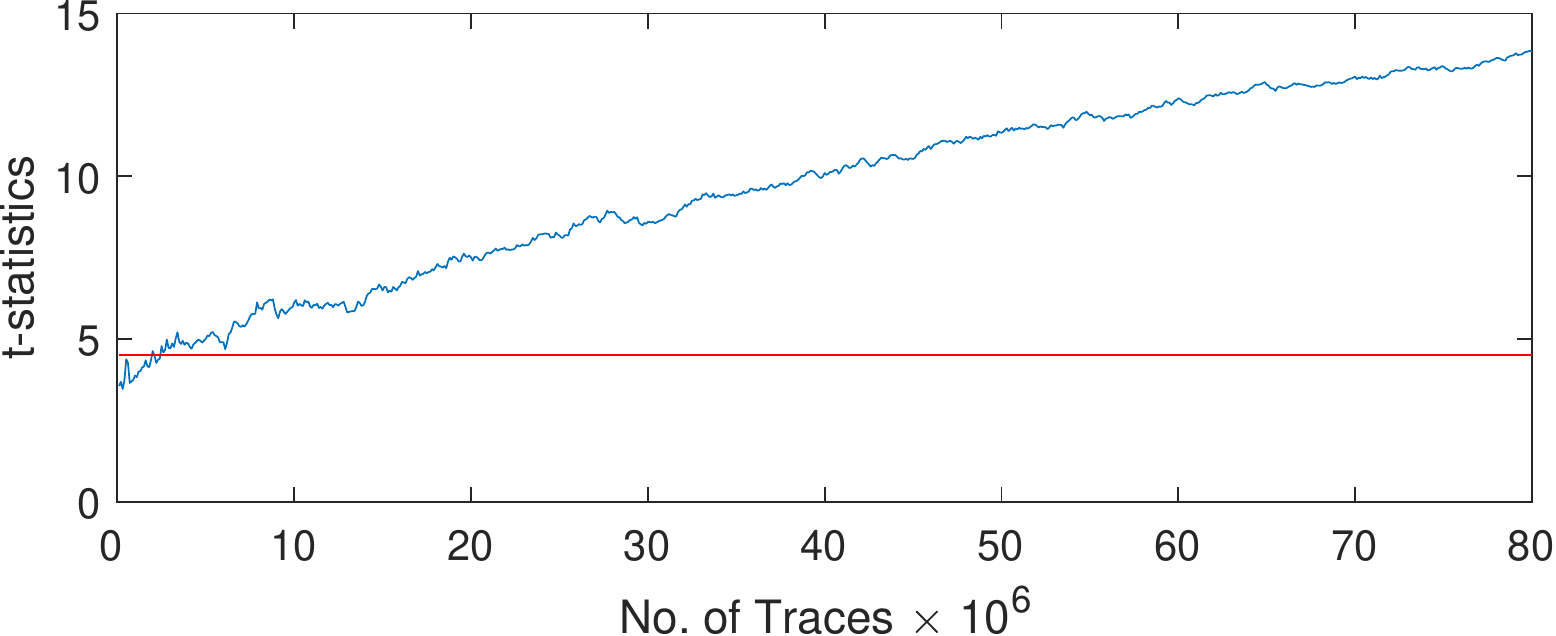}
	\label{fig:65nm_PRNG_ON_Trojan_ON_1st_prog}}\\
\subfigure{
\includegraphics[width=0.45\textwidth]{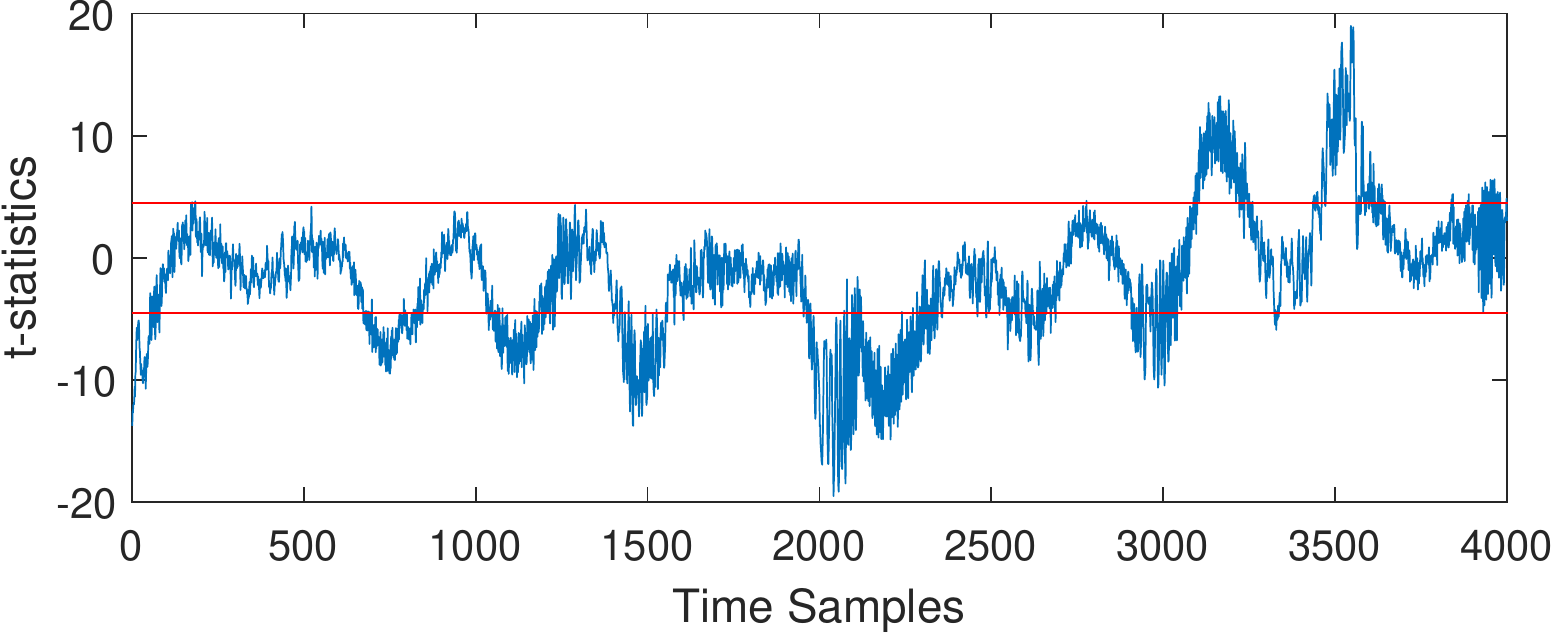}
\label{fig:65nm_PRNG_ON_Trojan_ON_2nd}}\hfill
\subfigure{
	\includegraphics[width=0.45\textwidth]{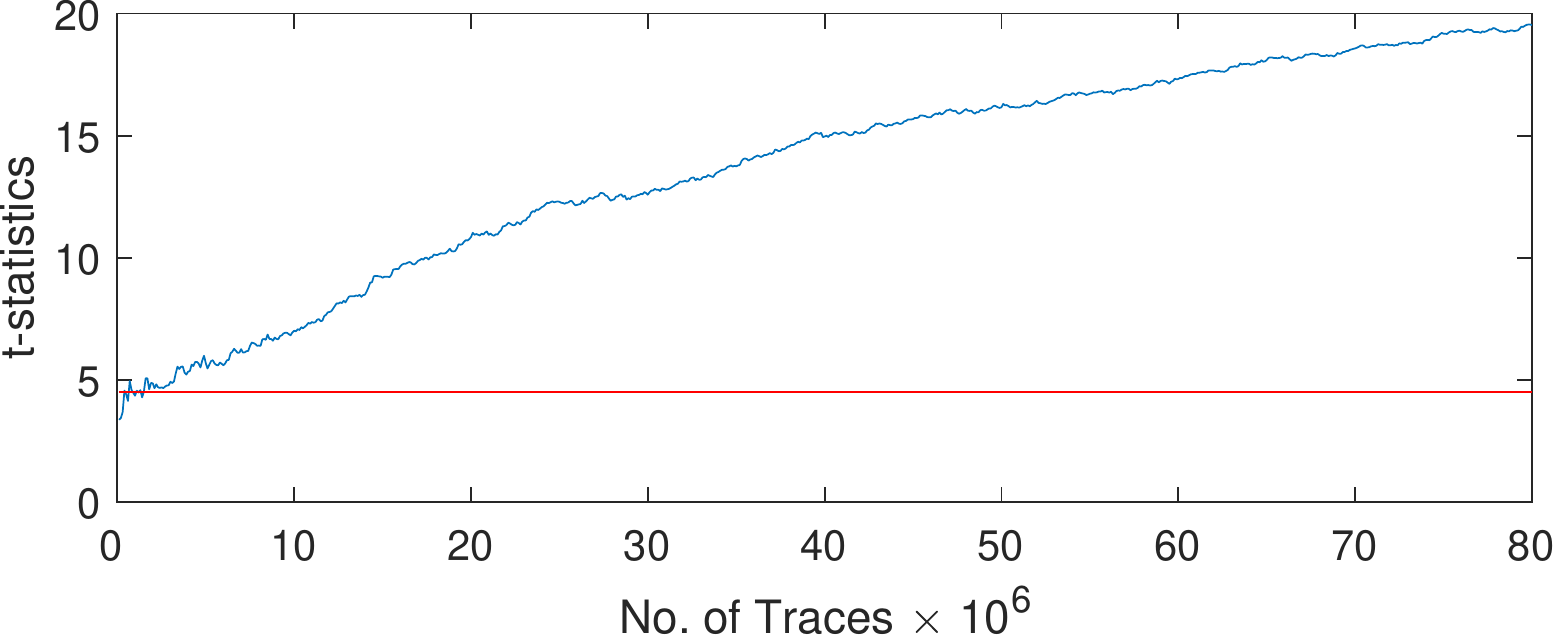}
	\label{fig:65nm_PRNG_ON_Trojan_ON_2nd_prog}}\\
\subfigure{
\includegraphics[width=0.45\textwidth]{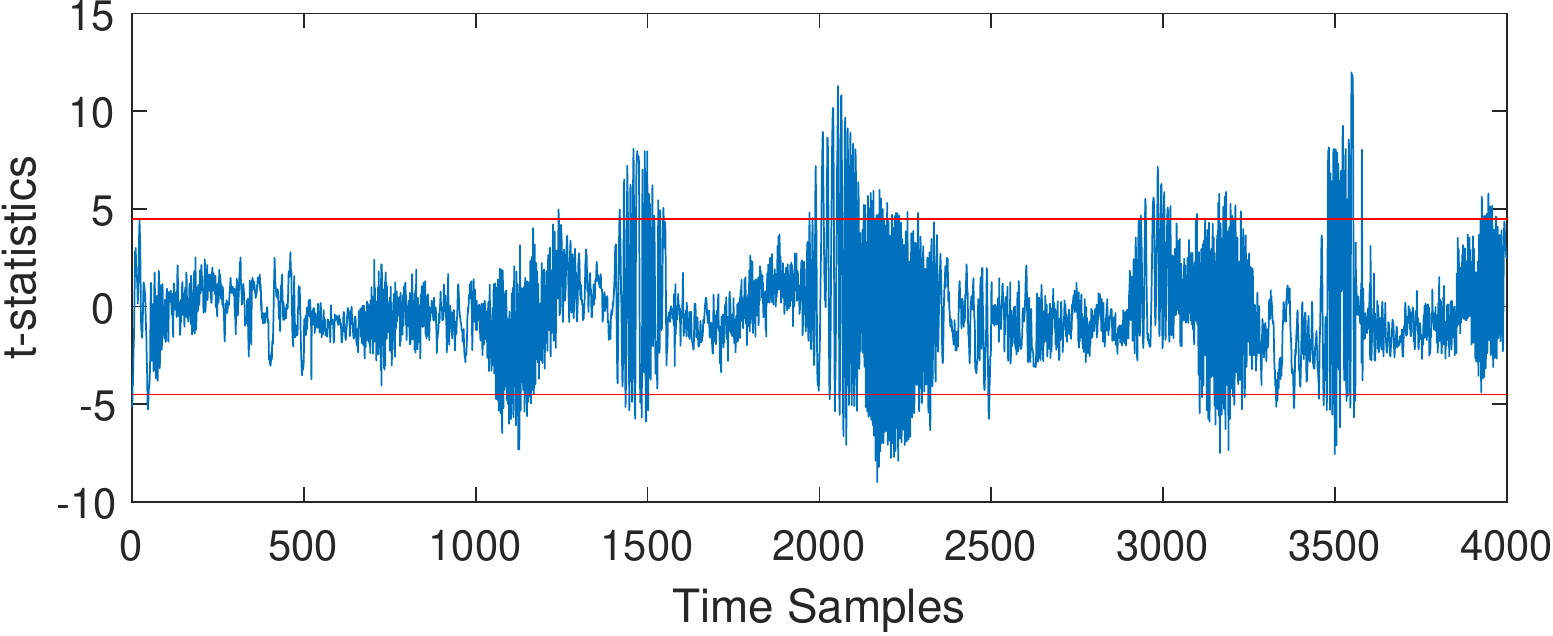}
\label{fig:65nm_PRNG_ON_Trojan_ON_3rd}}\hfill
\subfigure{
\includegraphics[width=0.45\textwidth]{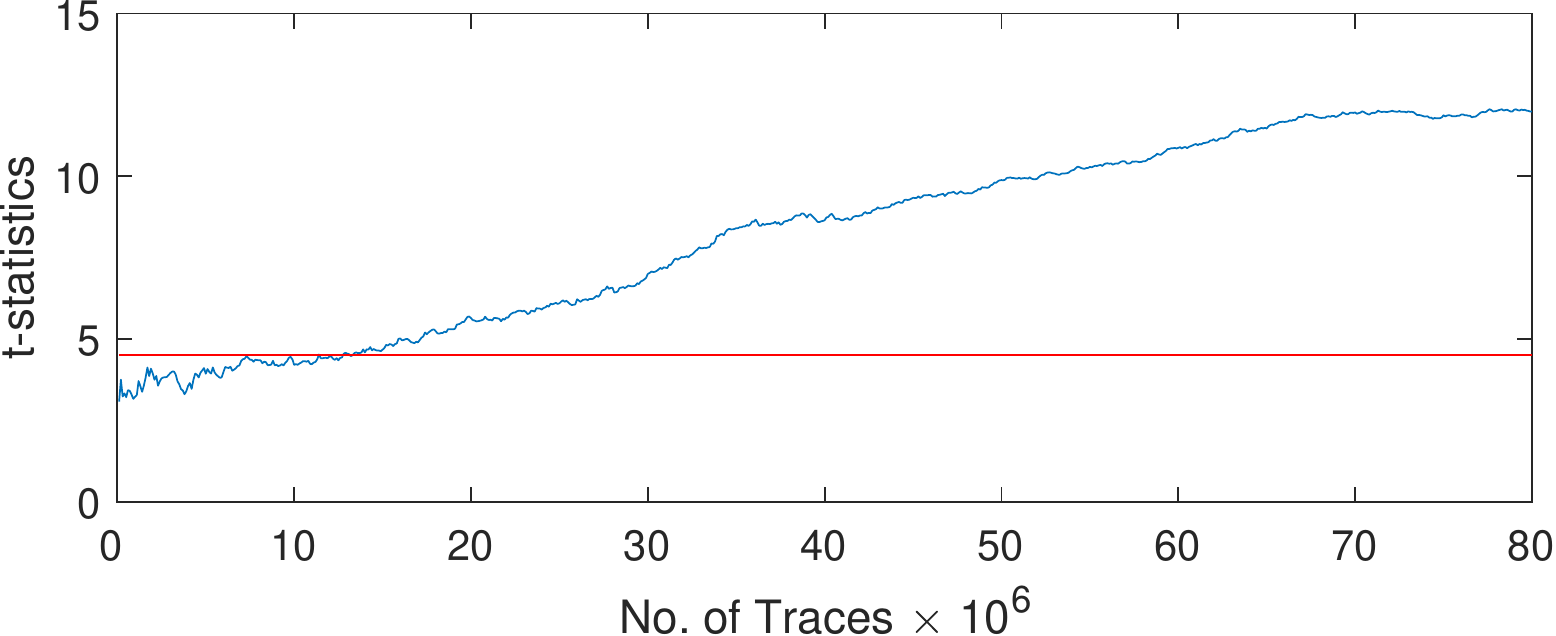}
\label{fig:65nm_PRNG_ON_Trojan_ON_3rd_prog}}
\caption{65\,nm ASIC, PRNG on, clock frequency 50~MHz (Trojan triggered), $t$-test results with 80 million traces (left), absolute maximum $t$-value over the number of traces (right).}
\label{fig:65nm_PRNG_ON_Trojan_ON}
\end{figure*}

When measuring at 50 MHz, however, i.e., triggering the Trojan, significant leakage can be detected in all moments, as apparent in Figure~\ref{fig:65nm_PRNG_ON_Trojan_ON}.

\begin{figure*}[h!]
\centering
\vspace{-.15 in}
\subfigure{
\includegraphics[width=0.99\columnwidth]{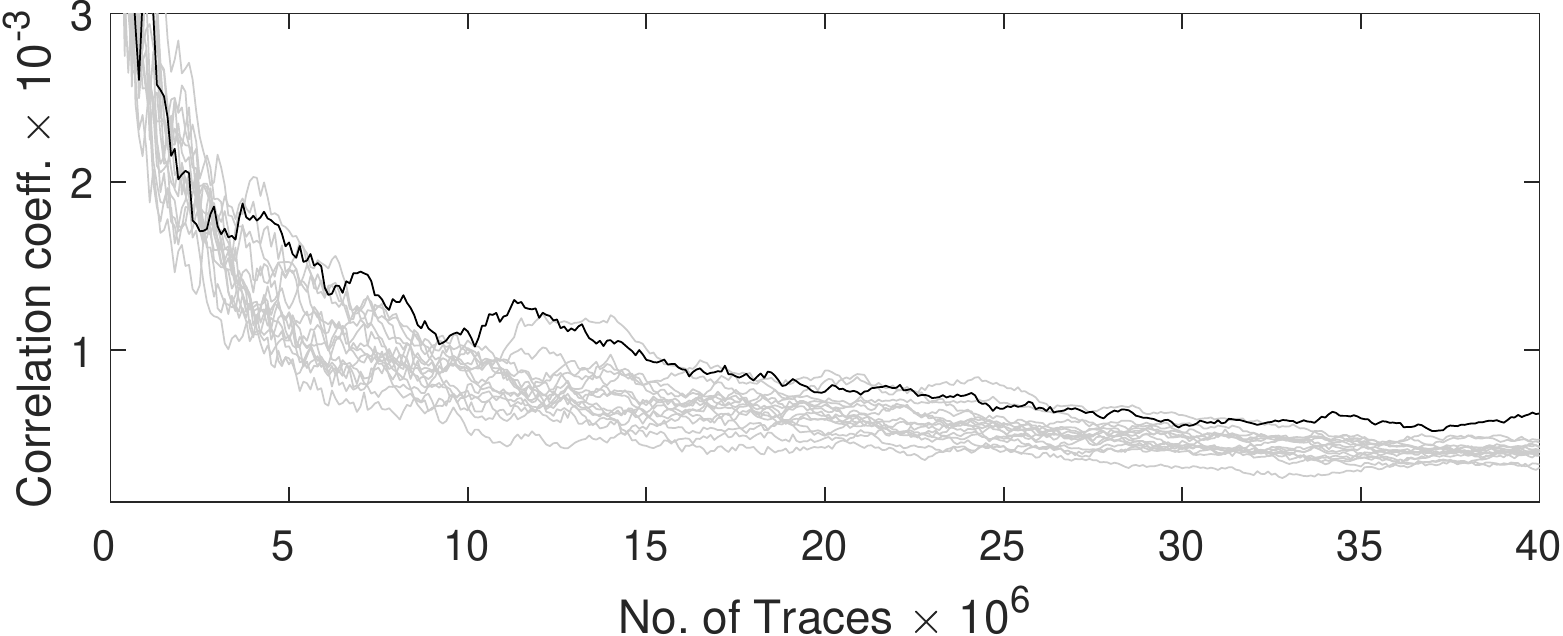}
\label{fig:65nm_CPA_prog}}\hfill
\subfigure{
\includegraphics[width=0.99\columnwidth]{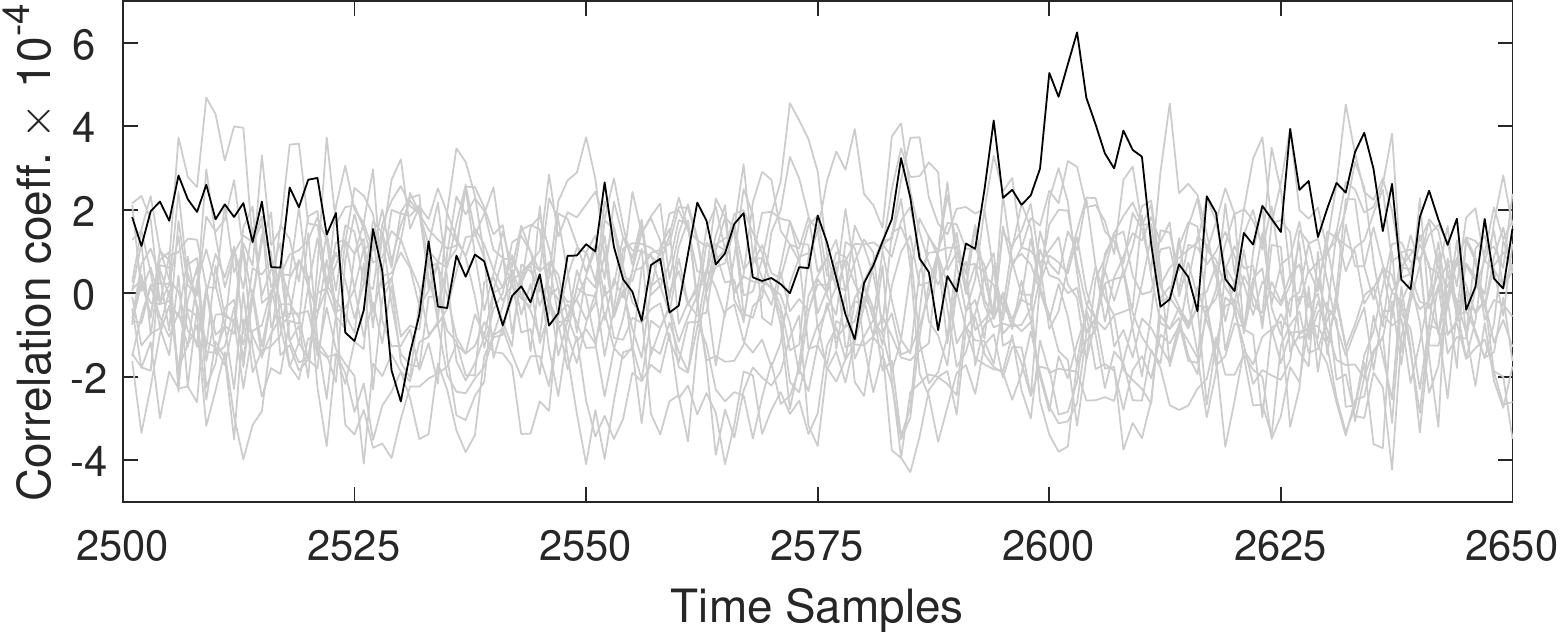}\label{fig:65nm_CPA_sample}}
\caption{65\,nm ASIC, PRNG on, clock frequency 50~MHz (trojan triggered), CPA results targeting a key nibble based on an S-Box output bit with 40 million traces (right), absolute maximum correlation coefficient over the number of traces (left).}
\label{fig:65nm_CPA}
\vspace{-0.2 in}
\end{figure*}

The successful CPA in~\ref{fig:65nm_CPA} targeting a key nibble based on an S-Box output bit using 40,000,000 traces confirms that the leakage is indeed exploitable.

%% file: conclude.tex
\section{Conclusions}
\label{sec:conclude}

We show how a parametric hardware Trojan with very low overhead can be inserted into SCA-resistant designs.
The presented Trojan is capable of being integrated into both ASIC and FPGA platforms.
This kind of parametric Trojan is very hard to be detected, since it bears the potential to be inserted without the addition or removal of any logic into or from the target design. Thus, even in a white-box scenario the Trojan remains stealthy and is unlikely to be detected by an evaluation lab.

The underlying concept is to lengthen certain paths of a combinatorial logic realizing a non-linear function under the foundations of threshold implementation, in such a way that by violating their delay (i.e., by running the device at a high frequency) the uniformity property of the utilized masking scheme is not fulfilled anymore. 
Our case study based on two ASIC prototypes, while admittedly suffering from some shortcomings, shows clearly that increasing the clock frequency triggers the malicious threshold implementation design to start leaking exploitable information through side channels.
Hence, the Trojan adversary can activate the Trojan and make use of the exploitable leakage, while the design can pass SCA evaluations when the Trojan is not triggered.
To the best of our knowledge, compared to the previous works in the areas of side-channel hardware Trojans, our construction is the only one which is applied on a first-order provably-secure SCA countermeasure, and is parametric with very low (or even no) overhead.

In fact, by decreasing the supply voltage the same effect can be seen. 
As a message of this paper, overclocking and -- at the same time -- power supply reduction should be internally monitored to avoid such an SCA-based Trojan being activated.
Related to this topic we should refer to~\cite{DBLP:journals/tvlsi/EndoLHSOFNKDA15}, where the difficulties of embedding a ``clock frequency monitor'' in presence of supply voltage changes are shown.

%% file: main.bbl
\begin{thebibliography}{10}
\providecommand{\url}[1]{#1}
\csname url@samestyle\endcsname
\providecommand{\newblock}{\relax}
\providecommand{\bibinfo}[2]{#2}
\providecommand{\BIBentrySTDinterwordspacing}{\spaceskip=0pt\relax}
\providecommand{\BIBentryALTinterwordstretchfactor}{4}
\providecommand{\BIBentryALTinterwordspacing}{\spaceskip=\fontdimen2\font plus
\BIBentryALTinterwordstretchfactor\fontdimen3\font minus
  \fontdimen4\font\relax}
\providecommand{\BIBforeignlanguage}[2]{{%
\expandafter\ifx\csname l@#1\endcsname\relax
\typeout{** WARNING: IEEEtran.bst: No hyphenation pattern has been}%
\typeout{** loaded for the language `#1'. Using the pattern for}%
\typeout{** the default language instead.}%
\else
\language=\csname l@#1\endcsname
\fi
#2}}
\providecommand{\BIBdecl}{\relax}
\BIBdecl

\bibitem{DBLP:conf/iccad/LinBP09}
L.~Lin, W.~Burleson, and C.~Paar, ``{MOLES: Malicious off-chip leakage enabled
  by side-channels},'' in \emph{ICCAD 2009}.\hskip 1em plus 0.5em minus
  0.4em\relax {ACM}, 2009, pp. 117--122.

\bibitem{DBLP:conf/ches/LinKGPB09}
L.~Lin, M.~Kasper, T.~G{\"{u}}neysu, C.~Paar, and W.~Burleson, ``{Trojan
  Side-Channels: Lightweight Hardware Trojans through Side-Channel
  Engineering},'' in \emph{CHES 2009}, ser. Lecture Notes in Computer Science,
  vol. 5747.\hskip 1em plus 0.5em minus 0.4em\relax Springer, 2009, pp.
  382--395.

\bibitem{DBLP:conf/crypto/Kocher96}
P.~C. Kocher, ``{Timing Attacks on Implementations of Diffie-Hellman, RSA, DSS,
  and Other Systems},'' in \emph{CRYPTO 1996}, ser. Lecture Notes in Computer
  Science, vol. 1109.\hskip 1em plus 0.5em minus 0.4em\relax Springer, 1996,
  pp. 104--113.

\bibitem{dpa_kocher}
P.~C. Kocher, J.~Jaffe, and B.~Jun, ``{Differential Power Analysis},'' in
  \emph{CRYPTO 1999}, ser. Lecture Notes in Computer Science, vol. 1666.\hskip
  1em plus 0.5em minus 0.4em\relax Springer, 1999, pp. 388--397.

\bibitem{DBLP:journals/jce/KasperMBMGPB12}
M.~Kasper, A.~Moradi, G.~T. Becker, O.~Mischke, T.~G{\"{u}}neysu, C.~Paar, and
  W.~Burleson, ``{Side channels as building blocks},'' \emph{J. Cryptographic
  Engineering}, vol.~2, no.~3, pp. 143--159, 2012.

\bibitem{DBLP:conf/ches/BeckerRPB13}
G.~T. Becker, F.~Regazzoni, C.~Paar, and W.~P. Burleson, ``{Stealthy
  Dopant-Level Hardware Trojans},'' in \emph{CHES 2013}, ser. Lecture Notes in
  Computer Science, vol. 8086.\hskip 1em plus 0.5em minus 0.4em\relax Springer,
  2013, pp. 197--214.

\bibitem{DBLP:conf/ches/PoppKZM07}
T.~Popp, M.~Kirschbaum, T.~Zefferer, and S.~Mangard, ``{Evaluation of the
  Masked Logic Style MDPL on a Prototype Chip},'' in \emph{CHES 2007}, ser.
  Lecture Notes in Computer Science, vol. 4727.\hskip 1em plus 0.5em minus
  0.4em\relax Springer, 2007, pp. 81--94.

\bibitem{DBLP:journals/tvlsi/MoradiKEP12}
A.~Moradi, M.~Kirschbaum, T.~Eisenbarth, and C.~Paar, ``{Masked Dual-Rail
  Precharge Logic Encounters State-of-the-Art Power Analysis Methods},''
  \emph{{IEEE} Trans. {VLSI} Syst.}, vol.~20, no.~9, pp. 1578--1589, 2012.

\bibitem{yang2016a2}
K.~Yang, M.~Hicks, Q.~Dong, T.~Austin, and D.~Sylvester, ``A2: Analog malicious
  hardware,'' in \emph{Security and Privacy (SP), 2016 IEEE Symposium
  on}.\hskip 1em plus 0.5em minus 0.4em\relax IEEE, 2016, pp. 18--37.

\bibitem{hou2018r2d2}
Y.~Hou, H.~He, K.~Shamsi, Y.~Jin, D.~Wu, and H.~Wu, ``R2d2: Runtime reassurance
  and detection of a2 trojan,'' in \emph{2018 IEEE International Symposium on
  Hardware Oriented Security and Trust (HOST)}.\hskip 1em plus 0.5em minus
  0.4em\relax IEEE, 2018, pp. 195--200.

\bibitem{DBLP:conf/ches/BogdanovKLPPRSV07}
A.~Bogdanov, L.~R. Knudsen, G.~Leander, C.~Paar, A.~Poschmann, M.~J.~B.
  Robshaw, Y.~Seurin, and C.~Vikkelsoe, ``{PRESENT: An Ultra-Lightweight Block
  Cipher},'' in \emph{CHES 2007}, ser. Lecture Notes in Computer Science, vol.
  4727.\hskip 1em plus 0.5em minus 0.4em\relax Springer, 2007, pp. 450--466.

\bibitem{DBLP:conf/hldvt/ChakrabortyNB09}
R.~S. Chakraborty, S.~Narasimhan, and S.~Bhunia, ``{Hardware Trojan: Threats
  and emerging solutions},'' in \emph{HLDVT 2009}.\hskip 1em plus 0.5em minus
  0.4em\relax {IEEE} Computer Society, 2009, pp. 166--171.

\bibitem{DBLP:conf/host/JinM08}
Y.~Jin and Y.~Makris, ``{Hardware Trojan Detection Using Path Delay
  Fingerprint},'' in \emph{HOST 2008}.\hskip 1em plus 0.5em minus 0.4em\relax
  {IEEE} Computer Society, 2008, pp. 51--57.

\bibitem{DBLP:conf/dft/WangSTP08}
X.~Wang, H.~Salmani, M.~Tehranipoor, and J.~F. Plusquellic, ``{Hardware Trojan
  Detection and Isolation Using Current Integration and Localized Current
  Analysis},'' in \emph{DFT 2008}.\hskip 1em plus 0.5em minus 0.4em\relax
  {IEEE} Computer Society, 2008, pp. 87--95.

\bibitem{DBLP:conf/ches/GhandaliBHP16}
S.~Ghandali, G.~T. Becker, D.~Holcomb, and C.~Paar, ``{A Design Methodology for
  Stealthy Parametric Trojans and Its Application to Bug Attacks},'' in
  \emph{CHES 2016}, ser. Lecture Notes in Computer Science, vol. 9813.\hskip
  1em plus 0.5em minus 0.4em\relax Springer, 2016, pp. 625--647.

\bibitem{DBLP:conf/crypto/BihamCS08}
E.~Biham, Y.~Carmeli, and A.~Shamir, ``{Bug Attacks},'' in \emph{CRYPTO 2008},
  ser. Lecture Notes in Computer Science, vol. 5157.\hskip 1em plus 0.5em minus
  0.4em\relax Springer, 2008, pp. 221--240.

\bibitem{biham2016bug}
------, ``{Bug attacks},'' \emph{Journal of Cryptology}, vol.~29, no.~4, pp.
  775--805, 2016.

\bibitem{DBLP:conf/fse/OswaldMPR05}
E.~Oswald, S.~Mangard, N.~Pramstaller, and V.~Rijmen, ``{A Side-Channel
  Analysis Resistant Description of the AES S-Box},'' in \emph{FSE 2005}, ser.
  Lecture Notes in Computer Science, vol. 3557.\hskip 1em plus 0.5em minus
  0.4em\relax Springer, 2005, pp. 413--423.

\bibitem{DBLP:conf/ches/MangardPO05}
S.~Mangard, N.~Pramstaller, and E.~Oswald, ``{Successfully Attacking Masked AES
  Hardware Implementations},'' in \emph{CHES 2005}, ser. Lecture Notes in
  Computer Science, vol. 3659.\hskip 1em plus 0.5em minus 0.4em\relax Springer,
  2005, pp. 157--171.

\bibitem{DBLP:conf/acns/CanrightB08}
D.~Canright and L.~Batina, ``{A Very Compact "Perfectly Masked" S-Box for
  AES},'' in \emph{ACNS 2008}, ser. Lecture Notes in Computer Science, vol.
  5037, 2008, pp. 446--459.

\bibitem{DBLP:conf/ches/MoradiME10}
A.~Moradi, O.~Mischke, and T.~Eisenbarth, ``{Correlation-Enhanced Power
  Analysis Collision Attack},'' in \emph{CHES 2010}, ser. Lecture Notes in
  Computer Science, vol. 6225.\hskip 1em plus 0.5em minus 0.4em\relax Springer,
  2010, pp. 125--139.

\bibitem{DBLP:journals/joc/NikovaRS11}
S.~Nikova, V.~Rijmen, and M.~Schl{\"{a}}ffer, ``{Secure Hardware Implementation
  of Nonlinear Functions in the Presence of Glitches},'' \emph{J. Cryptology},
  vol.~24, no.~2, pp. 292--321, 2011.

\bibitem{CarletDGM12}
C.~Carlet, J.~Danger, S.~Guilley, and H.~Maghrebi, ``{Leakage Squeezing of
  Order Two},'' in \emph{INDOCRYPT 2012}, ser. Lecture Notes in Computer
  Science, vol. 7668.\hskip 1em plus 0.5em minus 0.4em\relax Springer, 2012,
  pp. 120--139.

\bibitem{MaghrebiGD11}
H.~Maghrebi, S.~Guilley, and J.~Danger, ``{Leakage Squeezing Countermeasure
  against High-Order Attacks},'' in \emph{WISTP 2011}, ser. Lecture Notes in
  Computer Science, vol. 6633.\hskip 1em plus 0.5em minus 0.4em\relax Springer,
  2011, pp. 208--223.

\bibitem{DBLP:journals/ccds/BilginNNRTV15}
B.~Bilgin, S.~Nikova, V.~Nikov, V.~Rijmen, N.~Tokareva, and V.~Vitkup,
  ``{Threshold Implementations of Small S-boxes},'' \emph{Cryptography and
  Communications}, vol.~7, no.~1, pp. 3--33, 2015.

\bibitem{DBLP:journals/joc/PoschmannMKLWL11}
A.~Poschmann, A.~Moradi, K.~Khoo, C.~Lim, H.~Wang, and S.~Ling, ``{Side-Channel
  Resistant Crypto for Less than 2, 300 {GE}},'' \emph{J. Cryptology}, vol.~24,
  no.~2, pp. 322--345, 2011.

\bibitem{DBLP:conf/asiacrypt/BilginGNNR14}
B.~Bilgin, B.~Gierlichs, S.~Nikova, V.~Nikov, and V.~Rijmen, ``{Higher-Order
  Threshold Implementations},'' in \emph{ASIACRYPT 2014}, ser. Lecture Notes in
  Computer Science, vol. 8874.\hskip 1em plus 0.5em minus 0.4em\relax Springer,
  2014, pp. 326--343.

\bibitem{DBLP:BeyneB16}
T.~Beyne and B.~Bilgin, ``{Uniform First-Order Threshold Implementations},'' in
  \emph{{SAC} 2016}, ser. Lecture Notes in Computer Science, vol. 10532.\hskip
  1em plus 0.5em minus 0.4em\relax Springer, 2017, pp. 79--98.

\bibitem{DBLP:conf/africacrypt/BilginGNNR14}
B.~Bilgin, B.~Gierlichs, S.~Nikova, V.~Nikov, and V.~Rijmen, ``{A More
  Efficient AES Threshold Implementation},'' in \emph{AFRICACRYPT 2014}, ser.
  Lecture Notes in Computer Science, vol. 8469.\hskip 1em plus 0.5em minus
  0.4em\relax Springer, 2014, pp. 267--284.

\bibitem{DBLP:conf/eurocrypt/MoradiPLPW11}
A.~Moradi, A.~Poschmann, S.~Ling, C.~Paar, and H.~Wang, ``{Pushing the Limits:
  {A} Very Compact and a Threshold Implementation of {AES}},'' in
  \emph{EUROCRYPT 2011}, vol. 6632.\hskip 1em plus 0.5em minus 0.4em\relax
  Springer, 2011, pp. 69--88.

\bibitem{DBLP:conf/crypto/ReparazBNGV15}
O.~Reparaz, B.~Bilgin, S.~Nikova, B.~Gierlichs, and I.~Verbauwhede,
  ``{Consolidating Masking Schemes},'' in \emph{CRYPTO 2015}, ser. Lecture
  Notes in Computer Science, vol. 9215.\hskip 1em plus 0.5em minus 0.4em\relax
  Springer, 2015, pp. 764--783.

\bibitem{DBLP:conf/ctrsa/GrossMK17}
H.~Gross, S.~Mangard, and T.~Korak, ``{An Efficient Side-Channel Protected
  {AES} Implementation with Arbitrary Protection Order},'' in \emph{CT-RSA
  2017}, ser. Lecture Notes in Computer Science, vol. 10159.\hskip 1em plus
  0.5em minus 0.4em\relax Springer, 2017, pp. 95--112.

\bibitem{DBLP:conf/eurocrypt/BiryukovCBP03}
A.~Biryukov, C.~D. Canni{\`{e}}re, A.~Braeken, and B.~Preneel, ``{A Toolbox for
  Cryptanalysis: Linear and Affine Equivalence Algorithms},'' in
  \emph{EUROCRYPT 2003}, ser. Lecture Notes in Computer Science, vol.
  2656.\hskip 1em plus 0.5em minus 0.4em\relax Springer, 2003, pp. 33--50.

\bibitem{DBLP:journals/tosc/BozilovBS17}
D.~Bozilov, B.~Bilgin, and H.~A. Sahin, ``{A Note on 5-bit Quadratic
  Permutations' Classification},'' \emph{{IACR} Trans. Symmetric Cryptol.},
  vol. 2017, no.~1, pp. 398--404, 2017.

\bibitem{DBLP:conf/asiacrypt/0001S16}
A.~Moradi and T.~Schneider, ``{Side-Channel Analysis Protection and Low-Latency
  in Action - - Case Study of {PRINCE} and Midori},'' in \emph{ASIACRYPT 2016},
  ser. Lecture Notes in Computer Science, vol. 10031, 2016, pp. 517--547.

\bibitem{DBLP:conf/ches/MoradiW15}
A.~Moradi and A.~Wild, ``{Assessment of Hiding the Higher-Order Leakages in
  Hardware - What Are the Achievements Versus Overheads?}'' in \emph{CHES
  2015}, ser. Lecture Notes in Computer Science, vol. 9293.\hskip 1em plus
  0.5em minus 0.4em\relax Springer, 2015, pp. 453--474.

\bibitem{DBLP:conf/sacrypt/SasdrichMG15}
P.~Sasdrich, A.~Moradi, and T.~G{\"{u}}neysu, ``{Affine Equivalence and Its
  Application to Tightening Threshold Implementations},'' in \emph{SAC 2015},
  ser. Lecture Notes in Computer Science, vol. 9566.\hskip 1em plus 0.5em minus
  0.4em\relax Springer, 2015, pp. 263--276.

\bibitem{DBLP:conf/ches/BilginBKMW13}
B.~Bilgin, A.~Bogdanov, M.~Knezevic, F.~Mendel, and Q.~Wang, ``{Fides:
  Lightweight Authenticated Cipher with Side-Channel Resistance for Constrained
  Hardware},'' in \emph{CHES 2013}, ser. Lecture Notes in Computer Science,
  vol. 8086.\hskip 1em plus 0.5em minus 0.4em\relax Springer, 2013, pp.
  142--158.

\bibitem{DBLP:conf/ches/BilginNNRS12}
B.~Bilgin, S.~Nikova, V.~Nikov, V.~Rijmen, and G.~St{\"{u}}tz, ``{Threshold
  Implementations of All 3 $\times$ 3 and 4 $\times$ 4 S-Boxes},'' in
  \emph{CHES 2012}, ser. Lecture Notes in Computer Science, vol. 7428.\hskip
  1em plus 0.5em minus 0.4em\relax Springer, 2012, pp. 76--91.

\bibitem{DBLP:conf/cardis/BilginDNNRA13}
B.~Bilgin, J.~Daemen, V.~Nikov, S.~Nikova, V.~Rijmen, and G.~V. Assche,
  ``{Efficient and First-Order {DPA} Resistant Implementations of Keccak},'' in
  \emph{CARDIS 2013}, ser. Lecture Notes in Computer Science, vol. 8419.\hskip
  1em plus 0.5em minus 0.4em\relax Springer, 2014, pp. 187--199.

\bibitem{DBLP:journals/tcad/BilginGNNR15}
B.~Bilgin, B.~Gierlichs, S.~Nikova, V.~Nikov, and V.~Rijmen, ``{Trade-Offs for
  Threshold Implementations Illustrated on {AES}},'' \emph{{IEEE} Trans. on
  {CAD} of Integrated Circuits and Systems}, vol.~34, no.~7, pp. 1188--1200,
  2015.

\bibitem{DBLP:conf/dsd/GrossWDE15}
H.~Gro{\ss}, E.~Wenger, C.~Dobraunig, and C.~Ehrenh{\"{o}}fer, ``{Suit up! -
  Made-to-Measure Hardware Implementations of ASCON},'' in \emph{DSD
  2015}.\hskip 1em plus 0.5em minus 0.4em\relax {IEEE} Computer Society, 2015,
  pp. 645--652.

\bibitem{DBLP:conf/ctrsa/Sasdrich0G17}
P.~Sasdrich, A.~Moradi, and T.~G{\"{u}}neysu, ``{Hiding Higher-Order
  Side-Channel Leakage - Randomizing Cryptographic Implementations in
  Reconfigurable Hardware},'' in \emph{CT-RSA 2017}, ser. Lecture Notes in
  Computer Science, vol. 10159.\hskip 1em plus 0.5em minus 0.4em\relax
  Springer, 2017, pp. 131--146.

\bibitem{DBLP:conf/itc/Smith85}
G.~L. Smith, ``{Model for Delay Faults Based upon Paths},'' in
  \emph{International Test Conference 1985}.\hskip 1em plus 0.5em minus
  0.4em\relax {IEEE} Computer Society, 1985, pp. 342--351.

\bibitem{DBLP:journals/tcad/GuptaKSS06}
P.~Gupta, A.~B. Kahng, P.~Sharma, and D.~Sylvester, ``{Gate-length biasing for
  runtime-leakage control},'' \emph{{IEEE} Trans. on {CAD} of Integrated
  Circuits and Systems}, vol.~25, no.~8, pp. 1475--1485, 2006.

\bibitem{eggersgluss2013improved}
S.~Eggersgl{\"u}{\ss}, R.~Wille, and R.~Drechsler, ``Improved sat-based atpg:
  More constraints, better compaction,'' in \emph{Proceedings of the
  international conference on computer-aided design}.\hskip 1em plus 0.5em
  minus 0.4em\relax IEEE Press, 2013, pp. 85--90.

\bibitem{ga2016matlab}
``{Genetic Algorithm},''
  \url{http://www.mathworks.com/discovery/genetic-algorithm.html}, [Accessed:
  2016-02-01].

\bibitem{ghandali2015low}
S.~Ghandali, B.~Alizadeh, and Z.~Navabi, ``{Low Power Scheduling in High-level
  Synthesis using Dual-Vth Library},'' in \emph{16th International Symposium on
  Quality Electronic Design (ISQED)}, 2015, pp. 507--511.

\bibitem{tang2005leakage}
X.~Tang, H.~Zhou, and P.~Banerjee, ``{Leakage Power Optimization With Dual-Vth
  Library In High-Level Synthesis},'' in \emph{42nd annual Design Automation
  Conference (DAC 2005)}, 2005, pp. 202--207.

\bibitem{DBLP:conf/asiacrypt/EnderG0P17}
M.~Ender, S.~Ghandali, A.~Moradi, and C.~Paar, ``{The First Thorough
  Side-Channel Hardware Trojan},'' in \emph{{ASIACRYPT} 2017}, ser. Lecture
  Notes in Computer Science, vol. 10624.\hskip 1em plus 0.5em minus 0.4em\relax
  Springer, 2017, pp. 755--780.

\bibitem{t_test2}
G.~{G}oodwill, B.~{J}un, J.~{J}affe, and P.~{R}ohatgi, ``{A testing methodology
  for side channel resistance validation},'' in \emph{{NIST} non-invasive
  attack testing workshop}, 2011,
  \url{http://csrc.nist.gov/news_events/non-invasive-attack-testing-workshop/papers/08_Goodwill.pdf}.

\bibitem{ttestCHES15}
T.~Schneider and A.~Moradi, ``{Leakage Assessment Methodology - A Clear Roadmap
  for Side-Channel Evaluations},'' in \emph{CHES 2015}, ser. Lecture Notes in
  Computer Science, vol. 9293.\hskip 1em plus 0.5em minus 0.4em\relax Springer,
  2015, pp. 495--513.

\bibitem{DBLP:journals/tc/ProuffRB09}
E.~Prouff, M.~Rivain, and R.~Bevan, ``{Statistical Analysis of Second Order
  Differential Power Analysis},'' \emph{{IEEE} Trans. Computers}, vol.~58,
  no.~6, pp. 799--811, 2009.

\bibitem{DBLP:journals/tvlsi/EndoLHSOFNKDA15}
S.~Endo, Y.~Li, N.~Homma, K.~Sakiyama, K.~Ohta, D.~Fujimoto, M.~Nagata,
  T.~Katashita, J.~Danger, and T.~Aoki, ``{A Silicon-Level Countermeasure
  Against Fault Sensitivity Analysis and Its Evaluation},'' \emph{{IEEE} Trans.
  {VLSI} Syst.}, vol.~23, no.~8, pp. 1429--1438, 2015.

\end{thebibliography}
